\documentclass[prd,preprintnumbers,floatfix,
nofootinbib,superscriptaddress]{revtex4}
\usepackage{float}
\usepackage{nicefrac}
\usepackage{slashed}
\usepackage{mathtools}
\usepackage{amsfonts} 
\usepackage{amssymb} 
\usepackage{amsmath} 
\usepackage{graphicx} 
\usepackage[caption=false]{subfig}
\usepackage{array} 
\usepackage{dcolumn} 
\usepackage{bm} 
\usepackage{latexsym} 
\usepackage{longtable} 
\usepackage{hyperref} 
\usepackage{verbatim}
\usepackage{epsfig}
\usepackage{xcolor}
\usepackage{color}
\usepackage{cancel}
\usepackage{braket}
\usepackage{xspace}
\usepackage[normalem]{ulem}
\usepackage[export]{adjustbox}
\newcommand{\Ac}[0]{{\mathcal{A}}}

\newcommand{\Gc}[0]{{\mathcal{G}}}
\newcommand{\Hc}[0]{{\mathcal{H}}}
\newcommand{\Ic}[0]{{\mathcal{I}}}
\newcommand{\Jc}[0]{{\mathcal{J}}}
\newcommand{\Kc}[0]{{\mathcal{K}}}
\newcommand{\Lc}[0]{{\mathcal{L}}}
\newcommand{\Mc}[0]{{\mathcal{M}}}

\newcommand{\Rc}[0]{{\mathcal{R}}}

\newcommand{\Wc}[0]{{\mathcal{W}}}
\newcommand{\Yc}[0]{{\mathcal{Y}}}

\newcommand{\Hb}[0]{{\mathbf{H}}}

\newcommand{\Wb}[0]{{\mathbf{W}}}
\newcommand{\pb}[0]{{\mathbf{p}}}
\newcommand{\kb}[0]{{\mathbf{k}}}

\newcommand{\cf}{cf.\xspace}
\newcommand{\eg}{e.g.\xspace}

\newcommand{\ie}{i.e.\xspace}

\newcommand{\df}[0]{{\rm{df}}}
\newcommand{\diff}[0]{{\rm{d}}}
\newcommand{\nn}[0]{\nonumber}

\renewcommand{\vec}[1]{\mathbf{#1}}

\begin{document}
\title{
On-shell representations of two-body transition amplitudes: \\
 single external current
}

\author{Ra\'ul A.~Brice\~no}
\email[e-mail: ]{rbriceno@jlab.org}
\affiliation{Thomas Jefferson National Accelerator Facility, 12000 Jefferson Avenue, Newport News, Virginia 23606, USA}
\affiliation{Department of Physics, Old Dominion University, Norfolk, Virginia 23529, USA}
\author{Andrew W. Jackura}
\email[e-mail: ]{ajackura@odu.edu}
\affiliation{Thomas Jefferson National Accelerator Facility, 12000 Jefferson Avenue, Newport News, Virginia 23606, USA}
\affiliation{Department of Physics, Old Dominion University, Norfolk, Virginia 23529, USA}
\author{Felipe G. Ortega-Gama}
\email[e-mail: ]{fgortegagama@email.wm.edu}
\affiliation{Thomas Jefferson National Accelerator Facility, 12000 Jefferson Avenue, Newport News, Virginia 23606, USA}
\affiliation{Department of Physics, William \& Mary, Williamsburg, Virginia 23187, USA}
\author{Keegan H. Sherman}
\email[e-mail: ]{ksher004@odu.edu}
\affiliation{Thomas Jefferson National Accelerator Facility, 12000 Jefferson Avenue, Newport News, Virginia 23606, USA}
\affiliation{Department of Physics, Old Dominion University, Norfolk, Virginia 23529, USA}

\preprint{JLAB-THY-20-3298}

\date{\today}
\begin{abstract}

This work explores scattering amplitudes that couple two-particle systems via a single external current insertion, $2+\mathcal{J}\to 2$. Such amplitudes can provide structural information about the excited QCD spectrum. 
We derive an exact analytic representation for these reactions.
From these amplitudes, we show how to rigorously define resonance and bound-state form-factors.
Furthermore, we explore the consequences of the narrow-width limit of the amplitudes as well as the role of the Ward-Takahashi identity for conserved vector currents.
These results hold for any number of two-body channels with no intrinsic spin, and a current with arbitrary Lorentz structure and quantum numbers.
This work and the existing finite-volume formalism provide a complete framework for determining this class of amplitudes from lattice QCD.

\end{abstract}
\maketitle
%

\section{Introduction \label{sec:intro}}

Resolving the hadronic spectrum has proven to be a significant challenge due to the non-perturbative nature of Quantum Chromodynamics (QCD). In the case of the lowest-lying spinless hadrons, the pseudoscalar pions can be readily identified as the pseudo-Goldstone bosons of chiral symmetry; however, the scalar hadrons are notoriously difficult to characterize. This is not surprising given the multitude of Fock states allowed to participate in this channel, i.e.\ quark-antiquark pairs, mesonic molecules, tetraquarks, glueball states, etc.~\footnote{A dedicated review from the PDG discusses tentative descriptions of scalar mesons below 2~GeV \cite{10.1093/ptep/ptaa104}.}  A satisfactory interpretation of these states demands for a more comprehensive understanding of the dynamics of QCD. 

For example, the determination of the mass and width of the $f_0(500)/\sigma$, the lightest QCD resonance, had been disputed since its discovery, and only recently has reached consensus  \cite{PELAEZ20161}. The difficulty to study this state arises in part due to its large decay width and the atypical shape of the cross section of its decay products, i.e.\ $\pi\pi$.
However, the nature of this state is not resolved from its mass and width alone, motivating the attention to other physical properties, like the charge radius or distribution functions, which naturally arise in transition amplitudes.

With this in mind, Ref.~\cite{PhysRevD.86.034003} calculated the $\pi\pi+H\to \pi\pi$ transition with unitarized chiral perturbation theory ($\chi$PT), where $H$ represents a scalar current. The scalar radius of the $\sigma$ was in turn estimated by analytically continuing the amplitude to the resonance position. The value found for this parameter supports an interpretation of this resonance as a compact state for pions at their physical mass, and a molecular $\pi\pi$ description if the quark masses are modified such that the pion mass is greater than $400$~MeV.~\footnote{Independent evidence of the non-compactness of the $\sigma$ for $m_\pi\sim 400$~MeV has also been observed in lattice QCD studies~\cite{Briceno:2016mjc, Briceno:2017qmb}.} This demonstrates that transition amplitudes can play a role in the description of resonances. The question that arises is how to determine these amplitudes directly from the dynamics of QCD, and the best current answer is lattice QCD.

Lattice QCD is a numerical implementation of the path integral in a finite volume, and can be used to calculate observables directly from QCD.
In the past decades the scope of the field has increased substantially, moving past the studies of stable ground states into the more interesting region of resonances and excited states.
Studies of excited and multiparticle states are challenging because of, among other things, the need for a formal connection between finite- and infinite-volume states, and matrix elements.
In the case of scattering amplitudes, the L{\"u}scher formalism and its extensions \cite{Luscher:1986pf,Rummukainen:1995vs,Kim:2005gf,Fu:2011xz,He:2005ey,Lage:2009zv,Bernard:2010fp,Briceno:2012yi,Hansen:2012tf,Feng:2004ua,Gockeler:2012yj,Briceno:2014oea,Morningstar:2017spu,2012PhRvD..85k4507L} have been tested and applied successfully in numerous processes, see the recent review~\cite{2018RvMP...90b5001B} and references therein. This includes determinations of the $\sigma$ mass and width \cite{Briceno:2016mjc, Briceno:2017qmb, Guo:2018zss} all the way to resonances that involve multiple coupled channels and partial waves~\cite{Dudek:2014qha, Wilson:2014cna,Dudek:2016cru, Woss:2018irj,Woss:2019hse,Moir:2016srx}, a remarkable example is the recent study of the $1^{-+}$ hybrid resonance \cite{2020arXiv200910034W}. 

Furthermore, the technology to compute \emph{transition matrix elements} involving excited states from the lattice has already been implemented and employed \cite{2015PhRvD..91k4501S}.
In addition to this, the seminal work in Ref.~\cite{Lellouch:2000pv} by Lellouch and L{\"u}scher laid the foundation to develop a general technique to match finite-volume matrix elements to $1+\Jc\to 2$ transition processes \cite{Briceno:2014uqa,Briceno:2015csa}, where $\Jc$ is some external local current, $1$ refers to a state of just a QCD-stable hadron and $2$ is an asymptotic state of two hadrons.
An application of this formalism was used to calculate the pion photoproduction in the $\pi+\gamma^\star\to\pi\pi$ process, from which the $\pi+\gamma^\star\to\rho$ transition form-factor was determined for heavier-than-physical pions by two distinct groups~\cite{Briceno:2016kkp,Alexandrou:2018jbt}.
Carrying on this effort, some of the authors developed a framework that addresses the finite-volume effects of amplitudes with two hadrons in the initial \emph{and} final states, i.e.~$2+\Jc\to2$~\cite{Briceno:2015tza,Baroni:2018iau}.
It is precisely through these amplitudes that \emph{elastic form-factors} of resonant or shallow bound states can be determined.

The purpose of this work is to complement this technique, relevant when translating finite-volume matrix elements into infinite-volume amplitudes, by deriving the universal analytic structure that the $2+\Jc\to2$ amplitudes receive from Lorentz symmetry and unitarity in an infinite-volume. This is especially important when evaluating the amplitude in the complex energy plane where resonances and bound poles reside. This framework is also applicable for the case of non-resonant amplitudes. In the latter case, understanding the analytic structure is critical in order to prevent the incorrect identification of kinematic singularities as dynamical poles.

We begin by considering processes with only one open two-hadron channel, in an arbitrary partial-wave $\ell$. Then, we show how the generalization to an arbitrary number of two-hadron channels is straightforward. Whenever possible we will ignore subtleties associated with the spin of the hadrons in the initial and final states, letting the total angular momentum of the two-particle states be equal to $\ell$. We will keep the masses of the hadrons to be distinct throughout. For the sake of generality, we leave the Lorentz structure, e.g.\ scalar, vector, etc., of the current as generic whenever possible.

The formalism we exploit relies on generic properties of a quantum field theory based on self-consistent integral equations for the off-shell $2\to2$, $1+\Jc \to 2$, and $2+\Jc\to2$ amplitudes. Our main results are presented in Sec.~\ref{sec:main}, where we summarize the \emph{on-shell} representation of each amplitude. After that, in Sec.~\ref{sec:resonances} we investigate the implication of our results for resonances. We use our formalism and the Ward-Takahashi Identity to show that the charge of a resonance is protected to be the sum of the charge of its decay products. We also investigate the  narrow-width limit of the resonance as a consistency cross-check.

The derivation of our results is presented in Sec.~\ref{sec:derivation}. First, in Sec.~\ref{sec:2to2} we recover the well-known analytic structure of the two-body scattering amplitude, which is also a direct consequence of unitarity. We use the fact that we are interested in a limited range of kinematics where only two-particle states can go on-shell. It is by separating the singularities that appear at each order in the two-particle loops that we can express the amplitudes to all-orders in terms of kinematic functions that contain all the non-analytic behavior, and real functions that encode the short-distance dynamics. Finally, we project the resulting equation on-shell, and partial-wave expand to yield amplitudes of definite angular momentum. After that, in Sec.~\ref{sec:1Jto2} we use this technique to recover the analytic form of the $1+\Jc \to 2$ amplitudes with any number of two-hadron coupled channels.

The derivation of the main result of this work is presented in Sec.~\ref{sec:2Jto2}, where we apply the aforementioned formalism to the $2+\Jc \to 2$ amplitude. Closely related techniques were used in Refs.~\cite{Briceno:2014uqa,Briceno:2015tza} to study the finite-volume analogues of these reactions. Non-trivial checks of this formalism have been carried out~\cite{Briceno:2019nns,Briceno:2020xxs}. We dedicate subsection~\ref{sec:comparison} to highlighting the novelty of our result with respect to what has been done in past work. Finally, we summarize our results in Sec.~\ref{sec:conclusion}.

\section{Analytic representation of amplitudes}
\label{sec:main}

The remainder of this work proceeds to present exact forms of two-body hadronic amplitudes involving a single current insertion. For the sake of completeness, we consider all amplitudes of the form $n+\Jc\to m$ with $n$~and~$m$ less or equal to two. We use all-orders perturbation theory to treat the hadronic contributions non-perturbatively within a generic effective field theory (EFT). 
Furthermore, since we focus on the on-shell behavior of amplitudes, our procedure is independent of the specifics about couplings or renormalization scheme, which are encoded into unknown short-distance functions.
In the absence of insertions of external currents, all-orders perturbation theory provides results that are consistent with unitarity constraints.~\footnote{Although evident for two-body systems, this was proven for three-particle systems in Refs.~\cite{Jackura:2019bmu, Briceno:2019muc}, where it was shown that previous results describing three-body amplitudes obtained using all-orders perturbation theory~\cite{Hansen:2015zga} and unitarity constraints~\cite{Jackura:2018xnx} were consistent.} In the presence of external currents, this provides a systematic procedure to asses the singularity structure of the resultant amplitudes. 

Here we present the final results and leave the derivation for Sec.~\ref{sec:derivation}. In arriving at these results, we make only two assumptions throughout the work. First, that the asymptotic particles considered, which will be referred to as hadrons,~\footnote{Even though our motivation is to understand reactions within QCD, we make no reference to the underlying theory.} carry no intrinsic spin. In other words, they can be either scalars or pseudoscalars. Second, we assume the energies are above the lowest-lying two-particle threshold and below the first unaccounted inelastic threshold, e.g. the three-particle threshold.
This means that the results hold for generic external currents and that the kinematics can be such that any number of two-particle states may go on-shell.

The number of classes of singularities and consequently the complexity of the amplitudes grows with the number of external current insertions and particles. However, the majority of these singularities are common across these amplitudes. As a result, the singularity structure of more complicated amplitudes can be written in terms of simpler ones representing subprocesses. 

\begin{figure}[t]
\begin{center}
\includegraphics[width=.8\textwidth]{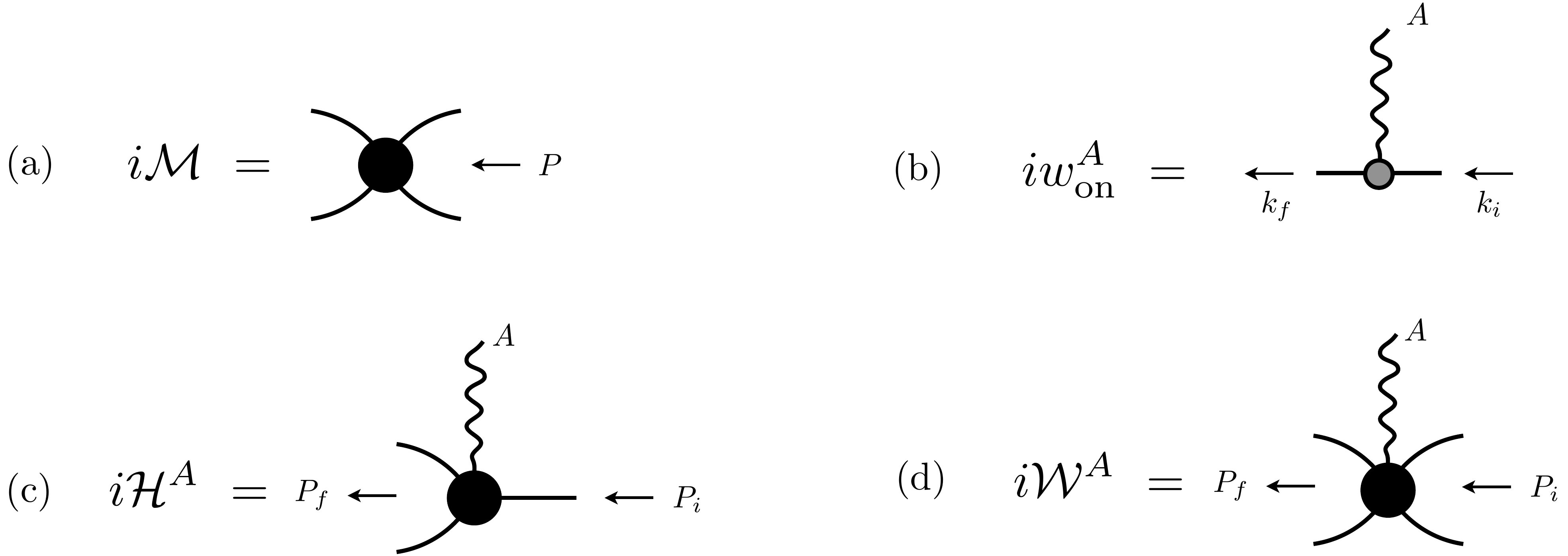}
\caption{Diagrammatic representation of the amplitudes considered in this work. Shown are the (a) $2\to  2$, (b) $1+\Jc\to 1$, (c) $1+\Jc\to 2$, and (d) $2+\Jc\to 2$ amplitudes along with momentum assignments.}
\label{fig:amplitudes}
\end{center}
\end{figure}

With this in mind, it is convenient to categorize the amplitudes according to the number of currents that are considered. We begin with the two-body scattering amplitude with no external currents, which we label as $\Mc$. First, we show the case of a single channel system, and then discuss the extensions to multiple scattering channels. In Sec.~\ref{sec:2to2}, we prove the well-known result that the on-shell partial-wave scattering amplitude can be written in the form
\begin{align}
\label{eq:2to2.M_on-shell}
i\Mc(s)
&=
i\Kc(s) \,
\frac{1}{1-i \rho \, \Kc(s)} \, ,
\end{align}
which is a matrix equation that has elements $\Mc_{\ell' m_{\ell'};\ell m_{\ell}}$, where $\ell$ is the angular momentum between the two particles defined in their center-of-momentum (CM) frame, and $m_{\ell}$ is its projection onto some fixed axis.~\footnote{Semi-colons in matrix elements separate initial and final state indices.}
Figure~\ref{fig:amplitudes}~(a) shows a diagrammatic representation of the amplitude.
Here $s=P^{2}$ is the usual Mandelstam invariant with $P$ being the four-momentum of the system, and $\Kc$ is the two-body $K$ matrix which is a real function in our kinematic region of interest.
In general, this function contains branch points associated with crossed-channel processes and multi-particle thresholds, but since these are   far from our kinematic region they can be described by smooth contributions. For the sake of brevity we will denote these as \emph{smooth} throughout the text.

Finally, $\rho$ is the phase space which is defined in the standard way for a single channel,
\begin{align}
\label{eq:ps}
\rho_{\ell' m_{\ell'};\ell m_{\ell}} 
&=
\delta_{\ell'\ell}\, \delta_{m_{\ell'}m_{\ell}}
\frac{\xi \, q^{\star}}{8\pi \sqrt{s}} \equiv \delta_{\ell'\ell}\, \delta_{m_{\ell'}m_{\ell}} \, \rho_0 \, ,
\end{align}
where $\xi$ is a symmetry factor which is defined to be $\xi = 1/2$ if the two scattering particles are identical and $\xi = 1$ otherwise, and $q^{\star}$ is the relative momentum between the two particles in their CM frame, 
\begin{align}
\label{eq:q_elastic}
q^{\star} = \frac{1}{2\sqrt{s}} \lambda^{1/2}(s,m_1^2,m_2^2) \, ,
\end{align}
where $\lambda(a,b,c) = a^2 + b^2 + c^2 - 2(ab + bc + ca)$ is the K\"all\'en triangle function, and $m_1$ and $m_2$ are the two masses in the channel considered.

Rotational invariance implies that the amplitude is diagonal in $\ell$ and independent of $m_{\ell}$, which reduces Eq.~\eqref{eq:2to2.M_on-shell} to a single algebraic relation for each partial-wave, i.e. $\Mc_{\ell' m_{\ell'};\ell m_{\ell}} = \delta_{\ell'\ell}\delta_{m_{\ell'}m_{\ell}} \Mc_{\ell}$ and similarly for $\Kc$.
The on-shell behavior of the scattering amplitude is fixed by $S$ matrix unitarity, which fixes the non-analytic behavior of the amplitude originating from direct channel pair production, as indicated by Eq.~\eqref{eq:ps}. 
However, kinematic singularities may remain due to the projection into the angular momentum basis, as discussed in Sec.~\ref{sec:2to2}, which requires that the amplitude $\Mc_{\ell}$ has a barrier suppression $\Mc_{\ell} \sim q^{\star\,2\ell}$ near threshold.
This implies that the $K$ matrix has the same threshold behavior.

For kinematics where multiple two-body channels are open, in Sec.~\ref{sec:2to2.coupled_channels} we show that these objects can easily be upgraded into matrices over the channel index. First, the masses of the particles would acquire an additional index to identify the individual particles in a given channel. We label the two particles in channel $``a"$ to have masses $m_{a1}$ and $m_{a2}$. The $K$ matrix and phase space factor would both be matrices in channel space with components $\Kc_{\ell' m_{\ell'};\ell m_{\ell}}\to \Kc_{a \ell' m_{\ell'};b \ell m_{\ell}}$ and $\rho_{\ell' m_{\ell'};\ell m_{\ell}}\to \rho_{a \ell' m_{\ell'};b \ell m_{\ell}}$, respectively. The phase space matrix $\rho$ would be diagonal in this space with elements defined as
\begin{align}
\label{eq:rho_cc}
\rho_{a \ell' m_{\ell'};b \ell m_{\ell}}
&=
\delta_{\ell'\ell}\, \delta_{m_{\ell'}m_{\ell}}\, \delta_{ab}
\frac{\xi_a \, q_a^{\star}}{8\pi \sqrt{s}} \, ,
\end{align}
where $q^{\star}_a =  \lambda^{1/2}(s,m_{a1}^2,m_{a2}^2)/2\sqrt{s}$ and $\xi_{a}$ is the symmetry factor for each channel $a$.
Therefore, Eq.~\eqref{eq:2to2.M_on-shell} becomes an enlarged matrix within this channel space, which has all the properties of the amplitudes as before, except the barrier suppression is now channel-dependent, 
$\Mc_{a\ell' m_{\ell'};b\ell m_{\ell}} \sim q_{a}^{\star\,\ell'} q_{b}^{\star\,\ell}$, where the dominant singular behavior is the lightest threshold.

%

\subsection{Amplitudes with a single current insertion}
\label{sec:results_one_current}
Having shown the well-known result for the two-body hadronic amplitude, we proceed with the description of transitions induced by an external local current insertion. As stated above, we make no assumptions about the quantum numbers of the current, which we denote as $\Jc$. In particular, the current can have an arbitrary Lorentz structure with indices $\mu_1\cdots\mu_N$. Furthermore, other indices can include quantum numbers associated with, for example, flavor-changing processes. For simplicity, we will adopt a notation similar to that in Ref.~\cite{Briceno:2019opb} where all of the quantum numbers of the current, including the Lorentz indices, are absorbed into a single index $A$. 

The simplest of these amplitudes is the one where an external current couples to a one-particle state of mass $m$, i.e.\ $1+\mathcal{J}\to1$, which we denote as $w_\mathrm{on}$ and show diagrammatically in Fig.~\ref{fig:amplitudes}~(b).
This amplitude is given by 
\begin{align}
	\label{eq:w_on}
    w^{A}_{\mathrm{on}}(k_{f},k_{i}) 
     = \sum_{j} K_j^{A}(k_f,k_i) \, f_j(Q^2) \, ,
\end{align}
where $k_{i}/k_{f}$ are the initial/final four-momentum of the single particle.
We write $w_{\mathrm{on}}$ in terms of kinematic functions, $K$, dictated by the Lorentz structure of the current and Lorentz invariant form-factors, $f$, which depend on \mbox{$Q^{2}=-(k_{f}-k_{i})^{2}$}. For a given current, there will always be a finite number of these form-factors, for which we will label the $j$-th form-factor as $f_j$.~\footnote{Since they do not possess Lorentz structure we drop the index $A$, but leave implicit their possible dependence on the current internal quantum numbers.}
The subscript ``on'' indicates that the form-factors are on-shell, while the kinematic tensor does not have to be.
The arbitrariness of the current allows for the particle species to change;
however, we focus only on kinematic regions where the form-factors are analytic functions of $Q^2$, i.e.\ above any pole of a state coupling to the current or particle production thresholds. When $k_i^2 = k_f^2 = m^2$, then Eq.~\eqref{eq:w_on} gives the single-particle matrix element $\bra{k_{f}} \Jc^A(x=0) \ket{k_{i}}$.

Moving up in complexity, the next amplitude we consider is one where the current generates a transition between a one- and two-particle state, i.e. $1+\mathcal{J}\to2$.
This amplitude, which we label as $\Hc$ and show in Fig.~\ref{fig:amplitudes}~(c), has been previously studied in Ref.~\cite{Briceno:2014uqa} for any number of channels. 
In Sec.~\ref{sec:1Jto2} we reproduce the finding that this amplitude has an on-shell form given by
\begin{align}
\label{eq:1Jto2.H_on-shell}
i\Hc^{A} (P_f,P_i)
&=
i\Mc (s_f) \mathcal{A}^{A} _{21}  (P_f,P_i) \, ,
\end{align}
where $P_{i}$ and $P_{f}$ are the four-momenta of the initial single-particle state and the final two-particle system, respectively, with $s_f = P_f^2$. The function $\Ac_{21}$ is \emph{real and smooth}~\footnote{up to barrier factors associated with partial-wave projections.} in $s_f$, with the same caveats described earlier for the $K$-matrix, and characterizes the short-distance dynamics. Additionally, it contains the same type of singularities in $Q^2$ that appear in $f_j$, these again can be described by smooth functions within our kinematic domain.
The indices on $\Ac_{21}$ refer to the number of hadrons coupling to this short-distance function. Unlike the $2\to 2$ amplitude, both $\Hc$ and $\Ac_{21}$ are Lorentz tensors which can be written in terms of kinematic pre-factors and energy-dependent form-factors, similar to the construction in Eq.~\eqref{eq:w_on}, that depend on the final state angular momentum $\ell$ as well as its projection $m_{\ell}$.

This expression makes explicit that $\Hc$ inherits the analytic structure of $\Mc$. 
For kinematics where a single channel is open, this is nothing more than the manifestation of Watson's final state theorem~\cite{Watson:1952}. As with $\Mc$, near threshold the angular momentum decomposition of $\Hc$ requires that $\Hc_{\ell m_{\ell}} \sim q_f^{\star\,\ell}$, where $q_f^{\star}$ is as in Eq.~\eqref{eq:q_elastic} with $s_f$.
In conjunction with $\Mc_{\ell} \sim q_f^{\star\,2\ell}$, this implies that any parameterization of the $\Ac_{21}$ amplitudes necessitates a barrier enhancement of the form $\Ac_{21,\ell m_{\ell}} \sim q_f^{\star\,-\ell}$ at threshold.
For multiple open channels, Eq.~\eqref{eq:1Jto2.H_on-shell} naturally extends such that $\Hc$ and $\Ac_{21}$ become vectors in channel space.
For further discussion on these aspects of $\Hc$ see Sec.~\ref{sec:1Jto2}.

\begin{figure}[t]
\begin{center}
\includegraphics[width=.7\textwidth]{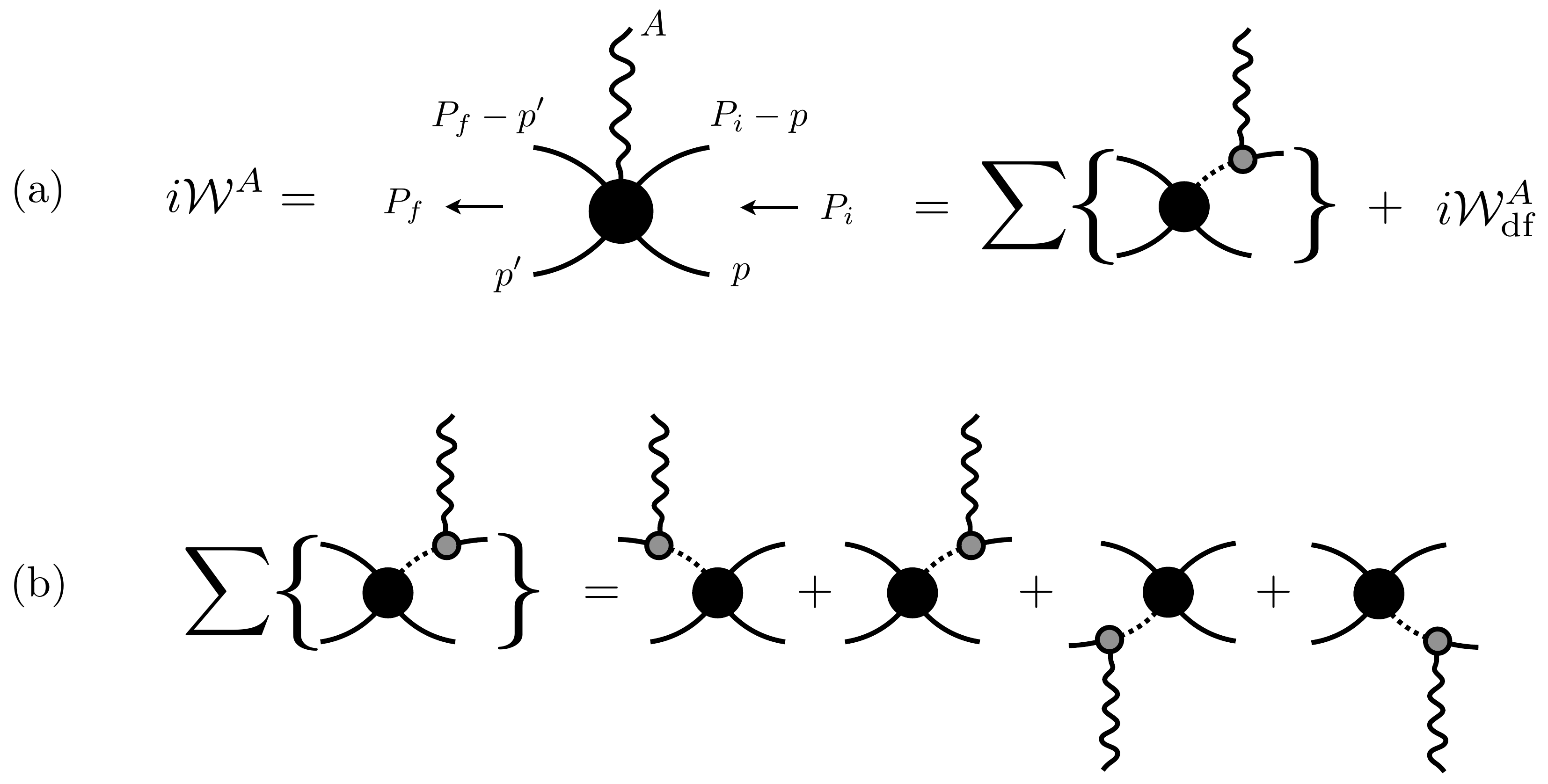}
\caption{(a) The relation between the full $2+\Jc\to 2$ amplitude and $\Wc_{\df}$ as defined in Eq.~\ref{eq:W_def}.  (b) Depicted are all allowed current insertions over the external legs. The symbol ``$\Sigma$" is used as a shorthand to express this sum. The dotted lines represent the pole piece $D$ of the propagator, the other objects are defined in Fig.~\ref{fig:amplitudes}.}
\label{fig:Sigma.Ex}
\end{center}
\end{figure}

The main original result presented here is the $2+\mathcal{J}\to2$ amplitude. 
These amplitudes were introduced in Refs.~\cite{Briceno:2015tza, Baroni:2018iau} with the goal of trying to obtain them using lattice QCD.
These studies were interested in finding a non-perturbative relation between the desired amplitudes and the finite-volume matrix elements that can be accessed via lattice QCD. In Sec.~\ref{sec:2Jto2} we derive the exact analytic form that these amplitudes must take.
We define the $2+\Jc\to 2$ amplitude, which we label as $\Wc$, via the matrix element
\begin{align}
\label{eq:2Jto2.W_ME_def}
\mathcal{W}^{A}(P_f,\hat{\mathbf{p}}_f'^{\star};P_i,\hat{\mathbf{p}}^{\star}_i)
& \equiv
\bra{P_f,\hat{\mathbf{p}}'^{\star}_f; \mathrm{out} } \mathcal{J}^{A}(0) \ket{P_i,\hat{\mathbf{p}}^{\star}_i ; \mathrm{in}}_{\rm conn} \, ,
\end{align}
where the initial asymptotic two-particle state depends on the total four-momentum $P_{i}$, as well as the orientation~$\hat{\mathbf{p}}^{\star}_i$ of the relative momentum between the two particles in their CM frame, and similarly for the final state defined in its own CM frame.
The subscript ``conn" highlights that we only consider connected diagrams, i.e.\ topologies where the hadrons do not interact with each other or with the current are not included. Figure~\ref{fig:amplitudes}~(d) shows a diagrammatic representation of the $\Wc$ amplitude, while Fig.~\ref{fig:Sigma.Ex}~(a) shows the momentum flow where we adopt the convention that the first and second particle have momenta $p$ and $P_i-p$ for the initial state, respectively, and $p'$ and $P_f-p'$ for the first and second particle in the final state, respectively. As mentioned before, we are focused on the kinematic region below three or more particle thresholds for both the initial and final two-particle states.
Additionally, we restrict the momentum transfer $Q^{2}=-(P_{f}-P_{i})^{2}$ of the current, such that we do not probe any multi-particle production threshold in the $Q^{2}$ channel.

In Sec.~\ref{sec:2Jto2} we derive that the exact analytic form can be separated into two types of terms depending on whether or not they contain single-particle poles associated with the current probing an external leg,
\begin{align}
\label{eq:W_def}
i\Wc^{A}(P_f,\hat{\mathbf{p}}_f'^{\star};P_i,\hat{\mathbf{p}}_i^{\star}) &= 
\sum \left\{
iw_{\mathrm{on}}^{A}\, iD\, i\overline{\Mc}
\right\}
+
i\Wc_{\df}^{A}(P_f,\hat{\mathbf{p}}_f'^{\star};P_i,\hat{\mathbf{p}}_i^{\star})\, .
\end{align}
Starting with the first term, which represents the case in which the current probes an external leg,
$w_{\mathrm{on}}$ is as defined in Eq.~\eqref{eq:w_on} and $D$ is the pole piece of the fully-dressed single-particle propagator which, for a particle with mass $m_{\alpha}$ with $\alpha=1,2$, can be written as
\begin{align}
\label{eq:pole_D}
iD_\alpha(k)=
\frac{i}{k^{2}-m_{\alpha}^{2}+i\epsilon}\, .
\end{align}
The $\overline{\Mc}$ was introduced in Refs.~\cite{Briceno:2015tza, Baroni:2018iau}, and is the full $2\to 2$ scattering amplitude which has additional barrier factors in its partial wave expansion to cancel out singularities of the spherical harmonics at threshold.
For the lowest partial wave, $\ell=0$, $\Mc$ and $\overline{\Mc}$ are identical. We give the exact definition of this in Sec.~\ref{sec:W1B} in Eq.~\eqref{eq:Mbar}. Finally, the symbol $\sum$ reminds one to \emph{sum} over all allowed insertions of the current over the external legs of the amplitude, illustrated in Fig.~\ref{fig:Sigma.Ex}~(b). For example, in the case where the current only couples to particle 2, only the first two diagrams of Fig.~\ref{fig:Sigma.Ex}~(b) contribute, written explicitly as
\begin{align}
\label{eq:Sumpart2}
\sum \left\{
iw_{\mathrm{on}}^{A}\, iD\, i\overline{\Mc}
\right\}
 &= 
iw_{\mathrm{on},2}^{A}(p_f',p_i') iD_2(p_i') i\overline{\Mc}({\pb}_i'^{\star},\hat{\mathbf{p}}_i^{\star})
+
i\overline{\Mc}(\hat{\mathbf{p}}_f'^{\star},{\pb}_f^{\star}) iD_2(p_f) iw^A_{\mathrm{on},2}(p_f,p_i) \, ,
\end{align}
where we use the notation $p_{i/f}^{(\prime)}\equiv  P_{i/f}-p^{(\prime)}$, and the three-vectors ${\pb}_{i}'^{\star}$ and ${\pb}_{f}^{\star}$ are the spatial part of the four-vectors $p^{\prime\mu}$ and $p^\mu$, respectively, when evaluated in the CM frame indicated by the subscript.
The quantity $w_{\mathrm{on},2}$ is the elastic matrix element of particle 2. Equation~\eqref{eq:Sumpart2} diverges whenever the four-vector in either $D$ goes on shell, for example when $P_f=P_i$.

So far we have discussed the single-pole contribution to $\Wc$, which can diverge for physical kinematics and as we have already seen, these singularities are completely described by simpler amplitudes. The more phenomenologically interesting component of $\Wc$ is the second term in Eq.~\eqref{eq:W_def}, which is appropriately labeled with a subscript ``$\df$'', meaning \emph{divergence free}. In Sec.~\ref{sec:2Jto2} we prove that it can be written in an on-shell partial-wave projected form as,
\begin{align}
i\Wc_{\df}^{A}(P_f,P_i) 
= 
\Mc (s_f) \, \left[ \,  i\Ac_{22}^{A}(P_f,P_i) 
+ \sum_{j,\alpha} if_{j,\alpha}(Q^2) \Gc_{j,\alpha}^{A}(P_f,P_i)  \, \right] \, \Mc(s_i) \, ,
\label{eq:Wdf_def}
\end{align}
where $\Ac_{22}$ is a \emph{real and smooth function} in both $s_i$ and $s_f$, up to barrier factors with the same caveats as $\Kc$ and $\Ac_{21}$. The symbol $f_{j,\alpha}$ is the $j$-th form-factor of the $\alpha$-th particle as defined in Eq.~\eqref{eq:w_on}, and $\Gc_{j,\alpha}$ is a kinematic function to be described shortly. Unlike the scattering amplitude Eq.~\eqref{eq:2to2.M_on-shell}, $\Wc_{\df}$ is in general a dense matrix in ($\ell$,$m_{\ell}$)-space since the current can inject angular momentum.
Similar to $\Ac_{21}$, defined in Eq.~\eqref{eq:1Jto2.H_on-shell}, $\Ac_{22}$ contains barrier enhancements near threshold, with this case being $\Ac_{22;\ell' m_{\ell'} ; \ell m_{\ell}} \sim ( q_f^{\star} )^{-\ell'}(q_i^{\star})^{-\ell}$.
This function is unknown, and can be parameterized with energy-dependent form-factors which can be determined, \eg via lattice QCD calculations using the formalism presented in Refs.~\cite{Briceno:2015tza, Baroni:2018iau}.

\begin{figure}[t]
\begin{center}
\includegraphics[width=.6\textwidth]{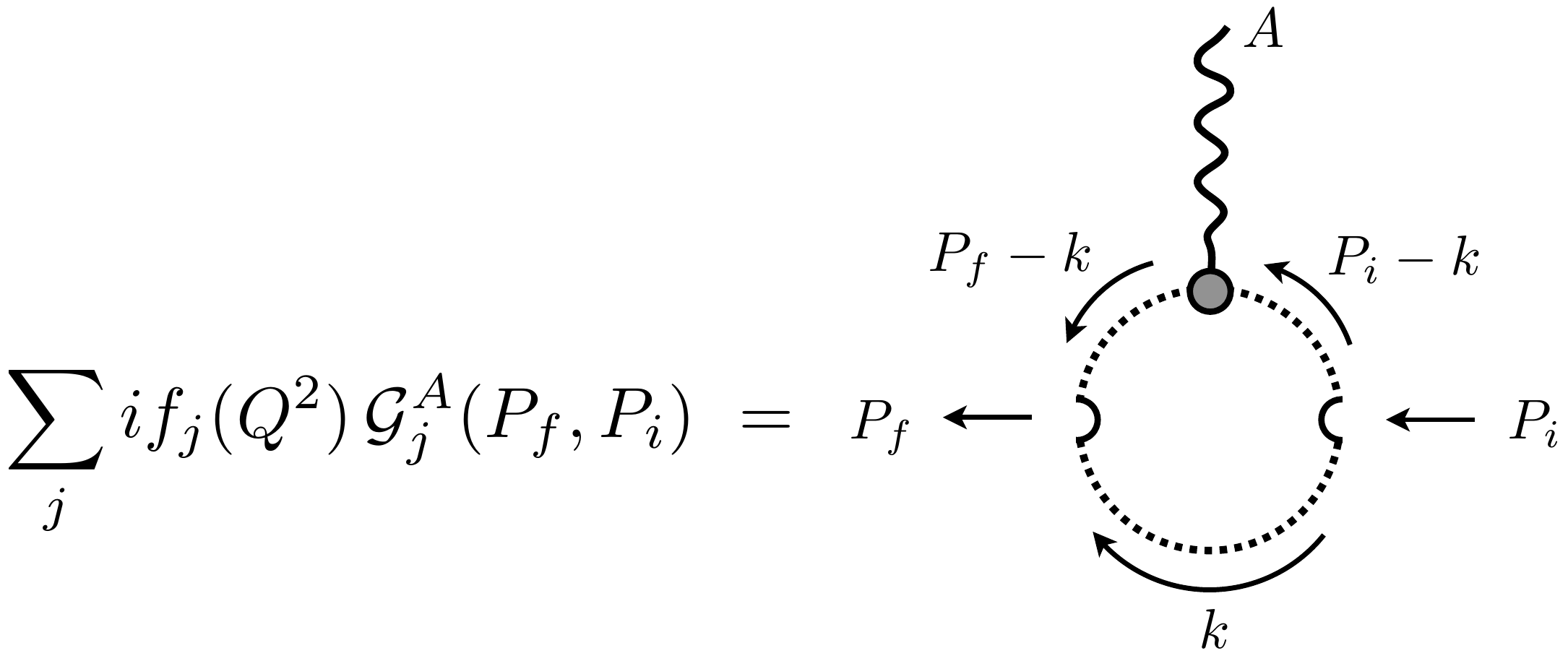}
\caption{ Triangle-function contribution to $\Wc_{\df}$, Eq.~\eqref{eq:Wdf_def}, written in terms of  the single-hadron form-factors ($f_j$) and the triangle loops ($\Gc_j$) defined in Eq.~\eqref{eq:2Jto2.G}. The gray circle and dashed lines were defined in Figs.~\ref{fig:amplitudes} and~\ref{fig:Sigma.Ex}, respectively. The open semi-circles represent the modified spherical harmonics, defined in Eq.~\eqref{eq:sph_w_barrier}, which have the angular momenta associated with the initial and final two-particle states.}
\label{fig:triangle}
\end{center}
\end{figure}

The only quantity not yet defined is the triangle function $\Gc_{j,\alpha}$, diagrammatically shown in Fig.~\ref{fig:triangle}, which occurs when the current probes either particle 1 or particle 2 in the intermediate state, and for example with $\alpha =2$ has matrix elements given by 
\begin{align}
\label{eq:2Jto2.G}
\Gc_{j,2; \ell' m_{\ell'} ; \ell m_{\ell}}^{A} (P_f,P_i) 
&\equiv 
\int\! \frac{\diff^4k}{(2\pi)^4}  
 \,  \frac{ \mathcal Y_{\ell' m_{\ell'}}^{*}({\mathbf{k}}_f ^{\star} )  \,\,\, iK_{j,2}^{A}( k_{f},k_{i})  \,\,\, \mathcal{Y}_{\ell m_{\ell}}({\mathbf{k}}_i^{\star} ) }{(k^2 - m_1^2 + i\epsilon)(k_f^2 - m_2^2 + i\epsilon) (k_i^2 - m_2^2 + i\epsilon)} \, ,
\end{align}
where $k_{i/f}\equiv P_{i/f}-k$ and $\mathbf{k}^{\star}_{i/f}$ is the spatial part of the four-vector $k^\mu$ in the initial/final CM frame, and $K$ are the kinematic functions defined in Eq.~\eqref{eq:w_on}.
The symbol $\Yc_{\ell,m_{\ell}}$ contains a spherical harmonic multiplied by the necessary barrier factor to cancel its singular behavior as $k^{\star}\equiv|\mathbf{k}^{\star}| \to 0$~\cite{Briceno:2015tza, Baroni:2018iau},
\begin{equation}
\label{eq:sph_w_barrier}
\mathcal{Y}_{\ell m_{\ell}}({\mathbf{k}} ^{\star} )
=
\sqrt{4\pi} \, Y_{\ell m_{\ell}}(\hat{\mathbf{k}} ^{\star} )
 \, \left(  \frac{k^{\star}}{q^{\star}} \right)^{\ell} \,.
\end{equation}
The triangle function when $\alpha =1$ is written as is shown in Eq.~\eqref{eq:2Jto2.G} but with 1 and 2 switched. In general, Eq.~\eqref{eq:2Jto2.G} is UV divergent and requires some regularization procedure. For a given scheme, $\Ac_{22}$ will compensate this choice such that $\Wc_{\df}$ remains finite and scheme independent. Note that if the particles are identical, then there is no sum over the particle index $\alpha$ in Eq.~\eqref{eq:Wdf_def}.

In addition to having threshold singularities, this kinematic function also has a new class of singularities, known as the triangle singularities~\cite{Landau:1959fi}. The triangle singularities have a logarithmic behavior which we summarize here and give a full description in Appendix~\ref{app:trian}.
For example, in the case where $\alpha=2$, both the initial and final states are in $S$ wave, and for a scalar current with $K_{j}=1$, the triangle function is given by,
\begin{equation}\label{eq:G0000}
\,\Gc_{00;00}(P_f,P_i) = \frac{i}{32\pi \sqrt{(P_f\cdot P_i)^2-P_i^2P_f^2}}\,
\left[\log\left({\frac{1+z^\star_f +i\epsilon}{1-(z^\star_f+i\epsilon)}}\right)
+
\log\left({\frac{1+z^\star_i+i\epsilon}{1-(z^\star_i+i\epsilon)}}\right)\right]
+\dots\,,
\end{equation}
where the ellipsis represents non-singular terms, $z^{\star}_{f}$ is a function of $P_{i}$, $P_{f}$, and the masses of the external particles, defined in Eq.~\eqref{eq:zfstarcov} and $z^{\star}_{i}$ is the same function but with the labels $f$ and $i$ switched. The logarithmic singularities can be seen as divergences in the imaginary part, while the real part exhibits a discontinuity at the same energy. This energy is the point at which all three particles in the triangle are able to go on-shell simultaneously.

If there are multiple open scattering channels, then Eqs.~\eqref{eq:W_def} and~\eqref{eq:Wdf_def} generalize to matrices in channel space, e.g. $\Wc_{\ell' m_{\ell'};\ell m_{\ell}}\to\Wc_{a \ell' m_{\ell'};b \ell m_{\ell}}$ where $a$ and $b$ are channel indices. 
In general the two-hadron form-factor $\Ac_{22}$ is a dense matrix in channel space while $\sum_j f_j\Gc_j$ is a dense matrix if the current allows for species transmutation, for example flavor changing processes ($\pi + W^{\star} \to K$) or radiative transitions ($\eta+\gamma\to \eta'$). If the current does not change species, then $\sum_j f_j\Gc_j$ is diagonal in channel space, similar to the two-body phase space factor in Eq.~\eqref{eq:rho_cc}, since the off-diagonal elements of $f_{j;a;b}$ are zero. For example, assuming $\ell'=\ell=0$ and $\alpha=2$, then Eq.~\eqref{eq:2Jto2.G} is modified as
\begin{align}
\label{eq:2Jto2.G.mchannel}
\Gc_{j,2;a;b}^{A} (P_f,P_i) 
&\equiv 
\delta_{a1,b1}
\int\! \frac{\diff^4k}{(2\pi)^4}  
 \,  \frac{ iK_{j,2}^{A}( k_{f},k_{i})}{(k^2 - m_{a1}^2 + i\epsilon)(k_f^2 - m_{a2}^2 + i\epsilon) (k_i^2 - m_{b2}^2 + i\epsilon)},
\end{align}
where $\delta_{a1,b1}$ ensures that particle 1 appears in both channels.

It is worth emphasizing the significance of being able to write $\Wc$ in the form shown in Eqs.~\eqref{eq:W_def} and~\eqref{eq:Wdf_def}. This demonstrates that even if the amplitude $\Wc$ is complex and features singularities, if one has previously determined the $w_{\mathrm{on}}$ and $\Mc$ amplitudes, only one class of functions remains to be constrained, $\Ac_{22}$, which are purely real and smooth. In practice, given a finite amount of partial waves, these functions can be defined in terms of energy-dependent form-factors associated with the desired angular momentum states.
In turn, in the following section, we discuss the determination of resonant form-factors in terms of the two-hadron form-factors $\Ac_{22}$.

\section{Implication for resonance form-factors}
\label{sec:resonances}

The analytic structure presented in the previous section is independent of the dynamics underlying the amplitudes. In this section we consider the implication of the expressions presented for systems that feature a hadronic resonance. For simplicity, we allow for the resonance to couple to a single channel. In particular, we review how one can obtain the mass, decay width, and transition form-factors for a scalar resonance coupling to a scalar external current. Additionally, we explain how elastic form-factors of this resonance can be obtained from $\Wc_\df$. Although we consider the simplest possible systems, we stress that the following procedure is valid for higher partial waves, multiple open channels, and currents with any Lorentz structure.

We begin by providing the standard definition of resonances as complex-valued poles $(s_R)$ in the analytic continuation of the scattering amplitude onto the second Riemann sheet,
\begin{align}
\label{eq:poledef}
\lim_{s\to s_R} \Mc^{\rm II}(s)
=-\frac{c^2}{s-s_R},
\end{align}
where $c$ is the coupling to the asymptotic states, and the ${\rm II}$ superscript has been introduced to emphasize that this is a pole in the second sheet of the scattering amplitude. 
The pole location can be related to the resonance mass ($m_R$) and its decay width ($\Gamma_R$) via  $\sqrt{s_R}=m_R- i\Gamma_R /2$.
The amplitude on the second Riemann sheet is found by analytically continuing through the branch cut, which is demonstrated in Appendix \ref{app:analytic_continuations}. 
Using the on-shell form Eq.~\eqref{eq:2to2.M_on-shell}, we find that the second sheet amplitude is expressed as 
\begin{align}
\label{eq:resonance.second_sheet}
    \Mc^{\mathrm{II}}(s) = \frac{1}{\Kc^{-1}(s) + i \rho_0} \,,
\end{align}
where the sign flip on the phase space factor arises from continuing through the branch cut.

\begin{figure}[t]
\begin{center}
\includegraphics[width=.75\textwidth]{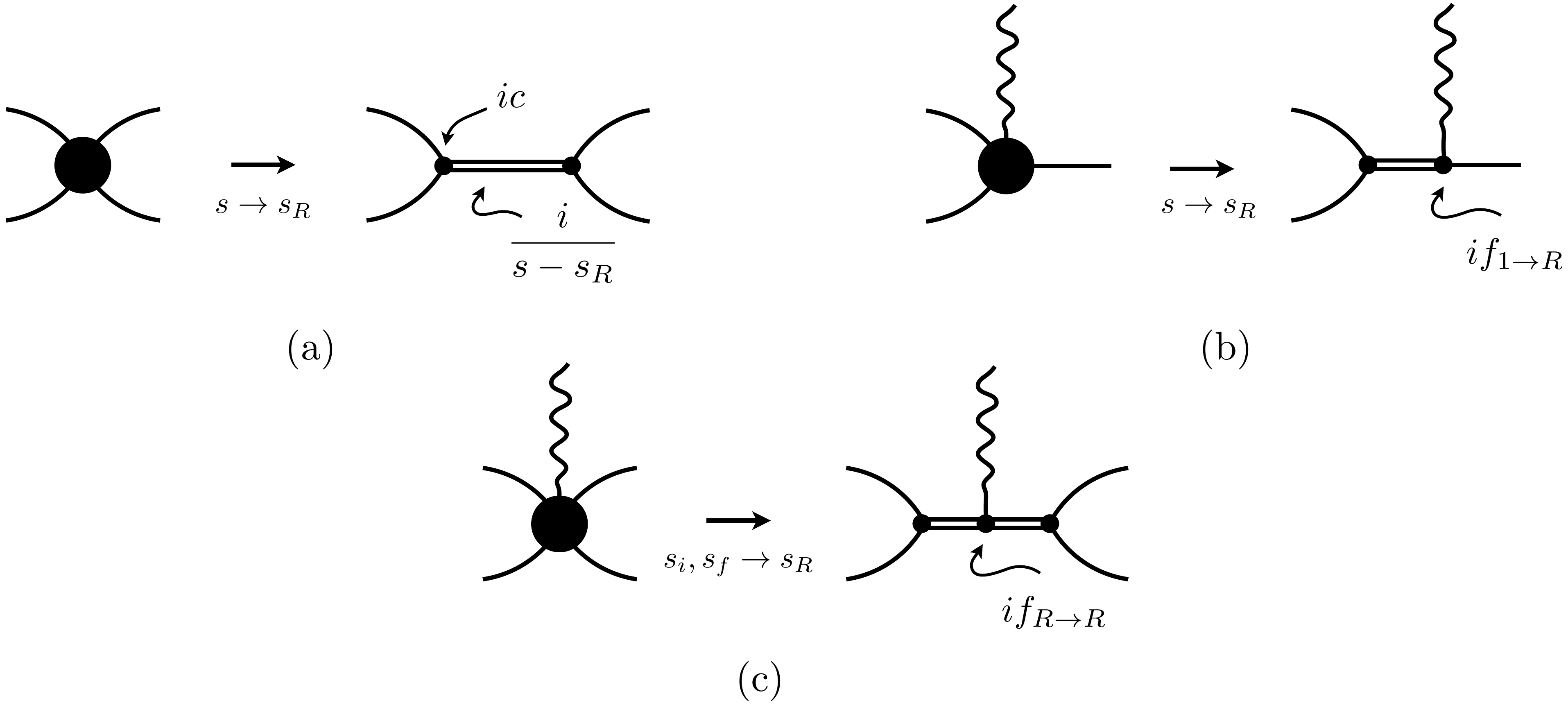}
\caption{(a) $2\to 2$, (b) $1+\Jc\to 2$, and (c) $2+\Jc\to 2$ amplitudes near the resonance pole, $s_R$. The residues at the pole define the coupling, $c$, transition form-factor, $f_{1\to R}$, and elastic resonance form-factor, $f_{R\to R}$.}
\label{fig:resonance}
\end{center}
\end{figure}

If there is a resonance present in $\Mc$, then both $\Hc$ and $\Wc_{\rm df}$ will inherit the singular behavior, as is evident from Eqs.~\eqref{eq:1Jto2.H_on-shell} and \eqref{eq:Wdf_def}, respectively. 
In the limit that one approaches the resonance pole, the relationship between the residues of the poles in these amplitudes and the desired transition and elastic form-factors can be obtained using the Lehmann-Symanzik-Zimmermann (LSZ) reduction procedure. This is illustrated diagrammatically in Fig.~\ref{fig:resonance}.
In our example of a scalar resonance and a scalar current, the transition amplitude $\Hc$ can be related to the transition form-factor, $f_{1\to R}(Q^2)$, via 
\begin{align}
\label{eq:scalarff1}
\lim_{s\to s_R} 
\Hc^{\rm II}(s,Q^2) =
-\frac{c \, f_{1\to R}(Q^2)}{s-s_R},
\end{align}
where again this is defined on the second sheet.
In Eq.~\eqref{eq:1Jto2.H_on-shell}, we wrote $\Hc$ in terms of $\Ac_{21}$. Because $\Ac_{21}$ has been defined so as to be non-singular, 
continuing $\Hc$ to the second sheet amounts to a continuation in $\Mc$ from Eq.~\eqref{eq:1Jto2.H_on-shell}, giving
\begin{align}
    \label{eq:resonance.1Jto2.second_sheet}
    \Hc^{\mathrm{II}}(s,Q^2) = \Mc^{\mathrm{II}}(s)  \Ac_{21}(s,Q^2)\, .
\end{align}
Note that for scalar currents, since $\Hc$ and $\Ac_{21}$ are Lorentz scalars, we adopt a convention that we write $s$ and $Q^2$ for their arguments, rather than $P_i$ and $P_f$.
In the case considered, $\Ac_{21}$ can be understood as an energy-dependent form-factor.  
We can rewrite the transition form-factor in terms of $\Ac_{21}$,
\begin{align}
f_{1\to R}(Q^2)
&=\lim_{s\to s_R}
\frac{(s_R-s)}{c}
\Mc^{\rm II}(s)\Ac_{21}(s,Q^2) \, ,
\nn\\
&=
\lim_{s\to s_R}
\frac{(s_R-s)}{c}
\left[\left(-\frac{c^2}{s-s_R}\right)\Ac_{21} (s, Q^2)\right] \, ,
\nn\\
&=
c\,\Ac_{21} (s_R,Q^2)\, .
\label{eq:fftrans}
\end{align}
Given $\Ac_{21} (s, Q^2)$ is real and has no nearby singularities, it is a convenient function to parameterize. This in part illustrates that if you know $\Ac_{21} (s, Q^2)$ as a function of energy, it is straightforward to determine the transition form-factor of the resonance of interest. This insight was used in, for example, recent exploratory determinations of the $\rho\to\pi$ electromagnetic form-factor from lattice QCD~\cite{Briceno:2015dca, Briceno:2016kkp, Alexandrou:2018jbt}.

Similarly, one can determine the elastic form-factor of the resonance, ${f_{R\to R}(Q^2)}$, from the residue of $\Wc$ via the LSZ reduction as
\begin{align}
\label{eq:scalarff2}
\lim_{s_i,s_f\to s_R} 
\Wc^{\rm II, II}(s_f,Q^2,s_i) &=
\lim_{s_i,s_f\to s_R} 
\Wc_{\df}^{\rm II, II}(s_f,Q^2,s_i) \, ,
\\
&=
\frac{-c}{s_f-s_R}
\,{f_{R\to R}(Q^2)}
\,
\frac{-c}{s_i-s_R}
\, ,
\end{align}
where $f_{R\to R}(Q^2)$ is defined at fixed $s_i = s_f = s_R$. 
The appearance of two ${\rm II}$ superscripts emphasizes that one must continue the amplitude in both $s_i$ and $s_f$ planes in order to evaluate at the resonance pole.
In the first equality, we have used the fact that the difference between $\Wc$ and $\Wc_\df$, given in Eq.~\eqref{eq:W_def}, only couples to either the initial or final resonance pole but not both. As a result, this is suppressed relative to the double pole.~\footnote{This procedure is followed in Ref.~\cite{PhysRevD.86.034003} for studying the $\sigma$ as well as in Refs.~\cite{Briceno:2019nns, Chen:1999vd, Kaplan:1998sz} for theories with bound states.}

From the on-shell representation Eq.~\eqref{eq:Wdf_def}, we find that the $\Wc_{\df}$ amplitude on the second Riemann sheet in both variables takes the form
\begin{align}
\label{eq:resonance.2Jto2.WII}
    \Wc_{\df}^{\mathrm{II},\mathrm{II}}(s_f,Q^2,s_i) = \Mc^{\mathrm{II}}(s_f) \left[ \, \Ac_{22}(s_f,Q^2,s_i) + f(Q^2) \Gc^{\mathrm{II},\mathrm{II}}(s_f,Q^2,s_i) \, \right] \Mc^{\mathrm{II}}(s_i) \, ,
\end{align}
where $\Gc^{\mathrm{II},\mathrm{II}}$ is the analytically continued triangle function which is
\begin{align}
    \label{eq:resonance.triangle_second_sheet}
    \Gc^{\mathrm{II},\mathrm{II}}(s_f,Q^2,s_i) = \Gc(s_f,Q^2,s_i) - 2i \, \mathrm{Im} \, \Gc(s_f,Q^2,s_i) \, .
\end{align}
In arriving at Eq.~\eqref{eq:resonance.2Jto2.WII}, it was necessary to use the fact that $\Ac_{22}$ is non-singular in the kinematic region considered. A detailed proof of Eq.~\eqref{eq:resonance.2Jto2.WII} is provided in App.~\ref{app:analytic_continuations}.

Using the all-order expression for $\Wc_{\rm df}$ in terms of the triangle loop and the energy-dependent form-factor given in Eq.~\eqref{eq:Wdf_def}, one finds
\begin{align}
\label{eq:fftelast}
{f_{R\to R}(Q^2)}
&\equiv 
\lim_{s_i,s_f\to s_R} 
 \frac{s_i-s_R}{-c}
\,\frac{s_f-s_R}{-c}
\,
\Mc^{\rm II}(s_{f}) \,
\left[
\Ac_{22}(s_f,Q^{2},s_i)
+
f(Q^2)\,\Gc^{\mathrm{II},\mathrm{II}}(s_f,Q ^2,s_i) 
\right]\,
\Mc^{\rm II}(s_i)
\nn\\
&= 
c^2
 \,
\left[
\Ac_{22}(s_{R},Q ^2,s_R)
+
f(Q^2)\,\Gc^{\mathrm{II},\mathrm{II}}(s_{R},Q ^2,s_R) 
\right]\,.
\end{align}
As previously stated, although the arguments presented here were for scalar currents for $S$ wave systems, the relations easily generalize to arbitrary currents, partial waves, and channels.
In the case of currents with non-trivial Lorentz structure, form-factors are accompanied by kinematic Lorentz tensors, which do not alter the analytic structure. 
For multiple scattering channels, one must take care on which sheet the amplitude is continued to, following the same methodology as presented in, for example, Ref.~\cite{Gribov:1962fx}.

In the special case of conserved vector currents, it was shown in Ref.~\cite{Briceno:2019nns} that current conservation via the Ward-Takahashi identity constrains the forward direction of the $2+\Jc\to 2$ amplitude. 
For example, the amplitude for a two hadron system consisting of one neutral and one charged particle must follow the relation
\begin{align}
\label{eq:resonance.WTI}
\lim_{P_i \to P_f}\Wc_{\df}^{\mu}(P_f,P_i) = 2P^{\mu} \mathrm{Q}_0 \frac{\partial}{\partial s} \Mc(s) \, ,
\end{align}
where $\mathrm{Q}_0$ is the charge of the particle.
This identity imposes further constraints on $\Ac_{22}$, namely that in the forward limit
\begin{align}
\Ac_{22}^{\mu}(P,P) = -2\mathrm{Q}_0 P^{\mu}  \frac{\partial}{\partial s}\Kc^{-1}(s) - \mathrm{Q}_0 \mathrm{Re}\,\Gc^{\mu}(P,P) \, ,
\end{align}
which follows directly from Eq.~\eqref{eq:Wdf_def} and noting that the imaginary part of $\Gc^{\mu}$ is proportional to $\partial \rho / \partial s$, ensuring that $\Ac_{22}$ is a real function.

If there is a resonance in this system, then at the resonance pole Eq.~\eqref{eq:resonance.WTI} further imposes that the form-factor of the resonance at $Q^2 = 0$ is the charge of the resonant state. We define the form-factor for a scalar resonance with a vector current in an analogous way to Eq.~\eqref{eq:scalarff2} as
\begin{align}
    \label{eq:resonance.ff_def}
    (P_f+P_i)_{\mu} f_{R\to R}(Q^2) & \equiv \lim_{s_i,s_f\to s_R} \frac{s_f-s_R}{c}\Wc_{\df,\mu}^{\mathrm{II},\mathrm{II}}(P_f,P_i) \frac{s_i-s_R}{c} \, .
\end{align}
Taking the $P_i\to P_f$ limit of Eq.~\eqref{eq:resonance.ff_def}, we then use Eq.~\eqref{eq:resonance.WTI} to find
\begin{align}
    2P^{\mu}f_{R\to R}(0) 
    & =
    2P^\mu \mathrm{Q}_0 \lim_{s_\to s_R} \frac{(s-s_R)^2}{c^2}
     \frac{\partial}{\partial s} \Mc^{\mathrm{II}}(s),
    \\
    & = 2P^{\mu} \mathrm{Q}_0 \, .
\end{align}
Therefore, we conclude that the resonance form-factor for a conserved vector current yields its charge at $Q^2 = 0$ as one may expect. 
The use of the Ward-Takahashi identity to impose additional constraints on two-hadron resonances has been explored, e.g. in Ref.~\cite{Bauer:2012at} for the Roper and in Ref.~\cite{Machavariani:2007fh} for the $\Delta$.

In practical lattice QCD calculations, renormalizing a conserved current by demanding that the form-factor at $Q^2 = 0$ of one of the stable hadrons is equal to its physical charge, ensures that the charge of the rest of the stable hadrons is recovered.
Since the normalization of form-factors of resonances and bound states is fixed by this same charge, according to both the analytic expression presented here as well as the finite volume framework in Ref.~\cite{Briceno:2015tza,Baroni:2018iau}, they do not require any additional renormalization.

\subsection{Recovering the narrow-width approximation  \label{sec:narrow_width}}

For processes where resonances may appear as intermediate states, it is advantageous to perform an expansion about their decay width in the limit where it becomes infinitesimally small. See Ref.~\cite{Carlson:2017lys} for a recent example of this. In this limit, the scattering amplitude acquires a pole at physical energy values, thus violating unitarity. This can be remedied by calculating corrections to the non-zero width. In order to gain further insight into $\Wc_{\df}$ we explore its behavior in the presence of an infinitesimally narrow resonance. Given Eq.~\eqref{eq:scalarff2}, we expect this to be dominated by a double pole structure.

For simplicity, we consider the case where all amplitudes are saturated by the $\ell=0$ partial wave and are dominated by a narrow resonance, this motivates the use of the $S$ wave Breit-Wigner parameterization,
\begin{align}
\tan \delta(s) &= \frac{\sqrt{s} \, \Gamma(s)}{m_0^2 - s} \, , \\
\Gamma(s) &= \frac{g^2}{6\pi} \frac{m_0^2}{s } q^\star\, 
= \frac{4\,g^2}{3\xi} \frac{m_0^2}{\sqrt{s} }\, \rho_0
\, ,
\end{align}
where $\delta(s)$ is the scattering phase shift, $\Gamma(s)$ is the energy-dependent width, and $m_0$ and $g$ are constants of the parameterization that in general do not possess direct physical interpretation. In the last equality, we used the phase-space definition given in Eq.~\eqref{eq:ps}. Finally, for a single partial wave the $K$ matrix relates to the scattering phase shift via, $\Kc^{-1} =  \rho_0 \cot \delta(s)$. Having this parameterization in place, we can analytically continue the corresponding amplitude to the resonant pole as discussed in Sec.~\ref{sec:resonances} and relate these parameters to the pole location and residue. 

Given the definition of $\Gamma$ above, it is clear that the narrow-width limit can be considered by expanding the Breit-Wigner amplitude about $g\approx0$. In this case, $\sqrt{s_R}=m_0 + \mathcal{O}(g^2)$. In what follows, we will expand the amplitudes to leading order in $g$. In addition to verifying the expected behavior of the amplitudes, it informs us how the generalized form-factors $\mathcal{A}_{21}$ and $\mathcal{A}_{22}$ behave as a function of $g$ at leading order.

At leading order in $g$, the purely hadronic amplitude is equal to
\begin{align}
\Mc(s)
&= \frac{4g^2m_0^2}{3\xi}\frac{1}{m_0^2 - s }
+\mathcal{O}(g^4)\, .
\end{align}
This has the same form as Eq.~\eqref{eq:poledef}, allowing us to identify the residue at the pole in terms of the parameters of the model. At leading order in $g$ we get
\begin{align}
c= \frac{2gm_0}{\sqrt{3\xi}}\, .
\label{eq:cg_rel}
\end{align}
Given this, we can explore the narrow-width limit for the remaining amplitudes by performing an expansion in $g$ and keeping the leading order term. To do this, it is useful to categorize the various building blocks of the aforementioned amplitudes in their leading behavior in $g$. For the purely hadronic ones it is evident that 
\begin{align}
\Mc(s)
&=\mathcal{O}({g}^{2}) \, .
\end{align}
For the $1+\Jc\to2$ transition amplitude it is slightly less obvious how it should scale. Focusing on scalar currents, we begin by anchoring their scaling in terms of the transition form-factor, given in Eq.~\eqref{eq:fftrans}. Keeping only the leading order behavior in $g$, we find
\begin{align}
 f_{1\to R}(Q^2)
&=
\frac{2m_0}{\sqrt{3\xi}}\,g\Ac_{21} (m_0^2,Q^2)\,,
\end{align}
where we used the relationship between $c$ and $g$, given in Eq.~\eqref{eq:cg_rel}. Given that the form-factor must in general be nonzero in the $g\to0$ limit, we find how $\Ac_{21} (s,Q^2)$ must scale with $g$. This, in combination with Eq~\eqref{eq:1Jto2.H_on-shell}, tells us the scaling of the other $1+\Jc\to2$ transition building blocks 
\begin{align}
\Ac_{21} (s,Q^2)&=\mathcal{O}({g}^{-1})\, ,
\nn\\
\Hc (s,Q^2)&=\mathcal{O}({g})\, .
\end{align}

We proceed to apply the same logic to the $2+\Jc\to 2$ amplitude. In other words, we first look at the elastic resonant form-factor, Eq.~\eqref{eq:fftelast}, and use the relationship between $g$ and $c$ to find,
\begin{align}
 {f_{R\to R}(Q^2)}
&= 
\frac{4g^2m_0^2}{3\xi}
 \,
\left[
\Ac_{22}(m^2_0,Q ^2,m^2_0)
+
f(Q^2)\,\Gc(m^2_0,Q ^2,m^2_0) 
\right]\,,
\end{align}
where we neglected higher-order corrections in $g$. 
Again, this is in general non-zero in the $g\to0$ limit, so we deduce that the term in brackets must scale as $1/g^{2}$. This scaling, of course, cannot come from either $f$ or $\Gc$. The former is the single particle form-factor, which in general has no knowledge of the resonance being considered. The latter is a purely kinematic function which contains no information of dynamical quantities, like $g$. 
Therefore neither can have any information of $g$, leaving us to conclude that 
\begin{align}
\Ac_{22}(m^2_0,Q ^2,m^2_0)=\mathcal{O}({g}^{-2}) \, .
\end{align}
As a result at leading order in $g$ the form-factor satisfies,
\begin{align}
{f_{R\to R}(Q^2)}
&= 
\frac{4g^2m_0^2}{3\xi}
 \,
\Ac_{22}(m^2_0,Q ^2,m^2_0)
\,.
\end{align}
Thus, from Eq.~\eqref{eq:Wdf_def} we find the leading order behavior in $g$ of the double pole contribution to $\Wc_{\rm df}$ to be
\begin{align}
\Wc_{\rm df} (P_f,Q ^2,P_i)
=\mathcal{O}({g}^{2}) \, .  
\end{align}

In conclusion, this suggests that for a narrow resonance within the Breit-Wigner parametrization, one should introduce 
\begin{align}
\Ac_{21} (s,Q^2)&= \frac{\widetilde{\Ac}_{21} (s,Q^2)}{g} \, ,
\nn\\
\Ac_{22}(s_f,Q ^2,s_i)
&=\frac{\widetilde{\Ac}_{22}(s_f,Q ^2,s_i)}{g^2}
\,,
\end{align}
where $\widetilde{\Ac}_{21} $ and $\widetilde{\Ac}_{22}$ do not scale with $g$.

It is worth commenting further as to why this is a sensible conclusion within an EFT point of view. One can always introduce an auxiliary field for a narrow resonance that couples to asymptotic states. 
Defining the coupling to the scattering states to be proportional to $g$, one immediately find that any loop  would be $\mathcal{O}(g^2)$. In other words, the $s$-channel loops appearing in $\Mc$ and $\Hc$ are $\mathcal{O}(g^2)$ suppressed, and these loops are the source of the resonance non-zero width. Similarly, in $\Wc_\df$ the $s$-channel loops, including the triangle one in Fig.~\ref{fig:triangle} are $\mathcal{O}(g^2)$ suppressed. This is consistent with the fact that $\Ac_{22}$ is enhanced relative to the triangle function by $1/g^2$.

\section{Derivation of on-shell representations}
\label{sec:derivation}

In this section, we present derivations of the on-shell representations presented in Sec.~\ref{sec:main}.
Our main tool relies on summing diagrams to all-orders, separating singularities induced by particle production in the physical region, from short-distance contributions.
All short-distance physics is absorbed into a set of unknown functions which can be determined from lattice QCD.
We first review the on-shell projection for the $2\to 2$ hadronic amplitude in Sec.~\ref{sec:2to2}, recovering the well-known $K$ matrix representation.
Following this, we turn to the transition amplitudes, first reviewing the known result for $1+\Jc\to 2$ processes in Sec.~\ref{sec:1Jto2}.
We then present the derivation of the main original result in this article, the $2+\Jc\to 2$ transition amplitude, in Sec.~\ref{sec:2Jto2}.

\subsection{Review of the \texorpdfstring{$2\to 2$}{2to2}  scattering amplitude}\label{sec:2to2}

We begin with the hadronic $2\to 2$ scattering amplitude in the kinematic region where only one channel, composed of two scalar particles with masses $m_1$ and $m_2$, is open. In Sec.~\ref{sec:2to2.coupled_channels}, we lift this assumption to accommodate any number of intermediate two-particle states.
It is convenient to consider the off-shell extension of this amplitude, $\Mc(p',p)$, where the initial state carries momenta $p$ and $P-p$ for particles 1 and 2, respectively, and the final state carries momenta $p'$ and $P-p'$, for particles 1 and 2, respectively. 
Note, the momenta of the initial/final state appear in the rightmost/leftmost part of the arguments of $\Mc(p',p)$. We will follow this convention throughout. 
We leave the dependence on the total conserved momentum $P$ in the amplitude $\Mc$ implicit for notational convenience.
The on-shell amplitude is recovered by placing the external legs on shell 
\begin{align}
\label{eq:2to2.OnShellPres}
\Mc(\hat{\mathbf{p}}'^{\star},\hat{\mathbf{p}}^{\star})
=
\Mc(p',{p})\bigg|_{p^2=p'^2=m_1^2;\,(P-p)^2=(P-p')^2 = m_2^2} \, ,
\end{align}
where $\star$ denotes CM coordinates, and $\mathbf{\hat{p}}^{\star}$ and $\mathbf{\hat{p}}'^{\star}$ are the orientations of particle 1 in the initial and final state in this frame, respectively. These are not fixed when placing the particles on their mass shell.

\begin{figure}[t]
\begin{center}
\includegraphics[width=.8\textwidth]{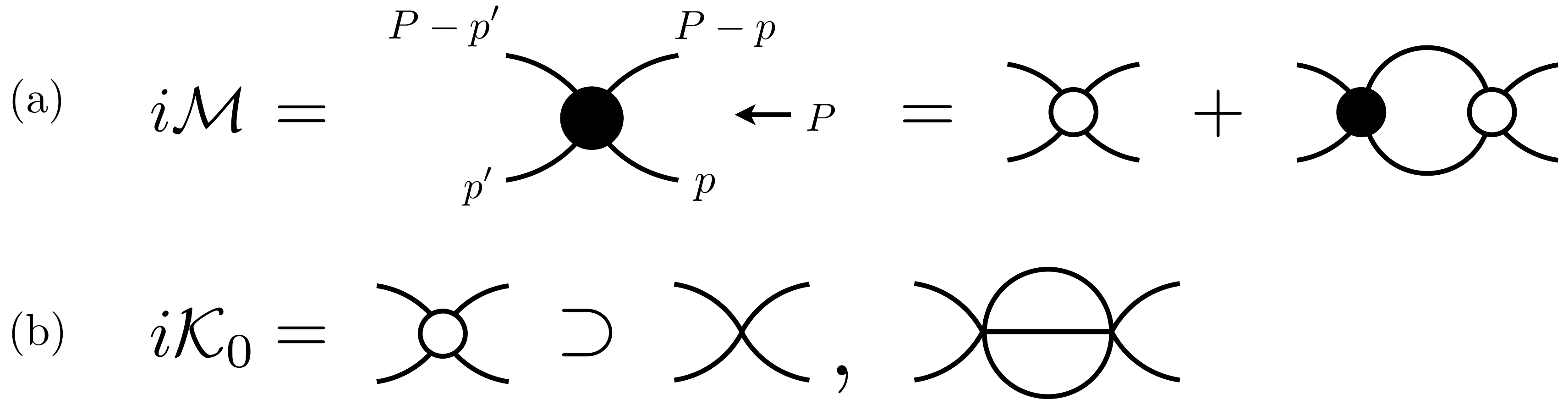}
\caption{(a) Self-consistent integral equation for the hadronic scattering amplitude $\Mc$ as given in Eq.~\eqref{eq:2to2.M_dse}. (b) Examples of diagrams contributing to the Bethe-Salpeter kernel, $\Kc_0$, which contains all $s$-channel two-particle irreducible diagrams.}
\label{fig:2to2.MandK}
\end{center}
\end{figure}

It can be shown, \eg summing to all-orders in perturbation theory, that $\Mc(p',p)$ satisfies the self-consistent integral equation
\begin{align}
\label{eq:2to2.M_dse}
i\Mc(p',{p})
&=
i\Kc_{0}(p',{p})
+
\xi\int \! \frac{\diff^{4}k}{(2\pi)^{4}} i\Mc(p',k)
i\Delta_{1}(k)
i\Delta_{2}(P-k)
i\Kc_{0}(k,p) \, ,
\end{align}
where $i\Kc_{0}$ is the Bethe-Salpeter kernel, which contains all $s$-channel two-particle irreducible diagrams, $\xi$ is the symmetry factor defined in the previous section, $i\Delta_1$ and $i\Delta_2$ are the fully dressed propagators for particles 1 and 2, respectively, and where the integral runs over the four-momentum of the intermediate state particles. This equation is depicted pictorially in Fig.~\ref{fig:2to2.MandK}. 
Note that Eq.~\eqref{eq:2to2.M_dse} can be written such that $\Mc$ and $\Kc_0$ in the second term are interchanged.
In the following manipulations, we work with Eq.~\eqref{eq:2to2.M_dse} as presented, but remark that the same procedure holds for the alternative where $\Mc$ and $\Kc_0$ are interchanged in the second term.

For each particle $\alpha=1,2$, we choose the propagators to have unit residue at the pole mass,
\begin{align}
\label{eq:2to2.propagators}
i\Delta_{\alpha}(k) & = \frac{i}{k^{2}-m_{\alpha}^{2}+i\epsilon} + iS_{\alpha}(k) \, , \\ 
& \equiv iD_{\alpha}(k) + iS_{\alpha}(k) \, ,
\end{align}
where $S_{\alpha}$ are non-singular at the pole.
For the kinematic region of interest, $i\Kc_{0}$ is non-singular and can be thought of as a smooth function.

We proceed to now separate the on-shell components from Eq.~\eqref{eq:2to2.M_dse}, exploiting the fact that in the elastic kinematic domain the only singularity the amplitude has is the two-particle intermediate state threshold.
In other words, we will make explicit the on-shell singularities required by $S$ matrix unitarity, and all off-shell contributions will be absorbed into some short-distance function to be determined, \eg from lattice QCD calculations.
We first iterate Eq.~\eqref{eq:2to2.M_dse} once by substituting the relation into itself, giving
\begin{align}
\label{eq:2to2.M_dse_iterated}
i\Mc(p',p)
& =
i\Kc_{0}(p',{p})
+
\xi\int\! \frac{\diff^{4}k}{(2\pi)^{4}} i\Kc_{0}(p',k) i\Delta_{1}(k) i\Delta_{2}(P-k) i\Kc_{0}(k,p) \nn \\
& + \xi\int\! \frac{\diff^{4}k'}{(2\pi)^{4}} \, \xi \int \! \frac{\diff^{4}k}{(2\pi)^4} i\Mc(p',k') i\Delta_{1}(k') i\Delta_{2}(P-k') i\Kc_{0}(k',k)  i\Delta_{1}(k) i\Delta_{2}(P-k) i\Kc_{0}(k,p) \, .
\end{align}

We now take the second term, as well as the second loop in the third term, and separate out the on-shell contribution, which amounts to identifying all components which can give an imaginary part.
In Appendix~\ref{app:bubble} we describe in detail the procedure we follow, which closely resembles that presented in Ref.~\cite{Kim:2005gf}.
Following the operations outlined in the aforementioned Appendix, we find
\begin{align}
\label{eq:2to2.loop_id_on-shell}
    \xi \int \! \frac{\diff^{4}k}{(2\pi)^4} i\Kc_{0}(p',k)  i\Delta_1(k)  i\Delta_2(P-k) i\Kc_{0}(k,p) & = i\Kc_{1}(p',p) + \int\! \frac{\diff \hat{\kb}^{\star}}{4\pi} i\Kc_{0}(p',\hat{\kb}^{\star}) \, \rho_0\, i\Kc_{0}(\hat{\kb}^{\star},p) \, , \nn \\
    & = i\Kc_{1}(p',p) + \sum_{\ell,m_{\ell}} i\Kc_{0, \ell m_{\ell}}(p') \, \rho_0 \, i\Kc_{0, \ell m_{\ell}}(p) \, ,
\end{align}
where $\Kc_1$ is a smooth function, and $\rho_0$ is the two-particle phase space defined in Eq.~\eqref{eq:ps} which is evaluated at $s$.
The quantities $\Kc_{0}(p',\hat{\kb}^{\star})$ and $\Kc_{0}(\hat{\kb}^{\star},p)$ are the kernels where the intermediate state is projected on-shell. The intermediate states are further decomposed into partial waves $\Kc_{0, \ell m_{\ell}}(p')$ and $\Kc_{0, \ell m_{\ell}}(p)$ defined as 
\begin{align}
\label{eq:2to2.pwk1}
\Kc_{0}(\hat{\mathbf{k}}^{\star} ,p) & = \sqrt{4\pi} \, \sum_{\ell, m_{\ell}} Y_{\ell, m_{\ell}} (\hat{\mathbf{k}}^{\star}) \, \Kc_{0,\ell m_{\ell}}(p) \, , \\[5pt]
\Kc_{0}(p', \hat{\mathbf{k}}^{\star} ) & = \sqrt{4\pi} \, \sum_{\ell, m_{\ell}}  \Kc_{0,\ell m_{\ell}}(p') \, Y_{\ell, m_{\ell}}^{*} (\hat{\mathbf{k}}^{\star}) \, ,
\label{eq:2to2.pwk2}
\end{align}
where $\ell$ is the angular momentum between the two particles defined in their CM frame, and $m_{\ell}$ is its projection onto some fixed axis.
Note that we only projected the intermediate on-shell states of each kernel, leaving the external kinematics off their mass-shell until after we iterated over all loops in the integral equation.

\begin{figure}[t]
\begin{center}
\includegraphics[width=.55\textwidth]{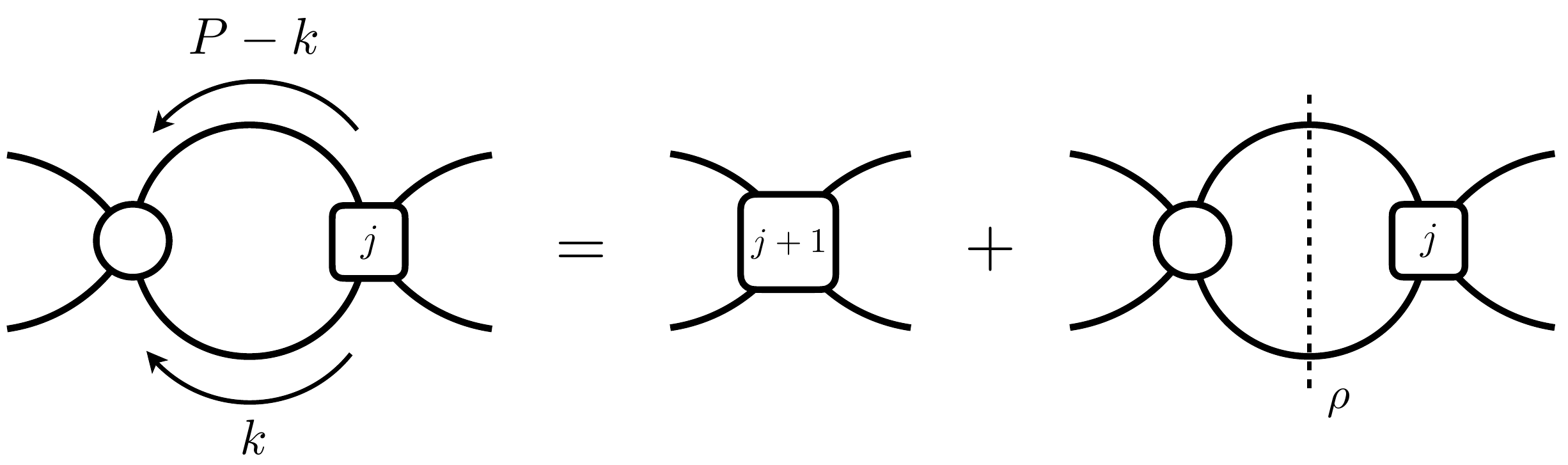}
\caption{Iterated loop identity as given in Eq.~\eqref{eq:2to2.loopIden}, where the open circle is the Bethe-Salpeter kernel $\Kc_0$, the box with label $j$ is the $j$-th kernel, and the dashed line indicates the two-particle phase space cut which places intermediate state particles on-shell.}
\label{fig:loop_id}
\end{center}
\end{figure}

Applying this loop identity, and combining terms using Eq.~\eqref{eq:2to2.M_dse}, $\Mc$ becomes
\begin{align}
\label{eq:2to2.M_1st_on-shell}
i\Mc(p',p)
&=
i\Kc_{0}(p',p)
+
\sum_{\ell,m_{\ell}} i\Mc_{\ell m_{\ell} }(p') \, \rho_0 \, i\Kc_{0, \ell m_{\ell}}(p) \nn \\
& 
+
i\Kc_{1}(p',p)
+
\xi \int \frac{\diff^{4}k}{(2\pi)^4} i\Mc(p',k) i\Delta_1(k) i\Delta_2(P-k) i\Kc_1(k,p) \, ,
\end{align}
where in the second term, we recovered $\Mc$ using Eq.~\eqref{eq:2to2.M_dse}, where its intermediate state is projected on-shell and into partial waves similar to Eq.~\eqref{eq:2to2.pwk2}.
We see that the third and fourth terms form the same structure as our starting point Eq.~\eqref{eq:2to2.M_dse}, with a new kernel $\Kc_1$. 
Therefore, we can insert Eq.~\eqref{eq:2to2.M_dse} into the fourth term, and repeat the on-shell separation as before, yielding new terms that go like Eq.~\eqref{eq:2to2.loop_id_on-shell} with the rightmost $\Kc_0$ replaced by $\Kc_1$, and $\Kc_1$ replaced by $\Kc_2$.
This pattern continues for every loop with the new kernel defined by the previous separation.
We therefore define the iterated loop identity
\begin{align}
\label{eq:2to2.loopIden}
    \xi \int \! \frac{\diff^4 k}{(2\pi)^4} \, i\Kc_{0}(p',k) i\Delta_1(k) i\Delta_2(P-k) i\Kc_{j}(k,p) = i\Kc_{j+1}(p',p) + \sum_{\ell,m_{\ell}} i\Kc_{0,\ell m_{\ell}}(p') \, \rho_0 \, i\Kc_{j,\ell m_{\ell}}(p) \, ,
\end{align}
which is structurally identical to that of Eq.~\eqref{eq:2to2.loop_id_on-shell}, except for the involvement of the $j$-th and $(j+1)$-th kernels.
The iterated loop identity is shown diagrammatically in Fig.~\ref{fig:loop_id}.

Repeated use of Eqs.~\eqref{eq:2to2.M_dse} and~\eqref{eq:2to2.loopIden} to all orders allows us to write $\Mc$ as
\begin{align}
i\Mc(p',p)
&=
 \sum_{j=0}^{\infty}i\Kc_{j}(p',p)
+
\sum_{\ell,m_{\ell}}i\Mc_{\ell m_{\ell}}(p') \, \rho_0\, \sum_{j=0}^{\infty}i\Kc_{j,\ell m_{\ell}}(p) \, , \\
&\equiv
i\Kc(p',p)
+
\sum_{\ell,m_{\ell}} i\Mc_{\ell m_{\ell}}(p') \, \rho_0 \, i\Kc_{\ell m_{\ell}}(p) \, ,
\label{eq:2to2.Mmat}
\end{align}
where in the last line we defined the $K$ matrix as the sum of each iterated kernel. 
Since the intermediate state is now on-shell, we project the initial and final states on their mass shell, and project the initial and final states via the partial-wave expansion,
\begin{align}
\label{eq:2to2.PWE}
\Mc(\hat{\mathbf{p}}'^{\star},\hat{\mathbf{p}}^{\star} ) = 4\pi \sum_{\ell',m_{\ell'}}\sum_{\ell,m_{\ell}} Y_{\ell' m_{\ell'}}(\hat{\mathbf{p}}'^{\star}) \Mc_{\ell' m_{\ell'}; \ell m_{\ell}}(s) Y_{\ell m_{\ell}}^{*}(\hat{\mathbf{p}}^{\star}) \, ,
\end{align}
with a similar expansion for $\Kc$.
Substituting Eq.~\eqref{eq:2to2.PWE} into Eq.~\eqref{eq:2to2.Mmat} we recover Eq.~\eqref{eq:2to2.M_on-shell} as a matrix in angular momentum space.
The partial wave expansion induces kinematic singularities at threshold since the spherical harmonics become singular as $q^{\star} \to 0$.
Therefore, $\Mc_{\ell'm_{\ell'};\ell m_{\ell}}\sim q^{\star\,\ell'+\ell}$ in order to compensate for this.
As introduced in Sec.~\ref{sec:main}, rotational invariance allows us to write the amplitude as $\Mc_{\ell'm_{\ell'};\ell m_{\ell}} = \delta_{\ell'\ell}\delta_{m_{\ell'}m_{\ell}} \, \Mc_{\ell}$.

As a final remark, the on-shell form Eq.~\eqref{eq:2to2.M_on-shell} explicitly shows the singularities required by unitarity of the $S$ matrix, which for partial waves states that the discontinuity of the amplitude across the real $s$ axis must satisfy
\begin{align}
    \label{eq:2to2.unitarity}
    \mathrm{Disc} \, \Mc_{\ell} = 2i\, \mathrm{Im} \Mc_{\ell} = 2i\,\rho_0 \lvert \Mc_{\ell} \rvert^2 \, ,
\end{align} 
for energies greater than the production threshold, $s \ge s_{\mathrm{th}} \equiv (m_1 + m_2)^2$.
The equivalence between the discontinuity and the imaginary part holds from the Schwartz reflection principle, i.e.\ $\mathcal{M}(s^*)=\mathcal{M}^*(s)$.

\subsubsection{Arbitrary number of channels}
\label{sec:2to2.coupled_channels}

If one considers an arbitrary number of strongly interacting channels, then the previous discussions are extended such that the amplitudes are matrices in channel space. 
Let $a$, $b$, and $c$ be the channel index, which ranges from 1 to $N_{\mathrm{ch}}$, where $N_{\mathrm{ch}}$ indicates the number of channels.
The self-consistent integral equation for the scattering amplitude is then
\begin{align}
\label{eq:CC}
i\Mc_{a;b}(p',{p})
&=
i\Kc_{0,a;b}(p',{p})
+
\sum_{c = 1}^{N_{\mathrm{ch}}}\xi_c\int \! \frac{\diff^{4}k}{(2\pi)^{4}} \, i\Mc_{a;c}(p',k)
i\Delta_{1c}(k)
i\Delta_{2c}(P-k)
i\Kc_{0,c;b}(k,p) \, ,
\end{align}
where the $1c$ and $2c$ indices on the propagators indicate particle 1 and 2 in channel $c$, respectively.
Following the same steps to project the system to its on-shell representation, with the simple extension that each kernel is a matrix in channel space, we arrive at the expression
\begin{align}
    \label{eq:2to2.M_on-shell.cc}
    i\Mc_{a;b}(s) = i\Kc_{a;b}(s) + \sum_{c,d = 1}^{N_{\mathrm{ch}}} i\Mc_{a;c}(s) \,\rho_{c;d}\, i\Kc_{d;b}(s) \, ,
\end{align}
where angular momentum indices are left implicit while we explicitly show the indices for channel space.

\subsection{ The \texorpdfstring{$1+\Jc\to 2$}{1Jto2} transition amplitude}
\label{sec:1Jto2}

\begin{figure}[t]
\begin{center}
\includegraphics[width=.8\textwidth]{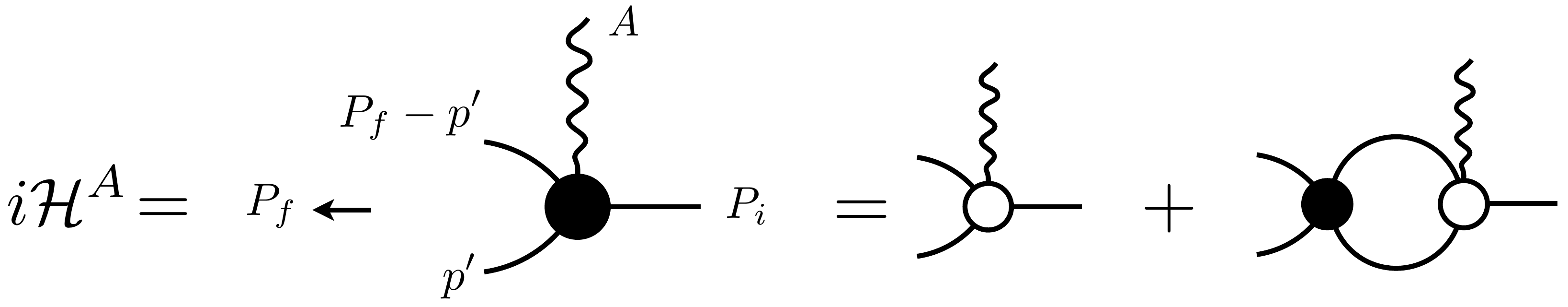}
\caption{Diagrammatic representation of the $1+\Jc\to 2$ all-orders equation defined in Eq.~\eqref{eq:1Jto2.H_dse}, where the open circle representing $\Hb_0$ contains all two-particle irreducible diagrams in the $s_f = P_f^2$ channel.}
\label{fig:H}
\end{center}
\end{figure}

In performing the on-shell projection of the $2\to2$ amplitude in the previous section, we effectively separated the short and long distance contributions of this amplitude. A similar separation can be made for the $1+\mathcal{J}\to2$ transition amplitude, $\Hc$. This can be done while placing no restrictions on the current, except that it is local. As a result, we consider an arbitrary external and local current.
The final two-particle state has momenta $p'$ and $P_f-p'$ for particles 1 and 2, respectively, while the initial state has only a single particle with momentum $P_i$, and associated mass $M$. The external current carries a momentum transfer squared $Q^2 = - (P_f - P_i)^2$. 
We again consider the off-shell extension $\Hc(p',P_f;P_i)$. As explained in Sec.~\ref{sec:results_one_current}, we will label the current and the subsequent amplitudes with a single superscript $A$, which encodes all identifiers of the current.

\begin{figure}[t]
\begin{center}
\includegraphics[width=.7\textwidth]{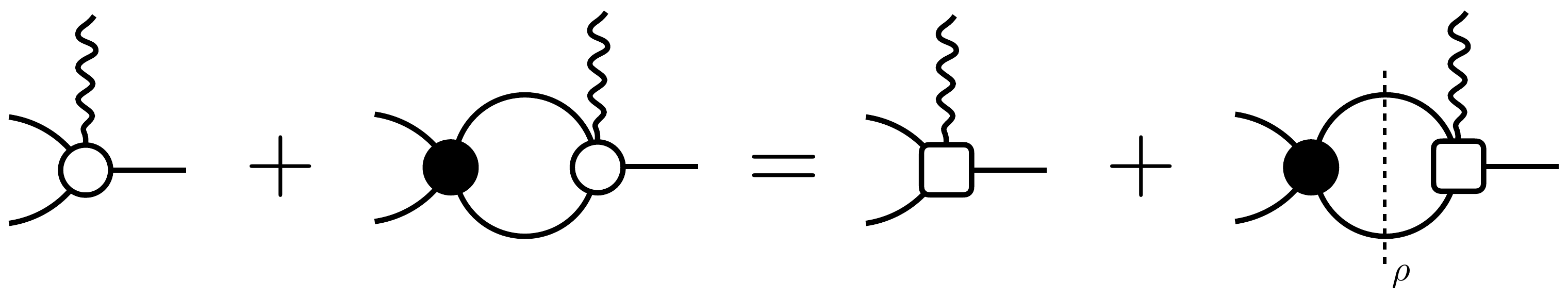}
\caption{On-shell projection of $\Hc$, where the left hand side is Eq.~\eqref{eq:1Jto2.H_dse} and the right hand side is Eq.~\eqref{eq:1Jto2.Hmat_V4}. The white box indicates the real function $\mathbf{H}$.}
\label{fig:Honshell}
\end{center}
\end{figure}

The amplitude satisfies the following self-consistent integral equation~\cite{Briceno:2014uqa},
\begin{align}
\label{eq:1Jto2.H_dse}
i\Hc^{A}(p',P_f;P_i) = i\Hb_{0}^{A}(p',P_f;P_i) + \int \! \frac{\diff^{4}k}{(2\pi)^{4}} \, i\Mc(p',k) i\Delta_{1}(k) i\Delta_{2}(P_f-k) i\Hb_{0}^{A}(k,P_f;P_i) \,.
\end{align}
Here $\Hb_{0}$ is a non-singular, smooth function of $s_f$ in the kinematic region of interest similar to $\Kc_{0}$, while $\Delta$ and $\Mc$ are as before.
Diagrammatically, this is shown in Fig.~\ref{fig:H}.
We now substitute the self-consistent relation for the $\Mc$ amplitude, given in Eq.~\eqref{eq:2to2.M_dse}, into Eq.~\eqref{eq:1Jto2.H_dse}. We then proceed as before separating out the on-shell behavior.
For this situation, we cast the iterated loop identity as
\begin{align}
\label{eq:1Jto2.iter_loop_id}
\xi \int\! \frac{\diff^4 k}{(2\pi)^4} \,  i\Kc_{0}(p',k) i\Delta_1(k) i\Delta_2(P_f-k) i\Hb_{j}^{A}(k,P_f;P_i)
&=
i\Hb_{j+1}^{A} (p',P_f;P_i)
+ 
\sum_{\ell,m_{\ell}} i\Kc_{0,\ell m_{\ell}}(p') \, \rho_0 \, i\Hb_{j,\ell m_{\ell}}^{A}(P_f;P_i) \, ,
\end{align}
where $\mathbf{H}_j$ is the $j$-th iterated kernel which absorbs all real contributions from the loop and the previous iterate kernel, and $\rho_0$ depends on $s_f$.
The subscript $j$ enumerates the number of $\Kc_0$ kernels inserted in the projection. 
Applying the same procedure as with $\Mc$ we arrive at
\begin{align}
\label{eq:1Jto2.Hmat_V4}
i\Hc_{\ell m_{\ell}}^{A}(P_f,P_i)
&=
i\Hb_{\ell m_{\ell}}^{A}(P_f,P_i)
+
i\Mc_{\ell}(s_f) \, \rho_0 \, i \Hb_{\ell m_{\ell}}^{A}(P_f,P_i) \, , 
\end{align}
where $\mathbf{H}$ is the infinite sum of all iterated kernels defined by Eq.~\eqref{eq:1Jto2.iter_loop_id} and is a real function in the kinematic domain of interest.
The on-shell projection is illustrated diagrammatically in Fig.~\ref{fig:Honshell}.
Note that the partial-wave expansion for $\Hc$ is only on the final state
\begin{align}
\Hc^{A}(p',P_f;P_i) = \sqrt{4\pi} \sum_{\ell , m_{\ell}} Y_{\ell m_{\ell} }(\hat{\mathbf{p}}'^{\star}_f) \Hc_{\ell m_{\ell}}^{A}(P_f,P_i) \, ,
\end{align}
where the subscript $f$ on $\hat{\pb}^{\prime \star}_f$ indicates that these angles are evaluated in the CM frame of the final state.

Using Eq.~\eqref{eq:2to2.M_on-shell} for $\Mc$, we can write the on-shell representation for $\Hc$ as
\begin{align}
\label{eq:1Jto2.H_amp}
i\Hc_{\ell m_{\ell}}^{A}(P_f,P_i) = \frac{1}{1 - \Kc_{\ell}(s_f)  \, i\rho_0 }  i\mathbf{H}_{\ell m_{\ell}}^{A}(P_f,P_i) \, .
\end{align}
At this stage, we make note that $\Hc$ is an analytic function in the complex $s_f$ plane except for the branch cut associated with the pair production of the intermediate state and potential bound state poles.
The $K$ matrix could in principle have poles in $s_f$ for physical energies, which do not appear in the scattering amplitude as poles on the real axis. One can show using all order perturbation theory that $\mathbf{H}$ must have these same poles in $s_f$. If this were not the case, the unphysical poles would correspond to zeros of the $\Hc$.
This motivates us to introduce a parameterization
\begin{align}
\label{eq:1Jto2.Hprod}
i\Hb_{\ell m_{\ell}}^{A}(P_f,P_i) = i\Kc_{\ell}(s_f)\, \Ac_{21,\ell m_{\ell}}^{A}(P_f,P_i) \, ,
\end{align}
where $\Ac_{21}$ is a smooth function in the allowed kinematic domain, except for barrier factors near threshold. 
Combining this with Eq.~\eqref{eq:1Jto2.H_amp}, we arrive at Eq.~\eqref{eq:1Jto2.H_on-shell}.
As mentioned in Sec.~\ref{sec:main}, $\Hc$ and $\Ac_{21}$ are Lorentz tensors which can be expanded in energy-dependent form-factors.

The form of the on-shell amplitude satisfies Watson's final state theorem~\cite{Watson:1952}, meaning that the phase of the amplitude $\Hc$ is equal to that of $\Mc$. 
This is a consequence of the unitarity condition for $\Hc$, $\mathrm{Disc} \, \Hc_{\ell m_{\ell}} = 2i\,\rho_0 \Mc_{\ell}^{*} \Hc_{\ell m_{\ell}}$, from which one immediately identifies Eq.~\eqref{eq:1Jto2.H_on-shell} as a solution.

Our results can be generalized to accommodate any number of two-body scattering channels. 
Using the results for the coupled-channel scattering amplitude, and extending Eq.~\eqref{eq:1Jto2.H_dse} for $N_{\mathrm{ch}}$ channels,
\begin{align}
i\Hc_{a}^{A}(p',P_f;P_i) = i\mathbf{H}_{0,a}^{A}(p',P_f;P_i) + \sum_{b=1}^{N_{\mathrm{ch}}} \xi_b \int \! \frac{\diff^{4}k}{(2\pi)^{4}} \, i\Mc_{a;b}(p',k) i\Delta_{b1}(k) i\Delta_{b2}(P_f-k) i\mathbf{H}_{0,b}^{A}(k,P_f;P_i) \, ,
\end{align}
the preceding arguments can be made to show that its on-shell representation has the same analytic structure, 
\begin{align}
    \label{eq:1Jto2.H_on-shell_cc}
    \Hc_{a}^{A}(P_f,P_i) = \sum_{b=1}^{N_{\mathrm{ch}}} \Mc_{a;b}(s_f) \Ac_{21,b}^{A}(P_f,P_i) \, .
\end{align}
which agrees with Eq.~\eqref{eq:1Jto2.H_on-shell} and the expressions presented in Ref.~\cite{Briceno:2014uqa}.
Here, $\Hc_a$ and $\Ac_{21,a}$ are elements of a vector in channel space for some given initial state, and $\Mc_{a;b}$ is defined in Eq.~\eqref{eq:2to2.M_on-shell.cc}.

All arguments above were for the case when the current interacts with an initial single-particle state, $1+\Jc\to 2$, and are the same manipulations if one considers the case where the current is ejected in the final state, $2 \to 1 + \Jc$.
Moreover, the previous derivation can be adapted to the case with no initial hadrons, \ie $\Jc\to 2$, by replacing the kernel $\mathbf{H}_0$ with another one encoding the short-distance dynamics of pair creation. Therefore the on-shell amplitude of both processes has the same analytic structure. This is because the $\Jc\to 2$ reaction is analogous to a $1+\Jc\to2$ transition where the initial state happens to have vanishing momentum.

\subsection{The \texorpdfstring{$2+\Jc\to 2$}{2Jto2} transition amplitude}
\label{sec:2Jto2}

Here we present an all orders calculation of the $2+\Jc\to 2$ amplitude, $\Wc$, where an external current with arbitrary Lorentz structure injects momentum into the initial two particle system. 
We consider the case where the species of the initial and final state particles are the same.
In this case, the initial state particles have momenta $p$ and $P_i-p$ for particles 1 and 2, respectively, while the final state has $p'$ and $P_f - p'$ for particles 1 and 2, respectively.
Therefore, the external current carries a momentum transfer squared $Q^2 = -(P_f-P_i)^2$.
The $\Wc$ amplitude is defined via the matrix element as in Eq.~\eqref{eq:2Jto2.W_ME_def}, and is the on-shell limit of the off-shell extension
\begin{align}
\mathcal{W}^{A}(P_f,\hat{\mathbf{p}}_f'^{\star};P_i,\hat{\mathbf{p}}_i^{\star}) 
& = \mathcal{W}^{A}(P_f,p';P_i,p)\bigg|_{p^2=p'^2=m_1^2;\,(P_i-p)^2=(P_f-p')^2 = m^{2}_2} \, .
\end{align}
Generally, the current can also be flavor changing, as in the $1+\Jc\to 2$ case.
We first focus on the case where the current is not flavor changing, so that the initial, intermediate, and final states are the same species.
Furthermore, for the following derivation we consider the case where particle 1 is neutral with respect to the external current, \ie the current only interacts with particle 2.
At the end of this section, we comment on the extension in which both particles interact with the current, including the case where the particles are identical.
Additionally, we show the result for an arbitrary number of channels to which the current can couple.

\begin{figure}[t]
\begin{center}
\includegraphics[width=.8\textwidth]{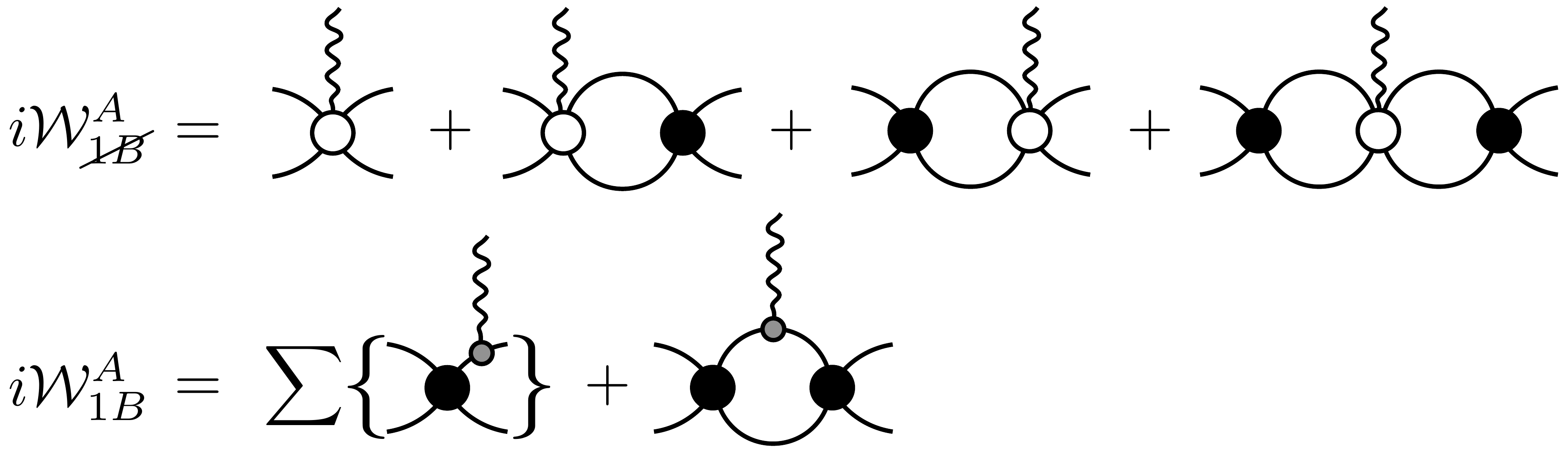}
\caption{Diagrammatic representations of the all-orders equations for $\Wc_{\cancel{1B}}$ given in Eq.~\eqref{eq:2Jto2.Wno1B_dse} and $\Wc_{1B}$ in Eq.~\eqref{eq:2Jto2.W1B_dse}, where the open circle is the kernel $i\Wb_{0|0}$.}
\label{fig:W}
\end{center}
\end{figure}

We proceed as before, only now we have to separate out the on-shell behavior from both the initial and final state two-particle scatterings. 
However, we encounter a new feature which is not present in the previous case, namely the triangle diagram topology in which the current interacts with a single particle only.
This introduces new kinematic singularities in addition to those present already in the two-particle loop.
Considering the elastic region for both the initial and final states, using all-orders perturbation theory the amplitude can be separated into two topologically distinct classes of amplitude,
\begin{align}
\label{eq:2Jto2.W_sum_dse}
i\mathcal{W}^{A}
&=
i\mathcal{W}_{\cancel{1B}}^{A}
+
i\mathcal{W}_{1B}^{A} \, ,
\end{align}
where the subscript $1B$ stands for topologies where the current couples to a single hadron (one-body), cf. Fig.~\ref{fig:W}.

Each amplitude contains a distinct kernel, which represents all short range physics which cannot go on-shell in the kinematic domain of our interest.
First, the kernel contained in the amplitude  $\mathcal{W}_{\cancel{1B}}$ we denote as $\Wb_{0|0}$, and it represents a short-range $2+\Jc \to 2$ transition amplitude, which is two-particle irreducible in both the $s_i$ and $s_f$ channels, where $s_i = P_i^2$ and $s_f=P_f^2$. 
Like the kernel $\Hb_0$, the subscript `0' denotes the absence of two-particle dressings from $\Kc_0$, with the vertical line representing a distinction between initial and final states.
These kernels are smooth, non-singular functions in the elastic region of both the initial and final states. 
Dressing this kernel with all two-particle scattering to all orders in the strong interactions, one can show that $\mathcal{W}_{\cancel{1B}}$ obeys the following equation,
\begin{align}
\label{eq:2Jto2.Wno1B_dse}
i\mathcal{W}_{\cancel{1B}} ^{A} (P_f,p';P_i,p)
& =
i\Wb_{0|0} ^{A}  (P_f,p';P_i,p) \nn \\
& +
\xi \int \! \frac{\diff^4 k}{(2\pi)^4} i\Mc(p',k) i\Delta_1(k) i\Delta_2(k_f) i\Wb_{0|0}^{A} (P_f,k;P_i,p) \nn \\
& +
\xi \int \! \frac{\diff^4 k}{(2\pi)^4} i\Wb_{0|0} ^{A}  (P_f,p';P_i,k) i\Delta_1(k) i\Delta_2(k_i) i\Mc(k,p) \nn \\
& +
\xi \int \!  \frac{\diff^4 k'}{(2\pi)^4} \xi \! \int\! \frac{\diff^4 k}{(2\pi)^4}  i\Mc(p',k') i\Delta_1(k') i\Delta_2(k_f') i\Wb_{0|0}^{A}(P_f,k';P_i,k) i\Delta_1(k) i\Delta_2(k_i) i\Mc(k,p) \, ,
\end{align}
where we remind the reader of the shorthand notation $k_i \equiv P_i - k$, and similarly for the final state and $k'$ momenta.
Equation~\eqref{eq:2Jto2.Wno1B_dse} is represented diagrammatically in Fig.~\ref{fig:W}.

The second type of kernel we consider is the off-shell extension of the single hadron transition amplitude $w$, which can be expressed in terms of the off-shell extended form-factors $f_j$ and kinematic functions $K_j$ as in Ref.~\cite{Baroni:2018iau},
\begin{align}
\label{eq:2Jto2.w1B_kernel}
w^{A}(k_f, k_i) = \sum_{j} K_{j}^{A}(k_{f},k_{i}) \,  f_{j}(Q^2,k_f^2,k_i^2) \, ,
\end{align}
where $k_i$ is the momentum flow from the initial state, and $k_f$ the momentum flow from the final state, giving the same invariant momentum transfer squared as before.
Since we consider here the case where the external current couples only to particle 2, all quantities in Eq.~\eqref{eq:2Jto2.w1B_kernel} refer to this particle.
This restriction can be trivially lifted by considering two sets of kernels $w_1$ and $w_2$, each coupling to particles 1 and 2, respectively, which we discuss later.

We consider an illustrative example the explicit decomposition of Eq.~\eqref{eq:2Jto2.w1B_kernel} for the following two cases. If the current is a scalar and the initial and final particles are identical scalars, then the sum only includes a single form-factor with no kinematic prefactor. If instead the current is a conserved vector current, the sum again only has one term and the prefactor is $K^{\mu}(k_{f},k_{i})=(k_{f}+k_{i})^\mu$.

Physical single-hadron form-factors are defined by the on-shell limit of this kernel, \ie~by continuing Eq.~\eqref{eq:2Jto2.w1B_kernel} to $k_i^2 = k_f^2 = m_2^2$, which is equivalent to the single hadron matrix element
\begin{align}
     w^{A}(k_f,k_i) \Big\rvert_{k_f^2 = m_2^2,\, k_i^2 = m_2^2}  
    & = w_{\mathrm{on}}^{A}(k_f,k_i) \, ,
\end{align}
where in this case we define the on-shell limit as affecting the form-factors only, leaving the kinematic tensors unaffected.
The on-shell form-factor is denoted with a single argument $Q^2$, differing from the off-shell extension where it depends on both the initial and final invariant mass
\begin{align}
    f_{j}(Q^2) = \lim_{ \substack{ k_i^2 \to m_2^2 \\ k_f^2 \to m_2^2 }} f_j(Q^2,k_f^2,k_i^2) \, .
\end{align}

There are three sets of diagrams that include the kernel $w$ which contribute to the all-orders relations for $\Wc_{1B}$.
Two of them involve the case where the current probes either the initial or final state, and the third set of diagrams include the case where the current probes one of the propagators in an intermediate state.
Summing to all orders, one finds the relation
\begin{align}
\label{eq:2Jto2.W1B_dse}
i\mathcal{W}_{1B}^{A} (P_f,p';P_i,p)
& =
iw^{A}(p_f', p_i') i\Delta_2(p_i') i\Mc(p',p) \nn  
+
i\Mc(p',p) i\Delta_2(p_i) iw^{A}(p_f,p_i) \nn \\[5pt]
& +
\int \! \frac{\diff^4k}{(2\pi)^4}  i\Mc(p',k) i\Delta_1(k) i\Delta_2(k_f) iw^{A}(k_f,k_i) i\Delta_2(k_i) i\Mc(k,p) \,.
\end{align}
Combining Eqs.~\eqref{eq:2Jto2.Wno1B_dse},~\eqref{eq:2Jto2.W1B_dse}, and~\eqref{eq:2Jto2.W_sum_dse} leads to a complete all-orders description of the $2+\Jc \to 2$ amplitude in terms of general one- and two-particle irreducible kernels.
Diagrammatically, $\mathcal{W}_{\cancel{1B}}$ and $\mathcal{W}_{1B}$ are shown in Fig.~\ref{fig:W}.

\subsubsection{On-shell projecting \texorpdfstring{$\Wc_{\cancel{1B}}$}{Wno1B} }

We now project Eq.~\eqref{eq:2Jto2.Wno1B_dse} such that intermediate states are on their mass-shell.
This case follows the same manipulations as in $\Hc$, except here there are two-body dressings on both the initial and final states.
We split the on-shell projection into two steps. Initially, consider the first and second terms of Eq.~\eqref{eq:2Jto2.Wno1B_dse}, where the final state is dressed with two-particle scattering processes.
The form of these two terms is similar to $\Hc$, and we can project these two terms to their on-shell form,
\begin{align}
    \label{eq:2Jto2.Wno1B.two_terms}
    i\Wb_{0|0}^{A} (P_f,p';P_i,p) & + \xi \int\!\frac{\diff^4 k}{(2\pi)^4} i\Mc(p',k) i\Delta_1(k) i\Delta_2(k_f)  i\Wb_{0|0} ^{A}  (P_f,k;P_i,p) \nn \\
    & = i\Wb_{\infty|0}^{A} (P_f,p';P_i,p) + \sum_{\ell,m_{\ell}} i\Mc_{\ell}(p') \, \rho_0 \, i\Wb_{\infty|0; \ell m_{\ell}} ^{A}  (P_f;P_i,p) \, , 
\end{align}
where the new kernel is defined as
\begin{align}
\label{eq:2Jto2.Wno1B_sumInfty}
    \Wb_{\infty|0}^{A} = \sum_{j = 0}^{\infty} \Wb_{j|0}^{A} \,,
\end{align}
and $\Wb_{j|0}$ is the $j$th iterated kernel for the final state, defined in the same way as Eq.~\eqref{eq:1Jto2.iter_loop_id} 
\begin{align}
\label{eq:2Jto2.Wno1B.iter_loop_id}
\xi \int\! \frac{\diff^4 k}{(2\pi)^4} \,  i\Kc_{0}(p',k) i\Delta_1(k) i\Delta_2(k_f) i\Wb_{j|0}^{A}(k,P_f;P_i)
\nn\\
&\hspace{-2cm}=
i\Wb_{j+1|0}^{A} (p',P_f;P_i)
+ 
\sum_{\ell,m_{\ell}} i\Kc_{0,\ell m_{\ell}}(p') \, \rho_0 \, i\Wb_{j|0;\ell m_{\ell}}^{A}(P_f;P_i) \,.
\end{align}
Again, the subscript $j$ indicates the number of absorbed kernels $\Kc_0$ from the on-shell projection.
In the second term, the intermediate-state particles are on-shell, and we expanded both the amplitude and the kernel into partial waves as in Eq.~\eqref{eq:2to2.pwk1}.
Recall that the external states remain off-shell, which means that this same decomposition holds for the third and fourth terms of Eq.~\eqref{eq:2Jto2.Wno1B_dse}, in which the kernels in Eq.~\eqref{eq:2Jto2.Wno1B.two_terms} are dressed with two-body scatterings on the initial state.
Next, the third and fourth terms represent a similar form to before, except we trade the final for initial state two-body dressings.
Repeating the same projection with this new kernel, now on the initial state, we arrive at the fully on-shell projected $\Wc_{\cancel{1B}}$ amplitude
\begin{align}
    \label{eq:2Jto2.Wno1B.on_shell}
    i\Wc_{\cancel{1B}; \ell' m_{\ell'} ; \ell m_{\ell}}^{A}(P_f,P_i) 
    = 
    \left[ \, 1 + i\Mc_{\ell'}(s_f) \, \rho_0 \, \right] \, i\Wb_{\cancel{1B}; \ell' m_{\ell'} ; \ell m_{\ell} }^{A}(P_f,P_i) \, \left[ \, 1 + \rho_0 \, i\Mc_{\ell}(s_i) \, \right] \, ,
\end{align}
where we introduce a new short-distance kernel $\Wb_{\cancel{1B}}$ which absorbs all the off-shell contributions from the intermediate state on-shell projections, defined by 
\begin{align}
    \label{eq:2Jto2.res.Wno1B}
    \Wb_{\cancel{1B}}^{A} \equiv \Wb_{\infty|\infty}^{A} = \sum_{j=0}^{\infty} \Wb_{\infty|j}^{A} \, ,
\end{align}
where $\Wb_{\infty|j}$ is defined via an iterate loop identity which we forgo writing since it is structurally identical to Eq.~\eqref{eq:2Jto2.Wno1B.iter_loop_id} except for the swapping of the kernels.
The phase space factors depend on the same total momentum as their respective adjacent factors of $\Mc$.
Additionally, we have placed the external states on their mass shell and expanded them into their respective partial-wave projections
\begin{align}
    \label{eq:2Jto2.Wno1B.pw}
    \Wc_{\cancel{1B}}^{A}(P_f,\hat{\mathbf{k}}_f^{\star} ;P_i, \hat{\mathbf{k}}^{\star}_i )  
    =
    4\pi \sum_{\ell', m_{\ell'}} \sum_{\ell,m_{\ell}} Y_{\ell' m_{\ell'}}(\hat{\mathbf{k}}_f^{\star} ) \,
    \Wc_{\cancel{1B} ; \ell' m_{\ell'} ; \ell m_{\ell} }^{A}(P_f,P_i) \,
    Y_{\ell m_{\ell}}^{*}(\hat{\mathbf{k}}^{\star}_i) \, ,
\end{align}
which holds for both $\Wc_{\cancel{1B}}$ and $\Wb_{\cancel{1B}}$.
Unlike the $2\to 2$ amplitude, the angular momentum between the hadrons is not conserved due to the insertion of the current, and thus is a dense matrix in this space.
Equation~\eqref{eq:2Jto2.Wno1B.on_shell} shows that the on-shell kernel is dressed by initial and final state rescatterings, similar to how the $1+\Jc\to 2$ amplitude is dressed in the final state, \cf Eq.~\eqref{eq:1Jto2.Hmat_V4}.

\subsubsection{On-shell projecting \texorpdfstring{$\Wc_{1B}$}{W1B}}\label{sec:W1B}

Moving on to $\mathcal{W}_{1B}$ we need to consider contributions due to the single hadron transition amplitude.
This leads to a new on-shell function, namely the triangle diagram.
We start with the first term of Eq.~\eqref{eq:2Jto2.W1B_dse} by using Eq.~\eqref{eq:2to2.M_dse} to obtain expressions where $w$ is attached to the kernel $\Kc_0$,
\begin{align}
\label{eq:2Jto2.wDeltaM}
    iw^{A}(p_f',p_i') i\Delta_2(p_i') i\Mc(p',p)
    & =
    iw^{A}(p_f',p_i') i\Delta_2(p_i') i\Kc_0(p',p) \nn \\
    & +
    iw^{A}(p_f',p_i') i\Delta_2(p_i') \, \xi \int\!\frac{\diff^4k}{(2\pi)^4} \, i\Kc_0(p',k) i\Delta_1(k) i\Delta_2( k_{i}) i\Mc(k,p) \, .
\end{align}

\begin{figure}[t]
\begin{center}
\includegraphics[width=.6\textwidth]{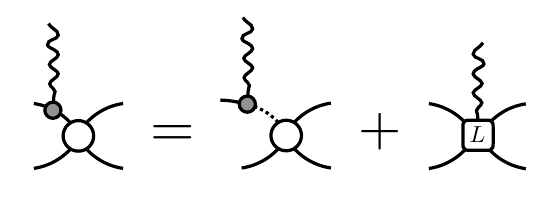}
\caption{Diagrammatic representation of the expansion shown in Eq.~\eqref{eq:2Jto2.W1b_Kmat_Rule1}. The dotted line in the first term represents the pole piece of the propagator, $D_{2}$, and the second term is a new smooth kernel we define as $\Wb_{L|0}$.}
\label{fig:W_L}
\end{center}
\end{figure}

Following similar steps outlined in detail in Ref.~\cite{Baroni:2018iau}, we use Eq.~\eqref{eq:2to2.propagators} to isolate the pole piece of the propagators.
We expand the kernels on either side of the pole term about the pole. Next, we project the kinematics of the single-particle form-factor(s), appearing in the definition of $w$, adjacent to the pole on-shell, however we leave the kinematic tensors $K_j$ off-shell.
We keep the on-shell kernels multiplying the pole term, and absorb all remaining contributions into a new smooth kernel which we define as $\Wb_{L|0}$, where $L$ denotes that the current absorbed was from the left.
These operations are summarized by the on-shell rule, diagrammatically shown in Fig.~\ref{fig:W_L},
\begin{align}
\label{eq:2Jto2.W1b_Kmat_Rule1}
    iw^{A}(p_f',p_i') i\Delta_2(p_i') i\Kc_0(p',p) = iw_{\mathrm{on}}^{A}(p_f',p_i') iD_2(p_i') \sum_{\ell m} \sqrt{4\pi}Y_{\ell m} (\hat{\pb}_i'^{\star})
    \left(\frac{p^{\prime\star}_i}{q^{\star}_i} \right)^\ell
    i\Kc_{0,\ell m}(p) + i\Wb_{L|0}^{A}(P_f,p';P_i,p)\,,
\end{align}
where we note that the final state for the kernel $\Kc_0$ has an additional barrier factor in its partial wave expansion as compared to Eq.~\eqref{eq:2to2.pwk1}. This factor ensures that no spurious threshold singularities arise since the partial-wave kernel behaves like $\Kc_{0,\ell m}(p) \sim q_i^{\star\ell}$, the relative momentum defined in Eq.~\eqref{eq:q_elastic} evaluated at $s_i$, while the spherical harmonics have a $(p^{\prime\star}_i)^{-\ell}$ behavior.
 In order to elucidate the validity of this expression, we note that the partial-wave decomposition of the off-shell kernel can be written as
 \begin{align}
i\Kc_0(p',p)
&=
\sum_{\ell' m_{\ell'},\ell m_{\ell}}
\sqrt{4\pi}
Y_{\ell' m_{\ell'}} (\hat{\pb}_i^{\prime\,\star})
i\Kc_{0,\ell' m_{\ell'};\ell m_{\ell}}(p_i^{\prime\,2},s_i)
\sqrt{4\pi}
Y^*_{\ell m_{\ell}} (\hat{\pb}_i^{\star}) \, ,
\nn\\
&=
\sum_{\ell' m_{\ell'},\ell m_{\ell}}
\Yc_{\ell' m_{\ell'}} ({\pb}_i'^{\star})
i\Kc_{0,\ell' m_{\ell'};\ell m_{\ell}}(s_i)
\sqrt{4\pi}
Y^*_{\ell m_{\ell}} (\hat{\pb}_i^{\star})
\nn\\
&\hspace{.5cm}+
\sum_{\ell' m_{\ell'},\ell m_{\ell}}
\Yc_{\ell' m_{\ell'}} ({\pb}_i^{\prime\,\star})
\left[
\left(\frac{q_i^{\star}}{p^{\prime\star}_i}\right)^{\ell'}
i\Kc_{0,\ell' m_{\ell'};\ell m_{\ell}}(p_i^{\prime\,2},s_i)
-
i\Kc_{0,\ell' m_{\ell'};\ell m_{\ell}}(s_i)
\right]
\sqrt{4\pi}
Y^*_{\ell m_{\ell}} (\hat{\pb}_i^{\star}) \, ,
\end{align}
where in the first equality we have made the dependence on $p_i'^2$ explicit for the off-shell kernel. In the last equality, we have added and subtracted the value of $\Kc_{0,\ell m;\ell' m'}$ when all particles are placed on-shell. The last term vanishes in the on-shell limit and does not have threshold singularities.

Substituting this into Eq.~\eqref{eq:2Jto2.wDeltaM} and simplifying using Eq.~\eqref{eq:2to2.M_dse}, we get
\begin{align}
\label{eq:2Jto2.W1Btermone}
    iw^{A}(p_f',p_i') i\Delta_2(p_i') i\Mc(p',p) 
    & = 
    iw_{\mathrm{on}}^{A}(p_f',p_i') iD_2(p_i') i\overline{\Mc}({\pb}_i'^{\star},p) \nn \\[5pt]
    & + 
    i\Wb_{L|0}^{A}(P_f,p';P_i,p) 
    +
    \xi \int\!\frac{\diff^4k}{(2\pi)^4} \, i\Wb_{L|0}^{A}(P_f,p';P_i,k) i\Delta_1(k) i\Delta_2(k_{i}) i\Mc(k,p) \, .
\end{align}
Here, $\overline{\Mc}$ indicates that the partial wave expansion of the $2\to 2$ amplitude contains a barrier factor~\cite{Briceno:2015tza, Baroni:2018iau}.
In this case, the partial-wave projection looks like
\begin{align}
	\label{eq:Mbar}
    \overline{\Mc}({\pb}_i'^{\star},\hat{\pb}_i)
    =
    \sqrt{4\pi} \sum_{\ell',m_{\ell'}}\sum_{\ell,m_{\ell}} \Yc_{\ell' m_{\ell'}}({\pb}_i'^{\star}) \Mc_{\ell' m_{\ell'} ; \ell m_{\ell}}(s_i) Y_{\ell m_{\ell}}^{*}(\hat{\pb}_i^{\star}) \, ,
\end{align}
From here on, all amplitudes which contain the overline follow this convention.

Following similar manipulations for the second term in $i\mathcal{W}_{1B}$, we obtain 
\begin{align}
\label{eq:2Jto2.W1Btermtwo}
i\Mc(p',p) i\Delta_2(p_f) iw^A(p_f,p_i)
& =
i\overline{\Mc}(p',{\pb}_f^{\star}) iD_2(p_f) iw^A_{\mathrm{on}}(p_f,p_i)
\nn \\
& +
i\Wb^A_{0|R}(P_f,p';P_i,p)
+
\xi \int i\Mc(p',k) i\Delta_1(k) i\Delta_2(k_f) i\Wb^A_{0|R}(P_f,k;P_i,p) \, ,
\end{align}
where $\Wb_{0|R}$ is a new non-singular kernel arising from the on-shell projection with the current on the right.

\begin{figure}[t]
\begin{center}
\includegraphics[width=.8\textwidth]{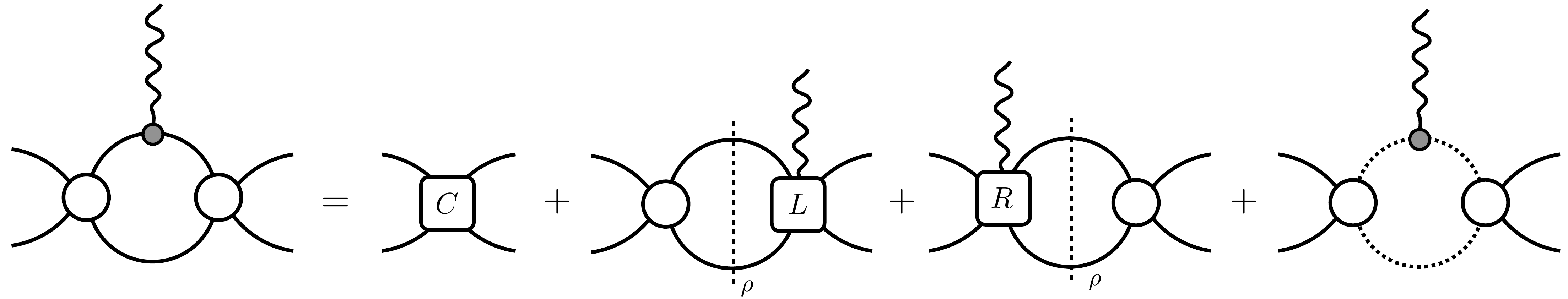}
\caption{Decomposition of triangle diagram into on-shell pieces involving new kernels $\Wb_{0|C|0}$, $\Wb_{L|0}$, $\Wb_{R|0}$.}
\label{fig:triangle_decomposition}
\end{center}
\end{figure}

For the final term in Eq.~\eqref{eq:2Jto2.W1B_dse}, we first use Eq.~\eqref{eq:2to2.M_dse} and look at the triangle diagram with $\Kc_0$ kernels on each side. In Appendix~\ref{app:trian}, we consider this loop in detail. We show that this can be written in terms of three singular pieces and a new non-singular contribution, which we label $\Wb_{0|C|0}$. Using the expression derived in Appendix~\ref{app:trian}, Eq.~\ref{eq:triangle_master}, as well as Eq.~\eqref{eq:2Jto2.W1b_Kmat_Rule1} and the equivalent for the $i\Kc_0 \, i\Delta_2 \, iw$ case, we find 
\begin{align}
\label{eq:2Jto2.Tri.Rule_C}
\int\!\frac{\diff^4 k}{(2\pi)^4} i\Kc_{0}(p',k) & i\Delta_1(k) i\Delta_2(k_f) iw^{A}(k_f,k_i) i\Delta_2(k_i) i\Kc_{0}(k,p) 
= 
i\Wb_{0|C|0}^{A}(P_f,p';P_i,p)   \nn \\
& +
\sum_{\ell',m_{\ell'}} i\Kc_{0,\ell' m_{\ell'}}(p') \, \rho_0 \, i\Wb_{L|0,\ell' m_{\ell'}}^{A}(P_f;P_i,p) 
+
\sum_{\ell,m_{\ell}} i\Wb_{0|R,\ell m_{\ell}}^{A}(P_f,p';P_i) \, \rho_0 \, i\Kc_{0,\ell m_{\ell}}(p) \nn \\
& +
\sum_{\ell',m_{\ell'}} \sum_{\ell, m_{\ell}} \Kc_{0,\ell' m_{\ell'}}(p') \sum_{j} if_j(Q^2)\,\Gc_{j;\ell' m_{\ell'} ; \ell m_{\ell}}^{A}(P_f,P_i) \Kc_{0,\ell m_{\ell}}(p) \, .
\end{align}
Equation~\eqref{eq:2Jto2.Tri.Rule_C} defines a new short-distance kernel $\Wb_{0|C|0}$ which absorbs all off-shell contribution. We also see two contributions from a two particle cut, and associated kernels $\Wb_{L|0}$ and $\Wb_{0|R}$, which are the same kernels we found in Eqs.~\eqref{eq:2Jto2.W1Btermone} and \eqref{eq:2Jto2.W1Btermtwo}. This decomposition is illustrated in Fig.~\ref{fig:triangle_decomposition}.

The final term of Eq.~\eqref{eq:2Jto2.Tri.Rule_C} involves $\Gc$, the triangle function, which is a purely kinematic function given in Eq.~\eqref{eq:2Jto2.G}. In Appendix~\ref{app:trian}, we present two different ways of defining this function. The first follows from Ref.~\cite{Baroni:2018iau}, and the second follows from the Cutkosky rules~\cite{Cutkosky:1960}. The difference between these two is a smooth function, and can be absorbed into the definition of $\Wb_{0|C|0}$.

Summing back the rescattering contributions using Eq.~\eqref{eq:2to2.M_dse}, and combining the result with Eqs.~\eqref{eq:2Jto2.W1Btermone} and \eqref{eq:2Jto2.W1Btermtwo}, we can on-shell project the remaining rescattering effects on the kernels $\Wb_{L|0}$, $\Wb_{0|R}$, and $\Wb_{0|C|0}$ in a manner similar in the preceding section. We arrive at the on-shell form for $\mathcal{W}_{1B}$
\begin{align}
\label{eq:2Jto2.W1B.on_shell}
i\mathcal{W}_{1B}^{A}(P_f,p';P_i,p)
& =
 iw_{\mathrm{on}}^{A}(p_f',p_i') iD_2(p_i') i\overline{\Mc}({\pb}_i'^{\star},p)
+
i\overline{\Mc}(p',{\pb}_f^{\star}) iD_2(p_f) iw^A_{\mathrm{on}}(p_f,p_i) \nn \\[5pt]
& +
\sum_{\ell',m_{\ell'}} \sum_{\ell,m_{\ell}} \Mc_{\ell' m_{\ell'}}(p') \sum_{j} if_{j}(Q^2) \Gc_{j; \ell' m_{\ell'} ; \ell m_{\ell}}^{A} (P_f,P_i) \Mc_{\ell m_{\ell}}(p) \nn \\
& +  4\pi \sum_{\ell',m_{\ell'}}\sum_{\ell, m_{\ell}} Y_{\ell' m_{\ell'}}(\hat{\pb}_f'^{\star}) \, [\, 1 +   i\Mc_{\ell'}(s_f) \, \rho_0 \, ]\,  i\Wb_{1B; \ell' m_{\ell'} ; \ell m_{\ell}}^{A}(P_f,P_i)  \, [\, 1+\rho_0 \,i\Mc_{\ell}(s_i) \, ] \, Y_{\ell m_{\ell}}^{*}(\hat{\pb}_i^{\star}) \, .
\end{align}
where $\Wb_{1B}$ is a new kernel which is defined as
\begin{align}
    \Wb_{1B}^{A} = \Wb_{\infty|C|\infty}^{A} + \Wb_{L|\infty}^{A} + \Wb_{\infty|R}^{A} \, ,
\end{align}
in which each of these kernels are defined to be fully dressed by the two-body kernels $\Kc_0$ which resulted from the on-shell projection, \eg as in Eq.~\eqref{eq:2Jto2.Wno1B_sumInfty}.
Both scattering amplitudes in the third term have on-shell kinematics for the intermediate states, and the last term has an identical structure as in Eq.~\eqref{eq:2Jto2.Wno1B.on_shell}, which we will exploit in the next section.

\subsubsection{Full on-shell result for \texorpdfstring{$\Wc$}{W}}
Finally, we can combine Eqs.~\eqref{eq:2Jto2.Wno1B.on_shell} and \eqref{eq:2Jto2.W1B.on_shell} into our on-shell expression for $\Wc$,
\begin{align}
\label{eq:2Jto2.res.W}
i\mathcal{W}_{1B}^{A}(P_f,p';P_i,p)
& =
 iw_{\mathrm{on}}^{A}(p_f',p_i') iD_2(p_i') i\overline{\Mc}({\pb}_i'^{\star},p)
+
i\overline{\Mc}(p',{\pb}_f^{\star}) iD_2(p_f) iw^A_{\mathrm{on}}(p_f,p_i) \nn \\[5pt]
& +
\sum_{\ell',m_{\ell'}} \sum_{\ell,m_{\ell}} \Mc_{\ell' m_{\ell'}}(p') \sum_{j}\left[ if_{j}(Q^2) \Gc_{j; \ell' m_{\ell'} ; \ell m_{\ell}}^{A} (P_f,P_i) \right] \Mc_{\ell m_{\ell}}(p) \nn \\
& +  4\pi \sum_{\ell',m_{\ell'}}\sum_{\ell, m_{\ell}} Y_{\ell' m_{\ell'}}(\hat{\pb}_f'^{\star}) \, [\, 1 +   i\Mc_{\ell'}(s_f) \, \rho_0 \, ]\,  i\Wb_{\ell' m_{\ell'} ; \ell m_{\ell}}^{A}(P_f,P_i)  \, [\, 1+\rho_0 \,i\Mc_{\ell}(s_i) \, ] \, Y_{\ell m_{\ell}}^{*}(\hat{\pb}_i^{\star}) \, ,
\end{align}
where $\Wb = \Wb_{\cancel{1B}} + \Wb_{1B}$.
The first two terms represent the case where the current probes an external particle, which yields a kinematic divergence from the pole term of the propagator.
As these terms do not yield physics involving short-range two-body dynamics, we define a divergence-free amplitude, $\Wc_{\df}$, which removes these two pole contributions,
\begin{align}\label{eq:W_and_Wdf}
i\Wc_{\df}^{A}(P_f,p';P_i,p) & \equiv i\Wc^{A}(P_f,p';P_i,p) 
- 
iw_{\mathrm{on}}^{A}(p_f',p_i') iD_2(p_i') i\overline{\Mc}({\pb}_i'^{\star},p)
-
i\overline{\Mc}(p',{\pb}_f^{\star}) iD_2(p_f) iw^A_{\mathrm{on}}(p_f,p_i) \, .
\end{align}
It is worth remarking that $\Wc_{\df}$ is the amplitude which naturally appears in the formalism for \emph{finite-volume} two-body matrix elements of local currents~\cite{Briceno:2015tza, Baroni:2018iau}. Of course, one can use the identity~\eqref{eq:W_and_Wdf} to obtain the full $\Wc$ amplitude.

Removing external leg contributions, which solely depend on previously known dynamical functions, we arrive at the expression
\begin{align}
    i\Wc_{\df; \ell' m_{\ell'} , \ell m_{\ell} }^{A}(P_f,P_i) & = \frac{1}{1-\Kc_{\ell'}(s_f) \, i\rho_0} \, i\Wb^A_{\ell' m_{\ell'} ; \ell m_{\ell}}(P_f,P_i) \, \frac{1}{1 - i\rho_0 \, \Kc_{\ell}(s_i) } \nn \\[5pt]
    & \hspace{2cm} +
    \Mc_{\ell'}(s_f) \sum_{j} if_j(Q^2) \Gc_{j;\ell' m_{\ell'} ; \ell m_{\ell} }^{A}(P_f,P_i) \Mc_{\ell}(s_i) \, .
\end{align}
In arriving at this expression, we have partial-wave projected $\Wc_{\df}$ using the expansion Eq.~\eqref{eq:2Jto2.Wno1B.pw}, and we used Eq.~\eqref{eq:2to2.M_on-shell} for the initial and final state scattering amplitudes.
As was the case for $\Hb$ in Sec.~\ref{sec:1Jto2}, $\Wb$ contains potential $K$ matrix poles. Because $\Wb$ depends on the energy of the initial and final states, it can have $K$ matrix poles associated with both of these states. To make this explicit, we introduce one final parameterization, which closely mirrors Eq.~\eqref{eq:1Jto2.Hprod},
\begin{align}
\label{eq:W_KFK}
\Wb^{A}_{\ell'm_{\ell'} ; \ell m_{\ell}}(P_f,P_i) = \Kc_{\ell'}(s_f) \, \Ac_{22; \ell' m_{\ell'} ; \ell m_{\ell} }(P_f,P_i) \, \Kc_{\ell}(s_i) \, .
\end{align}
Combining this with the previous equation, we arrive at our final result
\begin{align}
i\Wc_{\df ; \ell' m_{\ell'} , \ell m_{\ell} }^{A}(P_f,P_i) 
= 
\Mc_{\ell'} (s_f) \, \left[ \,  i\Ac_{22;\ell' m_{\ell'} ; \ell m_{\ell}}^{A}(P_f,P_i) 
+ \sum_{j} if_j(Q^2) \Gc_{j; \ell' m_{\ell'} ; \ell m_{\ell} }^{A}(P_f,P_i)  \, \right] \, \Mc_{\ell}(s_i) \, ,
\end{align}
where we have made explicit all partial wave indices.
This agrees with Eq.~\eqref{eq:Wdf_def} when considering that only particle 2 couples to the current.

\subsubsection{Generalizations for two charged species and identical particles}

We can easily generalize the above relations for the case where both particles are charged under interaction of the external current.
First, consider the case where the particles are distinguishable.
Each particle has an associated form-factor $f_{j,\alpha}$ where $\alpha = 1$ or $2$, and an associated kernel $w_{\alpha}$ given by Eq.~\eqref{eq:2Jto2.w1B_kernel}.
Our starting all-orders equation  Eq.~\eqref{eq:2Jto2.W1B_dse} is augmented with the addition of two more terms where the external leg of particle 1 is probed by the current and an additional triangle diagram.
These additional terms carry through the same on-shell projections as before, and we have the amplitude
\begin{align}
    \label{eq:2Jto2.gen.W_on-shell}
    i\Wc^{A}(P_f,p';P_i,p) & = \sum\Big\{ \, iw_{\mathrm{on}}\, iD\, i\overline{\Mc} \,\Big\} + i\Wc_{\df}^{A}(P_f,p';P_i,p) \, ,
\end{align}
where $\Wc_{\df}$ now has the form, as a matrix in angular momentum space,
\begin{align}
    \label{eq:2Jto2.gen.Wdf_on-shell}
    i\Wc_{\df}^{A}(P_f,P_i) = \Mc(s_f) \Bigg[\, i\Ac_{22}(P_f,P_i) + \sum_{j} if_{j,1}(Q^2) \Gc_{j,1}^{A}(P_f,P_i) + \sum_{j} if_{j,2}(Q^2) \Gc_{j,2}^{A}(P_f,P_i)  \,\Bigg] \Mc(s_i) \, ,
\end{align}
where $\Gc_{j,2}$ is the kinematic triangle function associated with particle 2, as defined in Eq.~\eqref{eq:2Jto2.G}, and $\Gc_{j,1}$ is the triangle function where particle 1 is switched with particle 2.

In the case of identical particles, there are still four terms in Eq.~\eqref{eq:2Jto2.gen.W_on-shell} associated with the current probing the external leg. 
In Eq.~\eqref{eq:2Jto2.gen.Wdf_on-shell}, however, there is only one triangle function contribution.

\subsubsection{Arbitrary number of channels}
\label{sec:2tJo2.coupled_channels}

For multiple two-particle scattering channels, $\Wc_{\df}$, $\Mc$, $\Ac_{22}$, and $\sum_j f_j\Gc_j$ become matrices in channel space,
where the latter is a dense matrix if the current allows for species transmutation.
Suppressing the angular momentum indices, one finds that the coupled-channel $\Wc_{\df}$ takes the form
\begin{align}
    \label{eq:2Jto2.Wcc}
    i\mathcal{W}_{\df ; a;b}^{A}(P_f,P_i) 
    = 
    \sum_{c,c' = 1}^{N_{\mathrm{ch}}} \Mc_{a;c'} (s_f) \, \left[ \,  i\Ac_{22;c';c}^{A}(P_f,P_i) 
    + \sum_{\alpha = 1}^{2}\sum_{j} if_{j,(c'\alpha); (c\alpha)}(Q^2) \Gc_{j; (c'\alpha); (c\alpha)}^{A}(P_f,P_i)  \, \right] \, \Mc_{c;b} (s_i) \, ,
\end{align}
where $a/b$ denote the external channel space indices, and the $c/c'$ indices are summing over the $N_{\mathrm{ch}}$ intermediate channels. For each intermediate channel, the current can couple to two legs in the triangle diagram. This is encoded in the $\alpha$ index which can take on either $1$ or $2$. The definition of the others labels, like $j$ and $A$, carry through from the previous expressions. In practice, for a specified current and in/out states, these expressions can be simplified further.
This is the most general form we obtain for Eq.~\eqref{eq:Wdf_def}.

\subsubsection{Comparison with existing formalism}\label{sec:comparison}

Having completed the derivation of our main result, Eq.~\eqref{eq:Wdf_def}, it is worth remarking on the comparison of this equation with existing formalism for $2+\Jc\to 2$ amplitudes. In short, previous work, which includes Refs.~\cite{PhysRevD.86.034003, Chen:1999vd, Kaplan:1998sz, Djukanovic:2013mka},~\footnote{For a study in 1+1D of the elastic form-factors of resonance, we point the reader to Ref.~\cite{Bruns:2019evf}.} considered the constructions of these amplitudes for a specific channel using a well motivated EFT. Furthermore, in all of these, the calculation was naturally performed to a finite order in the EFT expansion. This is to say, the goals and results of these studies were quite different than the one presented here. 

Nevertheless, what is true is that because our result is the most general to date, one can cast it in the form of previous existing formalism, while the converse is not true. As an illustrative example, we consider the result presented in Ref.~\cite{PhysRevD.86.034003}. This study used next-to-leading (NLO) order $\chi$PT to determine the $\pi\pi +\mathcal{J}\to \pi\pi$ scattering amplitude in the presence of an external scalar current. In order to assure that these amplitudes satisfy unitarity non-perturbatively, the authors follow a unitarization procedure where the $s$-channel diagrams are effectively upgraded in the power counting and summed to all orders. Special attention was placed on the scalar/isoscalar channel, where the $\sigma/f_0(500)$ resonance resides. By analytically continuing to the the $\sigma/f_0(500)$ pole, they were able to constrain the scalar radius of this state to be $\langle r^2 \rangle = (0.19\pm 0.02) - i (0.06\pm 0.02)~{\rm fm}^2$, suggesting it is a compact state.

The most general form of their result is in Eq.~(71) of Ref.~\cite{PhysRevD.86.034003}. The notation used in that reference is slightly different than the one adopted here. Therefore to make a comparison we first provide a relationship between the notations. The amplitude the authors considered, which they label $T_S$, does not include the single-particle poles. As result this can be interpreted as the analogue to $\Wc_{\df}$. The denominator of this has the same denominator as $\Mc(s_f)\Mc(s_i)$. More specifically the denominator of their expression can be written as $V_{\rm NLO}(s_f)V_{\rm NLO}(s_i)/[\Mc(s_f)\Mc(s_i)]$, where $V_{\rm NLO}$ is the Bethe-Salpeter kernel obtained at NLO in $\chi$PT. 

Within this scheme, the authors define the numerator at NLO in $\chi$PT. We label the numerator of Eq.~(71) of Ref.~\cite{PhysRevD.86.034003} as $N_{\rm NLO}$, instead of $W$ to avoid confusion. Given this, we can equate our main result Eq.~\eqref{eq:Wdf_def} to Eq.~(71) of Ref.~\cite{PhysRevD.86.034003}, to find 
\begin{align}
\left(  i\Ac_{22}  + if \cdot \, \Gc \right)
\approx
V_{\rm NLO}^{-1} \cdot N_{\rm NLO}\,\cdot V_{\rm NLO}^{-1} \, . 
\label{eq:matchChiPT}
\end{align}
Here we have dropped the kinematic arguments and the spherical harmonic indices of the functions. This is an approximate relation due to the fact that the right hand side is only defined perturbatively. Nevertheless, as illustrated in Ref.~\cite{PhysRevD.86.034003} going to NLO in the chiral expansion assures that $N_{\rm NLO}$ will encode the triangle singularity. In other words, the right hand side of Eq.~\eqref{eq:matchChiPT} has the same singularities as the left hand side. 

This example illustrates that given one EFT calculated to a sufficiently high order, one can recover the key features of our main result. Of course, the non-singular contributions and the exact prefactor of the singularities are in general non-perturbative dynamical functions. Presently the only tool available for obtaining these is lattice QCD following the formalism and procedure outlined in Refs.~\cite{Briceno:2015tza,Baroni:2018iau}.

\section{Conclusion}\label{sec:conclusion}
We have presented a model independent on-shell decomposition for transition amplitudes of two hadrons interacting with an external current.
Building off the known result for transitions involving $1+\Jc\to 2$ processes, we sum to all orders in the strong interaction, while working to leading order in the current insertion, to find an on-shell representation for $2+\Jc\to 2$ scattering.
The result is valid for any number of channels involving two spinless hadrons which couple to an arbitrary partial wave, as well as any structure for the external current.
Comparing to standard $2\to 2$ or $1+\Jc\to 2$ processes, the $2+\Jc\to 2$ transition amplitude contains, as well as the two-particle branch cut, an additional singular structure in the form of the triangle function.
The triangle function induces additional singularities stemming from on-shell intermediate states where one of the particles interacts with the external current.

We showed, given the on-shell $2+\Jc\to 2$ transition amplitude, we can rigorously define resonance form-factors by analytically continuing the amplitude to the unphysical Riemann sheet in both the initial and final state two-particle energies.
The transition amplitudes presented connect to the previously studied finite volume formalism~\cite{Briceno:2015tza,Baroni:2018iau} which links to matrix elements calculated with lattice QCD.
This allows us to ascertain structural information of resonant states, such as charge radii, in an EFT independent way.

We close by remarking that this work provides a key step towards understanding the analytic decomposition of more complicated transition amplitudes. In particular, one class of amplitudes that are pressing are those involving two-current insertions, $``{\rm in}"+\Jc^A\to ``{\rm out}" +\Jc^B $. These play an important role in our understanding of phenomena ranging from the nature of low-lying QCD states to extensions of the Standard Model. For instance, there is a growing demand to have reliable estimates of nuclear matrix elements pertinent for neutrino-less double beta decay \cite{Dekens:2020ttz}.~\footnote{For recent proposals and a detailed discussion for studying such amplitudes via lattice QCD, see Refs.~\cite{Briceno:2019opb, Christ:2015pwa,Feng:2020nqj,Davoudi:2020xdv,Davoudi:2020gxs}.} 

As one would expect, the analytic structure of  $``{\rm in}"+\Jc^A\to ``{\rm out}" +\Jc^B $ amplitudes will, in general, depend on the amplitudes for the allowed sub-processes. Depending on the nature of the $``{\rm in}"/``{\rm out}"$ states, these sub-processes will include those described by $\Mc, \Hc$, and $\Wc_\df$. This explains the claim made that understanding the analytic structure of $\Wc_\df$, among other things, is a necessary step towards constraining the aforementioned processes.

\section{Acknowledgements}

The authors would like to thank J. Dudek for useful comments on the manuscript, as well as M. Albaladejo, G. Blume, R. Edwards, M. Hansen, L. Leskovec, and the rest of the Hadron Spectrum Collaboration for useful discussions.
RAB and AWJ acknowledges support from U.S. Department of Energy contract DE-AC05-06OR23177, under which Jefferson Science Associates, LLC, manages and operates Jefferson Lab. 
RAB and KHS acknowledge support of the USDOE Early Career award, contract DE-SC0019229. 
FGO  acknowledges support from the U.S.\ Department of Energy contract DE-SC0018416 at William \& Mary and the JSA/JLab Graduate Fellowship Program. 
KHS acknowledges support by the U.S. Department of Energy, Office of Science Graduate Student Research (SCGSR) program. The SCGSR program is administered by the Oak Ridge Institute for Science and Education (ORISE) for the DOE. ORISE is managed by ORAU under contract number DE-SC0014664. All opinions expressed in this paper are the author’s and do not necessarily reflect the policies and views of DOE, ORAU, or ORISE.

\bibliographystyle{apsrev} 
\bibliography{bibi} 

\appendix

\section{Loop diagrams}

The derivations considered in this work rely on the correct identification of the analytic structure of the $s$-channel intermediate loop diagrams that contribute to the non-analytic part of the amplitudes. In this appendix we review the origin of these contributions. 

\subsection{Bubble loop}
\label{app:bubble}

We begin with the loop with no current insertions of Eq.~(\ref{eq:2to2.loop_id_on-shell}). This is the diagram that visually resembles a \emph{bubble} on the left hand side of Fig.~\ref{fig:loop_id}. It is well known that this loop leads to square-root singularities at the two-particle thresholds. We provide a derivation of this result for two reasons. The first is completeness. The second is the fact that the techniques used for the bubble diagram will also be used for the more complicated triangle diagram below.

We will use generic endcap functions $\Lc(P,k)$ and $\Rc(P,k)$, that can represent any of the different kernels used in our derivations. These are smooth for real energies and kinematics where two-particle states can go on-shell. Away from this limited region, they can have singularities, in particular associated with other thresholds. Here, $P$ denotes the total momentum of the system and $k$ is the momentum of one of the internal particles, this is illustrated in Fig.~\ref{fig:loop_id}. Furthermore, in general these will depend on the momenta of external states, which can be off-shell. Because the momenta of the external legs play no role in the expressions that follow, they will be left implicit.

Using these endcaps, the loop in full notation is written as
\begin{equation}
\Ic_0(P)=
\xi\int \frac{\diff^{4}k}{(2\pi)^{4}} i\Lc(P,k)
i\Delta_{1}(k)
i\Delta_{2}(P-k)
i\Rc(P,k) \, ,
\end{equation}
where the propagators $i\Delta_\alpha(k)$ and the symmetry factor $\xi$ are described in the main text.  

Our goal is not to provide an exact form of this integral, but rather to isolate its singular piece. To be able to evaluate the integral exactly would require having a closed form for the endcaps, which is not the case. Instead we recognize that the singularities of these diagrams emerge from intermediate particles going on-shell. As a result, the singular contribution will only depend on the endcaps evaluated on-shell. With this in mind, we define partial-wave projected on-shell endcaps that contain barrier factors, so as to avoid spurious singularities from the spherical harmonics,
\begin{align}
    i\overline{\Lc}(P,{\vec{k}}^{\star}) & \equiv \sum_{\ell,m_\ell}
    i\Lc_{\ell m_\ell}(P) \Yc^*_{\ell m_\ell}({\vec{k}}^{\star}) \,,\nn\\
    i\overline{\Rc}(P,{\vec{k}}^{\star}) & \equiv  \sum_{\ell,m_\ell}
    \Yc_{\ell m_\ell}({\vec{k}}^{\star})
    i\Rc_{\ell m_\ell}(P) \,,\label{eq:LR_onshell_bubble}
\end{align}
where the function $\Yc_{\ell m_\ell}$ is defined in Eq.~\eqref{eq:sph_w_barrier}. For values of ${\vec{k}}^{\star}$ putting the particles inside the loop on shell, these quantities are equal to the on-shell $\Lc$ and $\Rc$ respectively. The decomposition prescription of the off-shell kernels that we adopt in this case is
\begin{align}
i\Lc(P,k) &= i\overline{\Lc}(P,{\vec{k}}^{\star}) +[i\Lc(P,k)]\delta \, ,\nn\\
i\Rc(P,k) &= i\overline{\Rc}(P,{\vec{k}}^{\star}) +\delta[i\Rc(P,k)] \, ,
\label{eq:LR_onshell}
\end{align}
where in the first term on the rhs of each equation we have the kernel with the legs next to the loop evaluated on-shell, i.e.\ $k^2=m_1^2$ and $(P-k)^2=m_2^2$. The terms with a $\delta$ operator next to them are inspired by the notation introduced in Ref.~\cite{Briceno:2015tza}, in this case they will vanish when both internal legs of the loop are on-shell and therefore will be proportional to at least one factor of $(k^2-m_1^2)((P-k)^2-m_2^2)$. This means that a term with a $\delta$ operator times the propagators will be smooth. Furthermore, in order to obtain the singular contribution of the integral, we just need to consider the term with the propagators replaced by their singular pieces, 
\begin{equation}
\Ic_0(P) = \sum_{\ell,m_\ell,\ell^\prime,m_\ell^\prime}i\Lc_{\ell m_\ell}(P) 
\xi\int\! \frac{\diff^{4}k}{(2\pi)^{4}}\bigg( \Yc^*_{\ell m_\ell}(\hat{\vec{k}}^\star)
iD_{1}(k)
iD_{2}(P-k) \Yc_{\ell^\prime m_\ell^\prime}(\hat{\vec{k}}^\star)
\bigg)
i\Rc_{\ell^\prime m_\ell^\prime}(P) +\delta \Ic_0(P) \, ,
\end{equation}
where $\delta \Ic_0$ is purely smooth in the kinematic region of interest.

The singular contribution can be isolated from the first term above in a few different ways. First, one could evaluate the $k^0$ integral using Cauchy's theorem by closing the contour from below. The singular contribution would come from the pole at $\omega_{k1}\equiv\sqrt{\vec{k}^2+m_1^2}$ in $iD_1(k)$. The remaining three-dimensional integral can be written in spherical coordinates. The singular piece arises from the pole of the remaining propagator. By picking this contribution, one can evaluate the remaining spherical integral using the orthogonality relations of the spherical harmonics to find,
\begin{align}\label{eq:bubloopexp}
\Ic_0(P) 
&= \sum_{\ell,m_\ell}i\Lc_{\ell m_\ell}(P) \frac{\xi \, q^{\star}}{8\pi \sqrt{s}}i\Rc_{\ell m_\ell}(P) +\delta \Ic'_0(P) \, ,\nn\\
&= \sum_{\ell,m_\ell}i\Lc_{\ell m_\ell}(P)\, \rho_0\, i\Rc_{\ell m_\ell}(P) +\delta \Ic'_0(P) \, ,
\end{align}
where $\delta \Ic'_0$ contains all the non-singular terms and $\rho_0$ is the two-body phase space factor defined in Eq.~\eqref{eq:ps}. The application of this result is written diagrammatically in Fig.~\ref{fig:loop_id}. 

Alternatively, one can obtain the discontinuity using Cutkosky rules~\cite{Cutkosky:1960}, which amount to replacing the propagators with Dirac delta functions, 
\begin{equation}
    iD_1(k) \to 2\pi \delta(k^2-m_1^2)\theta(k^0)\,,\quad
    iD_2(P-k) \to 2\pi \delta((P-k)^2-m_2^2)\theta(P^0-k^0)\,.
\end{equation}
After doing the substitution, carrying out two integrals with the help of the Dirac deltas, and performing the remaining angular integration we obtain the well-known discontinuity of the bubble loop
\begin{equation}
    \frac{1}{2}\text{Disc}\, \Ic_0(P) =  \sum_{\ell,m_\ell}i\Lc_{\ell m_\ell}(P) \frac{\xi \, q^{\star}}{8\pi \sqrt{s}}\theta\left(s-s_\text{th}\right)i\Rc_{\ell m_\ell}(P)\,,
\end{equation}
which of course agrees with the result of Eq.~(\ref{eq:bubloopexp}).

We close by remarking that although we have in fact isolated the singular piece of the bubble diagram, which was our goal, there is a freedom as to how $\rho$ is defined. For example, one can could shift it by an overall real function, while simultaneously redefining $\delta \Ic'_0$, such that $\Ic_0$ is unchanged. One example of an alternative definition of $\rho$ includes the frequently used Chew-Mandelstam phase space~\cite{Chew:1960iv}. In the following section we will make this freedom explicit when defining the triangle singularity.

\subsection{Triangle loop\label{app:trian}}

Let us move on to the loop integral of Eq.~(\ref{eq:2Jto2.Tri.Rule_C}) that features a current insertion in one of the internal legs, the so-called triangle diagram,
\begin{equation}
\Ic_1(P_f,P_i,Q^2) = \int \frac{\diff^{4}k}{(2\pi)^{4}} i\Lc(P_f,k) i\Delta_1(k)i\Delta_2(k_f) iw(k_f,k_i) i\Delta_2(k_i) i\Rc(P_i, k) \, ,
\end{equation}
with $Q^2=-(P_f-P_i)^2$, and where we are using the same shorthand notations $k_i \equiv P_i - k$, $k_f \equiv P_f - k$ as in the main text. Again, we have left out the total energy dependence of the endcaps to simplify the notation, but it should be remembered that the current can insert momentum and the initial (final) four-momentum is $P_i$ ($P_f$). 

Although it is not absolutely necessary, we have simplified matters by assuming that the presence of the current only leads to elastic processes. In other words, the states with four-momenta $k_i$ and $k_f$ have the same mass. In general, this does not have to be the case, even when considering the insertion of electromagnetic current, but this can be readily generalized.

To extract the non-analytic behavior we will implement the second strategy that we used for the bubble loop. We perform the same separation of the on-shell part of the endcaps as in Eq.~\eqref{eq:LR_onshell_bubble} with the only modification that these now depend on different total momenta,
\begin{align}\label{eq:L_onshell_exp}
    i\overline{\Lc}(P_f,{\vec{k}}_f^{\star}) & \equiv \sum_{\ell,m_\ell}
    i\Lc_{\ell m_\ell}(P_f) \Yc^*_{\ell m_\ell}({\vec{k}}_f^{\star}) \,,\\
    i\overline{\Rc}(P_i,{\vec{k}}_i^{\star}) & \equiv  \sum_{\ell,m_\ell}
    \Yc_{\ell m_\ell}({\vec{k}}_i^{\star})
    i\Rc_{\ell m_\ell}(P_i) \,,\label{eq:R_onshell_exp}
\end{align}
where the subscript of the vectors indicates the frame where it is evaluated, since the initial and final CM frames can be different for a non-zero value of the current momentum insertion. The on-shell decomposition prescription follows from Eq.~\eqref{eq:LR_onshell} with the additional subscripts for the initial and final states,
\begin{align}
i\Lc(P_f,k) &= i\overline{\Lc}(P_f,{\vec{k}}^{\star}_f) +[i\Lc(P_f,k)]\delta \,,\nn\\
i\Rc(P_i,k) &= i\overline{\Rc}(P_i, {\vec{k}}^{\star}_i) +\delta[i\Rc(P_i,k)] \, .
\end{align}

In Ref.~\cite{Baroni:2018iau} it was realized that it is more convenient for the $\delta$ operator to not act on the whole current insertion $w$, but only on the scalar form-factors that contain the dynamics, such that we will need the Lorentz decomposition
\begin{equation}\label{eq:FF_Lor_dec}
iw(k_f,k_i) = \sum_j K_j(k_f,k_i)\,if_j(Q^2,k_f^2,k_i^2) \,,
\end{equation}
where $K_j$ are kinematic known functions of $P_f$, $P_i$ and $k$ that depend on the Lorentz structure of the current insertion. The on-shell expansion of these form-factors is
\begin{equation}\label{eq:FF_onshell_exp}
if_j(Q^2,k_f^2,k_i^2) = if_j(Q^2) + \delta[if_j(k_f^2,Q^2)] + [if_j(Q^2,k_i^2)]\delta +
\delta[if_j(k_f^2,k_i^2)] \delta \,,
\end{equation}
where the first term is the on-shell one-particle form-factor, i.e.\ the form-factor evaluated at $k_f^2=k_i^2=m^2_2$. The rest are terms that contain the off-shell behavior but vanish for values of $k$ when the initial or final state on-shell conditions are met, depending whether the term is acted by a $\delta$ operator from the left or the right, or both. The quantities
\begin{align}
    if_j(k_f^2,Q^2) &\equiv if_j(Q^2) + \delta[if_j(k_f^2,Q^2)]\,,\\
    if_j(Q^2,k_i^2) &\equiv if_j(Q^2) + [if_j(Q^2,k_i^2)]\delta\,,
\end{align}
will also simplify the clutter of the derivation. The explicit dependence of these quantities makes clear that they contain on-shell and off-shell information of the initial or final particle leg respectively. Finally, as done in Ref.~\cite{Briceno:2015tza} we introduce the following shorthand notation for endcaps with a current insertion in one of its external legs, and the divergent piece originating from the intermediate propagator subtracted,
\begin{align}
[i\Lc i\Delta_2if_j]_{\df}(P_f,k,Q^2) &=
i\Lc(P_f,k) i\Delta_2(k_f)if_j(k_f^2,Q^2)  - i\overline{\Lc}(P_f,{\vec{k}}^{\star}_f)iD_2(k_f)if_j(Q^2) \,,\\
[if_j i\Delta_2i\Rc]_{\df}(Q^2,P_i, k) &= if_j(Q^2,k_i^2)i\Delta_2(k_i) i\Rc(P_i,k) -if_j(Q^2) iD_2(k_i)i\overline{\Rc}(P_i,\vec{k}_i^{\star}) \, .
\end{align}
When the endcaps are $\Kc_0$, and each term is multiplied by $K_j$ and summed over $j$, these quantities correspond to the symbols $\Wb_{L|0}$ and $\Wb_{0|R}$ introduced in Eq.~\eqref{eq:2Jto2.W1b_Kmat_Rule1} of the main text. Given all these definitions, we can separate the integral $\Ic_1$ into terms with different analytic behavior
\begin{multline}\label{eq:I1lytsplit}
\Ic_1(P_f,P_i,Q^2) = \sum_j
\int \frac{\diff^4k}{(2\pi)^4}K_j\bigg (
i\overline{\Lc}(P_f,{\vec{k}}^{\star}_f)iD_1(k)iD_2(k_f)if_j(Q^2)iD_2(k_i)i\overline{\Rc}(P_i,{\vec{k}}^{\star}_i)\\
+[i\Lc i\Delta_2if_j]_{\df}(P_f,k,Q^2)iD_1(k)iD_2(k_i) 
i\Rc(P_i,k) \\
+i\Lc(P_f,k) iD_1(k)iD_2(k_f)[if_j i\Delta_2i\Rc]_{\df}(Q^2,P_i,k) +\dots
\bigg) \, ,
\end{multline}
where terms that do not appear explicitly are smooth analytic functions of the external momenta within the kinematic region of interest.

At this point there is not a unique choice about how to distill the singularities of these terms. Different choices would mean different functional forms of the analytic part of the amplitude, but all choices have to agree on the position and characteristics of any singularity in the amplitude. This freedom was emphasized at the end of the previous section. In here, we will write down two options, and the different smooth behaviors will be collected in the implicit definitions of smooth functions introduced below.
Both of these prescriptions agree that this integral can be diagrammatically represented by Fig.~\ref{fig:triangle_decomposition}.
The first, most conservative option, requires only the partial wave expansion of the endcap functions as we did before
\begin{multline}
\Ic_1(P_f,P_i,Q^2) =
\sum_{\ell,m_\ell,\ell',m^\prime_\ell} 
\Lc(P_f)_{\ell m_\ell}
\sum_j\big[if_j(Q^2)\Gc_{j;\ell m_\ell;\ell' m^\prime_\ell}(P_f,P_i)\big]
\Rc(P_i)_{\ell' m^\prime_\ell}
\\+ 
\sum_{\ell,m_\ell}\bigg(
[i\Lc i\Delta_2iw]_{\df,\ell m_\ell}(P_f,Q^2)\frac{\rho_i}{\xi}i\Rc_{\ell m_\ell}(P_i) +
 i\Lc_{\ell m_\ell}(P_f) \frac{\rho_f}{\xi}[iw i\Delta_2i\Rc]_{\df,\ell m_\ell}(Q^2,P_i)\bigg) 
+\delta \Ic_1(P_f,P_i,Q^2) \, ,
\label{eq:triangle_master}
\end{multline}
where $\delta \Ic_1$ is a residual, smooth function. The subscript in the phase space $\rho$ indicates whether $\rho_0$ is evaluated in Eq.~(\ref{eq:ps}) with $s_i$ or $s_f$. We have also introduced the partial-wave decomposition of the divergence-free kernels. This decomposition is similar to the ones from the simple kernels, except that we first multiply by the kinematic functions $K_j$, and sum over all $j$,
\begin{align}
  \sum_j K_j   \, [i\Lc i\Delta_2if_j]_{\df}(P_f,k,Q^2) & =[i\Lc i\Delta_2iw]_{\df}(P_f,\hat{\mathbf{k}}_i^{\star},Q^2) + [[i\Lc i\Delta_2iw]_{\df}(P_f,k,Q^2)]\delta \,,  \\
  [i\Lc i\Delta_2iw]_{\df}(P_f,\hat{\mathbf{k}}_i^{\star},Q^2) \,   & = \sqrt{4\pi}  \sum_{\ell, m_{\ell}} [i\Lc i\Delta_2iw]_{\df,\ell m_\ell}(P_f,Q^2) \, Y^*_{\ell m_{\ell}} (\hat{\mathbf{k}}_i^{\star})\,,
\end{align}
an analogous relation holds for $[iw i\Delta_2i\Rc]_{\df,\ell m_\ell}(P_i,Q^2)$.

This prescription for the isolation of the triangle singularity requires the calculation of the following matrix in angular momentum space
\begin{equation}\label{eq:Gfun}
\Gc_{j;\ell m_\ell;\ell' m^\prime_\ell}(P_f,P_i) = 
i\int \frac{\diff^4k\, }{(2\pi)^4}
K_j
\Yc_{\ell m_\ell}^*({\vec{k}}_f^{\star})D_1(k)D_2(k_f)D_2(k_i)\Yc_{\ell' m^\prime_\ell}({\vec{k}}_i^{\star})\,,
\end{equation}
which in general needs to be regularized for ultraviolet divergences, the needed counterterms would come from $\delta \Ic_1$ since they have to be analytic functions. For more details about this matrix, and how to efficiently evaluate and regularize it, the interested reader is referred to Ref.~\cite{Baroni:2018iau}.

The second option to obtain the singularities from Eq.~(\ref{eq:I1lytsplit}) is to use the Cutkosky rules to extract discontinuities of the loops, and verify with the Landau conditions \cite{Landau:1959fi} that all the singularities associated with the diagram are being described. In the case of the triangle loop, the Cutkosky rules require at least three cuts, one for each of the vertices. Since we are interested in the kinematic region where the current insertion energy is below the particle creation threshold, only two cuts contribute to the discontinuities associated with the triangle loop. These contain the branch-cuts associated with the initial/final two-particle states. 

As detailed discussion of Cutkosky rules can be found in \cite{VeltmanMartinus1994D:tp}, instead of jumping straight to the result, we provide some  of the  key  steps  that  arrive  one  to  the  final form. The two key conceptual points are the following. First, one must recognize that the discontinuity may only be obtained after replacing $D_1(k)$ by its imaginary piece. Second, having made this replacement, the discontinuity can be obtained by evaluating the difference across the $s_i$ and $s_f$ branch cuts. This can be made clear by introducing explicitly the dependence of the propagators on $\epsilon$, which will be fixed to be positive. For example, we intermediately replace $D_2(k_f)\to D_2(k_f,\epsilon)$, and take the limit as $\epsilon\to0$ afterwards. 

More explicitly, the discontinuity of the triangle loop is, %
\begin{align}
    \text{Disc}\,\Gc_{j;\ell m_\ell;\ell' m^\prime_\ell}(P_f,P_i) 
    &=
\int \frac{\diff^4k}{(2\pi)^3}K_j\Yc_{\ell m_\ell}^*({\vec{k}}^{\star}_f)
\delta(k^2-m^2_1)\theta(k^0)
 \bigg [
D_2(k_f,\epsilon)D_2(k_i,\epsilon)-D_2(k_f,-\epsilon)D_2(k_i,-\epsilon)\bigg]
\Yc_{\ell' m^\prime_\ell}({\vec{k}}_i^{\star})\,
\nn\\
    &=
\int \frac{\diff^3\vec{k}}{(2\pi)^3 2\omega_{k1}}K_j\Yc_{\ell m_\ell}^*({\vec{k}}^{\star}_f)
 2i \, \,{\rm Im}\bigg [
D_2(k_f,\epsilon)D_2(k_i,\epsilon) \bigg]
\Yc_{\ell' m^\prime_\ell}({\vec{k}}_i^{\star})
\bigg|_{k^0=\omega_{k1}}\,
\nn\\
   &=
- 2\pi i\int \frac{\diff^3\vec{k}}{(2\pi)^3 2\omega_{k1}}K_j\Yc_{\ell m_\ell}^*({\vec{k}}^{\star}_f)
 \, \,\bigg [
\delta(k_f^2-m^2_2)\theta(k_f^0)
\,{\rm P.V.} [D_2(k_i)]
\nn\\
&
\hspace{3.5cm}
+
{\rm P.V.} [D_2(k_f)]\,\,\delta(k_i^2-m^2_2)\theta(k_i^0)
\bigg] 
\Yc_{\ell' m^\prime_\ell}({\vec{k}}_i^{\star})
\bigg|_{k^0=\omega_{k1}}\,,
\end{align}
where we have used the following identity in the last step, 
\begin{equation}
    D_{\alpha}(k)={\rm P.V.}[D_{\alpha}(k)] - i\pi\delta(k^2-m_{\alpha}^2)\,.
\end{equation}

The three-vector integral is easiest to carry out in the CM frame where the second delta Dirac of each cut imposes the on-shell condition, i.e.\ $k^\star_{f}=q_{f}^\star$ or $k^\star_{i}=q_{i}^\star$ respectively. After that, we are left only with the angular integral
\begin{equation}
\text{Disc}\,\Gc_{j;\ell m_\ell;\ell' m^\prime_\ell}(P_f,P_i) =
-\frac{iq^\star_f\theta\left(s_f-s_\text{th}\right)}{4\pi \sqrt{s_f}}
{\rm P.V.}
\int \frac{\diff \Omega^\star_f}{4\pi}
K_j
\Yc^*_{\ell m_\ell}({\mathbf{k}}_f^{\star})
D_2(k_i)\Yc_{\ell^\prime m^\prime_\ell}({\mathbf{k}}_i^{\star})\bigg|_{k^0=\omega_{k1},k^\star_f=q^\star_f} +\dots\,,
\end{equation}
where the dots indicate the second term, which is identical to the first one by exchanging the initial and final labels. In this integral the pole of the remaining propagator follows the principal value integration prescription. The angular dependence of the numerator can be expressed in terms of a single spherical harmonic and a set of $B$ coefficients
\begin{equation}\label{eq:Bcoeff}
K_j
\Yc^*_{\ell m_\ell}({\mathbf{k}}_f^{\star})
\Yc_{\ell^\prime m^\prime_\ell}({\mathbf{k}}_i^{\star})
\bigg|_{k^0=\omega_{k1},k^\star_f=q^\star_f}
= \frac{\sqrt{4\pi}}{q^{\star\ell}_fq^{\star\ell^\prime}_i}\sum_{JM} B^{JM,f}_{j;\ell m_\ell;\ell^\prime m^\prime_\ell} q^{\star J}_f Y_{JM}(\hat{\mathbf{k}}_f^{\star})\,,
\end{equation}
where a similar decomposition but for the initial CM frame define the $B^{JM,i}$ coefficients. The $B^{JM,f}$ coefficients will include a relation from the spherical harmonics in the initial frame to 4-vectors, Lorentz boosts to the final frame, the relation from 4-vectors in the final frame to spherical harmonics, and the recombination of all the spherical harmonics into a single one. Explicit calculations of these coefficients are done in App.~B1 of Ref. \cite{Baroni:2018iau}, they will depend only on the value of $P_f$ and $P_i$. 

Note that there is a final spatial rotation $R(\theta,\phi)$ ambiguity of the Lorentz vectors in the final CM frame, we will choose the Lorentz transformation that leaves the spatial part of $P_i$ pointing in the $z$-direction. For instance, in the case of a vector current insertion, we would use the following relationship
\begin{equation}
k^\mu = [\Lambda_{-\boldsymbol \beta_f}]^\mu{}_\nu [R(\theta_i,\phi_i)]^\nu{}_\rho k^{\star\rho}_f = [\Lambda_{-\boldsymbol \beta_f} R(\theta_i,\phi_i)]^\mu{}_\nu k^{\star\nu}_f\,,
\end{equation}
where the direction $(\theta_i,\phi_i)$ refers to that of the the spatial part of the four vector $P_{i}^\mu$ when acted by the pure boost $[\Lambda_{\boldsymbol \beta_f}]$. This boost is such that when it acts on $P_{f}^\mu$, it provides $[\Lambda_{\boldsymbol \beta_f}]^\nu{}_\mu P_{f}^\mu = P^{\star\nu}_{f} =(E^\star_f,0)^\nu$.

Similarly, we need to make explicit the angular dependence of the remaining propagator in the respective CM frame. Since we have defined the transformation into the final CM frame such that $P_{i,f}^{\mu\star}=(P_{i,f}^{0\star},|\mathbf{P}_{i,f}^\star|\hat{\mathbf{z}})$, the propagator in this frame is equal to
\begin{align}
    D_2(k_i)
&= \frac{1}{(P^{0\star}_{i,f}-\omega^\star_{q1,f})^2-\mathbf{P}_{i,f}^{\star2}- q_{f}^{\star2}+2|\mathbf{P}_{i,f}^\star|q_{f}^{\star}\cos\theta^\star_f-m_2^2+i\epsilon} \, , \\
&= \frac{1}{2|\mathbf{P}_{i,f}^\star|q_{f}^{\star}(\cos\theta^\star_f-z^\star_f+i\epsilon)}\,,
\end{align}
where $z^\star_f$, in terms of Lorentz scalars, is given by
\begin{equation}
\label{eq:zfstarcov}
z^\star_f =
\frac{P_i\cdot P_f-s_i+(m_2^2-m_1^2)(1-P_i\cdot P_f/s_f)}{\frac{2q_f^\star}{\sqrt{s_f}}\sqrt{(P_f\cdot P_i)^2-P_i^2P_f^2}} \, .
\end{equation}
In arriving at this result, we have two identities to rewrite $\mathbf{P}_{i,f}^\star$ and other quantities above in a Lorentz invariant way, 
\begin{align}
    \sqrt{s_f}|\mathbf{P}_{i,f}^\star| &=
     \sqrt{s_i}|\mathbf{P}_{f,i}^\star| =\sqrt{(P_f\cdot P_i)^2-P_i^2P_f^2}\,,
     \nn\\
    2 P^{0\star}_{i,f} \omega^\star_{q1,f} &= P_i\cdot P_f \left(1-\frac{m_2^2-m_1^2}{s_f}\right) \,  .  
\end{align}
The quantity $\sqrt{(P_f\cdot P_i)^2-P_i^2P_f^2}$ is also equal to $\lambda^{1/2}(-Q^2,s_i,s_f)/2$ \cite{2007PhRvD..75a6001L}.

The azimuthal angle integral is proportional to $\delta_{M,0}$ since the only azimuthal dependence is contained in the spherical harmonics, while the polar angle integral is proportional to the Legendre functions of the second kind
\begin{equation}\label{eq:leg2type}
Q_J(z^\star_f)=\frac{1}{2}\text{P.V.} \int_{-1}^1 \frac{\diff z\, P_J(z)}{z_f^\star-z}\,.
\end{equation}
Special care has to be taken when evaluating $Q_J$ at threshold, or when the initial and final CM frame are the same, i.e.\ when $|\mathbf{P}_{i,f}^\star|=0$. Since $z^\star_f\propto 1/(q_f^\star |\mathbf{P}_{i,f}^\star|)$, in either of these cases, the argument of $Q_J$ diverges. However it can be found, either by using the series expansion of the Legendre functions of the second type, or the expansion of the fraction in Eq.~(\ref{eq:leg2type}) before performing the integration, that 
\begin{equation}
\lim_{x\to 0}    \frac{1}{x}Q_J\left(\frac{1}{x}\right) = \delta_{J,0} \,,
\end{equation}
so that even at $x=0$ a finite value can be assigned to the discontinuity.

Putting all the terms together we arrive to the discontinuity of the triangle diagram to be 
\begin{multline}\label{eq:DiscG}
    \text{Disc}\,\Gc_{j;\ell m_\ell;\ell' m^\prime_\ell}(P_f,P_i) =
    \frac{ i}{8\pi \sqrt{(P_f\cdot P_i)^2-P_i^2P_f^2}}
\sum_{J} \frac{\sqrt{2J+1}}{q^{\star\ell}_fq^{\star\ell^\prime}_i}
\bigg(q^{\star J}_f \theta\left(s_f-s_\text{th}\right) B^{J0,f}_{j;\ell m_\ell;\ell^\prime m^\prime_\ell} Q_J(z^\star_f)\\
+q^{\star J}_i
\theta\left(s_i-s_\text{th}\right)
B^{J0,i}_{j;\ell m_\ell;\ell^\prime m^\prime_\ell} Q_J(z^\star_i)
\bigg)
\,.
\end{multline}
In the case when the CM frame of the initial and final state coincide, i.e.\ $\boldsymbol \beta_i=\boldsymbol \beta_f$, the discontinuity only depends on $s_i$ and $s_f$, and simplifies to
\begin{equation}
\text{Disc}\,\Gc_{j;\ell m_\ell;\ell' m_\ell'}(s_f,s_i) =
\frac{i}{4\pi q^{\star\ell}_fq^{\star\ell'}_i}
\frac{\theta(s_f-s_\text{th})q^{\star}_f B^{00,f}_{j;\ell m_\ell;\ell' m_\ell'}
-\theta(s_i-s_\text{th})q^{\star}_iB^{00,i}_{j;\ell m_\ell;\ell' m_\ell'}}{(\sqrt{s_f}-\sqrt{s_i})(\sqrt{s_i s_f}+m_2^2-m_1^2)}
\,.
\end{equation}
Finally, for equal momenta of the initial and final states, i.e.\ $P_i=P_f=P$, the discontinuity only depends on $s=P^2$ and is equal to
\begin{equation}
  \text{Disc}\,\Gc_{j;\ell m_\ell;\ell' m_\ell'}(s) =
\frac{i\theta(s-s_\text{th})\sqrt{s}}{2\pi q^{\star\ell+\ell'}(s+m_2^2-m_1^2)}
\frac{\partial}{\partial s}\left(q^{\star} B^{00}_{j;\ell m_\ell;\ell' m_\ell'}\right)
\,.  
\end{equation}
The step function is able to move out of the derivative operator because the threshold behavior of $B^{00}_{j;\ell m_\ell;\ell' m_\ell'}$ is proportional to at least $\ell+\ell'$ powers of $q^\star$, see Eq.~(\ref{eq:Bcoeff}).

The matrix $\text{Disc}\,\Gc$ in angular momentum space is equal to twice the imaginary part of $\Gc$, modulo the phases inherited by the spherical harmonics encoded in the $B$ coefficients. Singularities in this matrix arise from the $Q_J$ functions, which generically feature a regular polynomial part $Q_J^r$, and a non-analytic piece proportional to a Legendre polynomial of the same $J$
\begin{equation}
    Q_J(x) = Q^r_J(x) + \frac{1}{2}P_J(x)
\log\left|{\frac{1+x}{1-x}}\right|\,.
\end{equation}
A careful inspection of the behavior of the $z^\star$ variables as a function of the external kinematics, and the matrix $\text{Disc}\, \Gc$ for the general case, yields that only when $z^\star_f=z^\star_i=1$ there is a logarithmic singularity, and that $\text{Disc}\,\Gc$ generically features square-root-type singularities at threshold. This is in agreement with the analysis of the Landau conditions of this diagram, see Ref.~\cite{2020PrPNP.11203757G} for a detailed review of this procedure. What is more, the imaginary part of the loop yields all the information about its singularities, i.e.\ its nature and their coefficients. Once we know the coefficient of each singularity, we can make a continuation of the $\text{Disc}\,\Gc$ function to also reproduce its real part.

\begin{figure}
\hfil
\subfloat[\label{fig:ztraj.1}A trajectory over the $E_i^\star$, $E_f^\star$ plane.]{\includegraphics{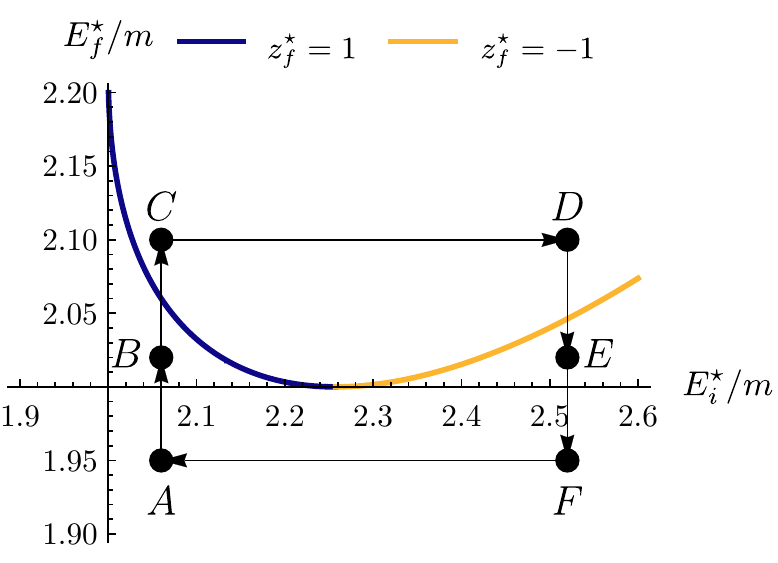}}
\hfil
\subfloat[\label{fig:ztraj.2}Phase of $\log((1+z_f^\star)/(1-z_f^\star))$ over the complex $z_f^\star$ plane.]{\includegraphics[width=.4762\textwidth]{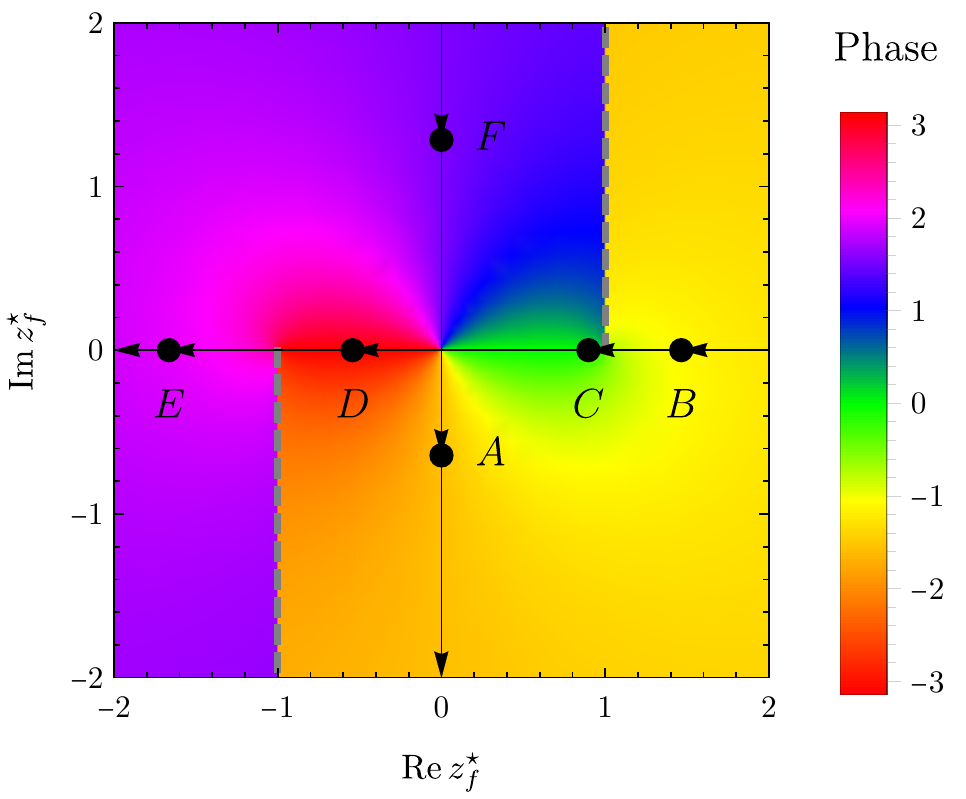}}
\hfil
\caption{\label{fig:ztraj}Behavior of the $z_f^\star$ variable in a closed trajectory over the $E_i^\star$, $E_f^\star$ plane for particles of equal mass $m$ and $|\vec{P}_{f,i}^\star|=(2\pi/6)m$.}
\end{figure}

However, given the behavior of the $z^\star$ variables as a function of the external kinematics, only a specific continuation around the branch points of the $Q_J$ function is consistent. To illustrate this we will describe the behavior of the $z_f^\star$ variable while moving in the trajectory shown in Fig.~\ref{fig:ztraj.1} in the $E_i^\star$, $E_f^\star$ plane with fixed $|\vec{P}_{f,i}^\star|=(2\pi/6) m$. 
To make this exercise more explicit, we rewrite $z^\star_f$, defined in Eq.~\eqref{eq:zfstarcov} in terms of $E_i^\star$, $E_f^\star$ and $|\vec{P}_{f,i}^\star|$
\begin{equation}
\label{eq:zfstarcov_v2}
z^\star_f =
\frac{E_f^\star\sqrt{E_i^{\star2}+|\mathbf{P}_{i,f}^\star|^2}-E_i^{\star2}+(m_2^2-m_1^2)\left(1-\frac{\sqrt{E_i^{\star2}+|\mathbf{P}_{i,f}^\star|^2}}{E_f^\star}\right)}
{2q_f^\star\,|\mathbf{P}_{i,f}^\star| } \, .
\end{equation}
Six points labeled $A$ through $F$ have been chosen on this plane. Correspondingly, the value of $z_f^\star$ at each of these points has been placed on the complex $z_f^\star$ plane on Fig.~\ref{fig:ztraj.2}. The color background on the latter figure represents the phase of the function $\log((1+z_f^\star)/(1-z_f^\star))$, which generates the Riemann sheet structure to all $Q_J$ functions. In this figure we have chosen to push the branch cut of $Q_J$, which conventionally runs from $-1$ to $1$, to run from $-1$ to infinity in the negative imaginary semiplane and then come back to $1$ through the positive imaginary semiplane. The branch cut is indicated by a gray dashed line.~\footnote{Since most software places the branch cut of the logarithms on the negative real axis, our choice of branch cut is implemented numerically with the function $\log(-i(z_f^\star+1))-\log(i(z_f^\star-1))$.} This choice, as shown shortly, will allow the variable $z_f^\star$ to remain on the same sheet for the values of $E_i^\star$ and $E_f^\star$ within our kinematic region of interest. 

To describe the behavior of $z_f^\star$ let us begin in the kinematic region when $E_f^\star$ is below threshold. By extending the domain of $q^\star_f$ below threshold within the physical Riemann sheet in the $s_f$ complex plane, one sees from Eq.~\eqref{eq:zfstarcov_v2} that $z_f^\star$ becomes purely imaginary, and takes positive or negative imaginary values depending on the value of $E_i^\star$. As a result, when moving from point $F$ to point $A$, $z_f^\star$ will pass through zero, motivating the choice to not have a branch cut there. When moving from point $A$ to point $B$, one must cross the threshold of the final two-particles states, at which points $z_f^\star$ diverges, see Eq.~\eqref{eq:zfstarcov_v2}. Given that there is no branch point at infinity in the $z_f^\star$ plane, one should remain in the same sheet when making this move from $A$ to $B$. This motivates having the negative imaginary infinity and the positive real infinity on the same side of the cut. In the trajectory $BCDE$ there are \textit{a priori} four options to go around the branch points, but only by going below the branch point at $1$ and above the branch point at $-1$, as shown in the figure, the points $E$ and $F$ will be connected to form a closed trajectory. This choice around the branch points can be encoded as an addition of $i\epsilon$ to the argument of $Q_J$. As a short hand we will introduce $Q^c_J$ as our choice of the analytic continuation of this function
\begin{equation}
    Q_J^c(x) = Q^r_J(x) + \frac{1}{2}P_J(x)
\log\left({\frac{1+x+i\epsilon}{1-(x+i\epsilon)}}\right)\,,
\end{equation}
where no absolute value of the argument of the logarithm is taken, and its range is extended into the complex plane.

With this information we can describe all the singular behavior of $\Gc$ analytically with
\begin{multline}\label{eq:SingG}
 \text{Sing}\,\Gc_{j;\ell m_\ell;\ell' m^\prime_\ell}(P_f,P_i) =
    \frac{ i}{16\pi \sqrt{(P_f\cdot P_i)^2-P_i^2P_f^2}}
\sum_{J} \frac{\sqrt{2J+1}}{q^{\star\ell}_fq^{\star\ell^\prime}_i}
\bigg(q^{\star J}_f
B^{J0,f}_{j;\ell m_\ell;\ell^\prime m^\prime_\ell}
Q^c_J(z^\star_f)
\\
+q^{\star J}_i
B^{J0,i}_{j;\ell m_\ell;\ell^\prime m^\prime_\ell}
Q^c_J(z^\star_i)
\bigg)
\,,
\end{multline}
in the case of arbitrary external kinematics, within our region of interest. Note the relative factor of 2 compared with the discontinuity, given in Eq.~\eqref{eq:DiscG}. This is because, as previously mentioned, the discontinuity is equal to twice the imaginary piece of $\Gc$. For the case where CM frame of final and initial state coincide it simplifies further to
\begin{align}
\text{Sing}\,\Gc_{j;\ell m_\ell;\ell' m_\ell'}(s_f,s_i) &=
\frac{i}{8\pi q^{\star\ell}_fq^{\star\ell'}_i}
\frac{q^{\star}_f B^{00,f}_{j;\ell m_\ell;\ell' m_\ell'}
-q^{\star}_iB^{00,i}_{j;\ell m_\ell;\ell' m_\ell'}}{(\sqrt{s_f}-\sqrt{s_i})(\sqrt{s_i s_f}+m_2^2-m_1^2)}
\,,\\
 \text{Sing}\,\Gc_{j;\ell m_\ell;\ell' m_\ell'}(s) &=
\frac{i\sqrt{s}}{4\pi q^{\star\ell+\ell'}(s+m_2^2-m_1^2)}
\frac{\partial}{\partial s}\left(q^{\star} B^{00}_{j;\ell m_\ell;\ell' m_\ell'}\right)
\,,
\end{align}
where the last equation holds for $s_f=s_i=s$.

Hence, our second option to express the analytic behavior of the triangle loop is
\begin{multline}
\Ic_1(P_f,P_i,Q^2) =
\sum_{\ell,m_\ell,\ell',m^\prime_\ell} 
i\Lc(P_f)_{\ell m_\ell}
\sum_j\big[if_j(Q^2)
\text{Sing}\,\Gc_{j;\ell m_\ell;\ell' m^\prime_\ell}(P_f,P_i)
\big]
i\Rc(p)_{\ell' m^\prime_\ell}
\\+ 
\sum_{\ell,m_\ell}\bigg(
[i\Lc i\Delta_2iw]_{\df,\ell m_\ell}(P_f,Q^2)\frac{\rho_i}{\xi}i\Rc_{\ell m_\ell}(p) +
 i\Lc_{\ell m_\ell}(P_f) \frac{\rho_f}{\xi}[iw i\Delta_2i\Rc]_{\df,\ell m_\ell}(Q^2,p)\bigg) \\
+\delta \Ic_1'(P_f,P_i,Q^2) \, ,
\end{multline}
where we capture the smooth contributions in $\delta \Ic_1'$. This is the main result from this section.

\begin{figure}[htbp!]
\centering
\includegraphics{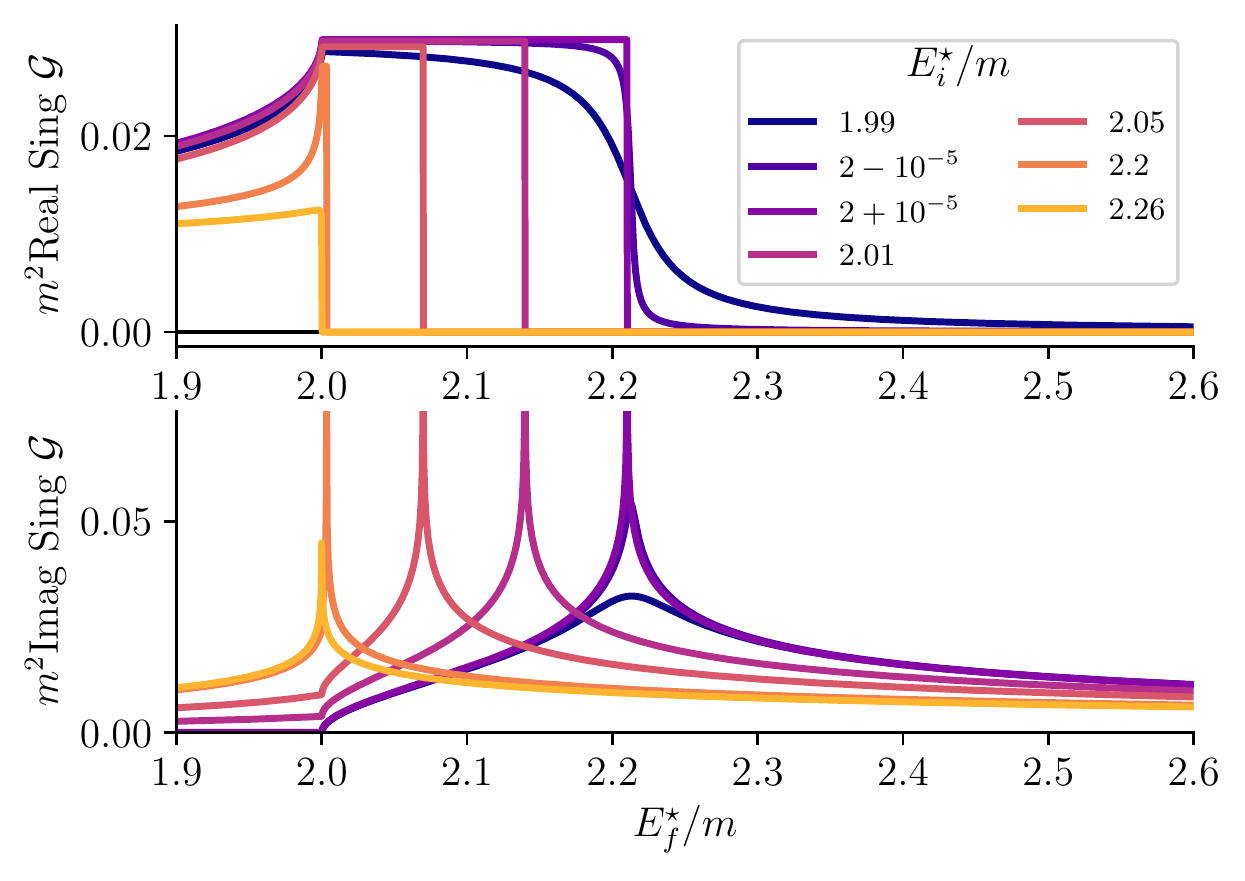}
\caption{Singularities of the scalar and $S$ wave case of the $\Gc$ loop integral as reproduced by the function $\text{Sing}\,\Gc_{00;00}$. The value of the spatial momentum $|\vec{P}_{f,i}^\star|=(2\pi/6)m$.}
\label{fig:DiscpStep}
\end{figure}

Finally, let us calculate two explicit examples of the $\text{Sing}\,\Gc_{j;\ell m_\ell;\ell' m_\ell'}$ loop. First, the case of a purely scalar kinematic function $K_j=1$, and the $S$ wave for both the initial and final partial waves. In that case, the integral Eq.~(\ref{eq:Gfun}) converges, and can be calculated analytically up to a single integration of a Feynman parameter. In this case the $B$ coefficients are simply equal to $B^{J0,f}_{00;00}=B^{J0,i}_{00;00}=\delta_{J,0}$ and the singular behavior of the triangle diagram is described by
\begin{equation}\label{eq:G0000_app}
\text{Sing}\,\Gc_{00;00}(P_f,P_i) = \frac{i}{32\pi \sqrt{(P_f\cdot P_i)^2-P_i^2P_f^2}}\,
\left[\log\left({\frac{1+z^\star_f +i\epsilon}{1-(z^\star_f+i\epsilon)}}\right)
+
\log\left({\frac{1+z^\star_i+i\epsilon}{1-(z^\star_i+i\epsilon)}}\right)\right]\,,
\end{equation}
A plot of $\text{Sing}\,\Gc_{00;00}$ as a function of $E_f^\star$ is shown in Fig.~\ref{fig:DiscpStep} for multiple values of $E_i^\star$ and spatial momenta $|\vec{P}_{f,i}^\star|=2\pi/6$. The logarithmic singularities can be seen as divergences in the imaginary part, while a step shows up in the real part at the same energy. The imaginary part coincides exactly to what is found by evaluating explicitly $\Gc$, while the real part only differs by smooth functions.  This can be verified, for instance, by comparing it to Fig.~7 of Ref.~\cite{Baroni:2018iau} where the loop integral was calculated with Feynman parameters and a final one-dimensional numerical integration.

The case $s_i=s_f=s$ is found as the limit of Eq.~\eqref{eq:G0000_app} to be
\begin{equation}
      \text{Sing}\,\Gc_{00;00}(s) =\frac{i}{32\pi \sqrt{s}q^{\star}}
\frac{s-m_2^2+m_1^2}{s}\,.
\end{equation}

The other case of interest is the vector current insertion with $K_j=k^\mu$. When thinking of the initial and final states to be in an $S$ wave, one needs
\begin{align}
B^{\mu;00,f(i)}_{00;00} &= [\Lambda_{-\boldsymbol \beta_{f(i)}} R(\theta_{i(f)},\phi_{i(f)})]^\mu{}_0\omega_{q1,f(i)}^\star\,, \qquad
B^{\mu;10,f(i)}_{00;00} = [\Lambda_{-\boldsymbol \beta_{f(i)}} R(\theta_{i(f)},\phi_{i(f)})]^\mu{}_k\frac{[\vec{\hat{z}}]^k}{\sqrt{3}} \, , \\
B^{\mu;00,f(i)}_{00;00} &= [\Lambda_{-\boldsymbol \beta_{f(i)}}]^\mu{}_0\omega_{q1,f(i)}^\star\,, \qquad
B^{\mu;10,f(i)}_{00;00} = [\Lambda_{-\boldsymbol \beta_{f(i)}} ]^\mu{}_k\frac{[\hat{\mathbf{P}}_{i,f(f,i)}^\star]^k}{\sqrt{3}} \, ,
\end{align}
and the singularities of $\Gc^\mu_{00;00}$ are captured by the function
\begin{multline}\label{eq:Gvecdisc}
\text{Sing}\,\Gc^\mu_{00;00}(P_f,P_i) =\frac{i}{32\pi \sqrt{(P_f\cdot P_i)^2-P_i^2P_f^2}}
\bigg(\log\left({\frac{1+z^\star_f +i\epsilon}{1-(z^\star_f+i\epsilon)}}\right)
\left([\Lambda_{-\boldsymbol \beta_{f}}]^\mu{}_0\omega_{q1,f}^\star + q^{\star}_f[\Lambda_{-\boldsymbol \beta_{f}}]^\mu{}_k[\hat{\mathbf{P}}_{i,f}^\star]^k z_f^\star\right)\\
+\log\left({\frac{1+z^\star_i+i\epsilon}{1-(z^\star_i+i\epsilon)}}\right)
\left([\Lambda_{-\boldsymbol \beta_{i}}]^\mu{}_0\omega_{q1,i}^\star +q^{\star}_i[\Lambda_{-\boldsymbol \beta_{i}}]^\mu{}_k[\hat{\mathbf{P}}_{f,i}^\star]^k
z_i^\star\right)
\\
-2 q^{\star}_f[\Lambda_{-\boldsymbol \beta_{f}}]^\mu{}_k[\hat{\mathbf{P}}_{i,f}^\star]^k
-2q^{\star}_i[\Lambda_{-\boldsymbol \beta_{i}}]^\mu{}_k[\hat{\mathbf{P}}_{f,i}^\star]^k
\bigg)\,.
\end{multline}
In the limit of $P_f=P_i=P$ the above function simplifies to
\begin{equation}
\text{Sing}\,\Gc^\mu_{00;00}(P)= \frac{i P^\mu}{16\pi s^{3/2} q^{\star}}
\left(2q^{\star 2} + m_1^2\right)\,.
\end{equation}

\section{Analytic continuations to unphysical sheets}
\label{app:analytic_continuations}

In this appendix, we review the analytic continuation of a two particle scattering amplitude to the unphysical sheet, and illustrate how the procedure extends to the transition amplitudes.
Physical scattering amplitudes on the real $s$ axis are boundary values of an analytic function, which has a discontinuity across the branch cut given by the unitarity relation Eq.~\eqref{eq:2to2.unitarity}.
Therefore, we can formally define the second sheet amplitude by continuing through the branch cut, cf. Eq.~\eqref{eq:2to2.unitarity}, using the boundary condition
\begin{align}
    \label{eq:app.boundary}
    \Mc^{\mathrm{II}}(s_{\pm}) = \Mc(s_{\mp}) \,,
\end{align}
where we have defined the short-hand notation $s_{\pm} = s\pm i \epsilon$ and assume that $\epsilon \to 0^{+}$. 
Using this short-hand, the unitarity relation can be expressed as
\begin{align}
\label{eq:app.2to2.unitarity.v2}
\Mc(s_+) - \Mc(s_-) = 2i\,\rho(s_+)  \Mc(s_-) \Mc(s_+) \, ,
\end{align}
where we have used the Schwartz reflection principle $\Mc^{*}(s) = \Mc(s^{*})$.
For technical convenience, we choose to continue the amplitude to the second sheet in the upper-half $s$ plane, i.e. $\Mc^{\mathrm{II}}(s_+) = \Mc(s_-)$.
We then use the Schwartz reflection principle to extend the result to the lower-half $s$ plane, which is nearest to the physical region assuming the usual $+i\epsilon$ prescription.
The result is identical if one chooses to continue to the lower-half plane directly, however we find this approach convenient to simplify the later derivation for the $2+\Jc\to 2$ amplitude.
Assuming a continuation to the upper-half plane, we now insert Eq.~\eqref{eq:app.boundary} into \eqref{eq:app.2to2.unitarity.v2}, and solve for $\Mc^{\mathrm{II}}$ to find
\begin{align}
    \label{eq:app.2to2.second_sheet}
    \Mc^{\mathrm{II}}(s) = \frac{1}{1 + 2i \, \rho(s) \Mc(s)} \Mc(s) \, ,
\end{align}
where we have used Cauchy's theorem to extend the domain from near the real axis to the upper-half complex plane.
One can make similar arguments for the lower-half plane, with an additional boundary condition for the phase space factor $\rho(s_+) = -\rho(s_-)$, finding the same form as Eq.~\eqref{eq:app.2to2.second_sheet}.

Alternatively, we may analytically continue the on-shell form Eq.~\eqref{eq:2to2.M_on-shell} directly, recognizing that the non-analyticity arises solely from the phase space factor.
Continuing this to the second sheet via the relation $\rho^{\mathrm{II}} = -\rho$, which is due to the square root branch cut, then we recover directly Eq.~\eqref{eq:resonance.second_sheet} which agrees with Eq.~\eqref{eq:app.2to2.second_sheet} when Eq.~\eqref{eq:2to2.M_on-shell} is substituted.

The analytic continuation for the $1+\Jc\to 2$ amplitude follows the same procedure as for the hadronic amplitude.
Focusing on the case for scalar currents with the hadrons in $S$ wave, the on-shell form Eq.~\eqref{eq:1Jto2.H_on-shell}
has a branch cut due to the hadronic scattering amplitude.
To obtain the resonance form-factors, we fix $Q^2$ to be real and analytically continue only in the $s$-plane.
The transition amplitude has the same boundary condition as Eq.~\eqref{eq:app.boundary}, i.e. $\Hc^{\mathrm{II}}(s_+,Q^2) = \Hc(s_-,Q^2)$.
From Eq.~\eqref{eq:1Jto2.H_on-shell}, we find that the analytic continuation to the unphysical sheet simply requires continuing $\Mc$, which gives Eq.~\eqref{eq:resonance.1Jto2.second_sheet}.
As with the $2\to 2$ amplitude, this can also be seen by using the unitarity relation
\begin{align}
    \mathrm{Im} \, \Hc(s,Q^2) = \Mc^{*}(s) \rho(s) \Hc(s,Q^2) \, ,
\end{align}
followed by writing the imaginary part as the discontinuity and imposing the boundary condition. 
We find that the second sheet $1+\Jc\to 2$ amplitude takes the form
\begin{align}
     \Hc^{\mathrm{II}}(s,Q^2) = \Hc(s,Q^2) - 2i \, \Mc^{\mathrm{II}}(s) \, \rho(s) \Hc(s,Q^2) \, ,
\end{align}
which after substitution of Eq.~\eqref{eq:app.2to2.second_sheet} and using the on-shell form \eqref{eq:1Jto2.H_on-shell}, we recover Eq.~\eqref{eq:resonance.1Jto2.second_sheet}.

In the case of the $2+\Jc\to 2$ amplitude $\Wc$, we have to analytically continue both the initial and final state invariant mass squares $s_i$ and $s_f$, respectively. 
It is sufficient to consider $\Wc_{\df}$ since this is the only contribution which can have both initial and final state resonance poles.
Since both variables are continued, we impose the boundary condition
\begin{align}
\label{eq:app.2Jto2.WII}
\Wc_{\df}^{\mathrm{II},\mathrm{II}}(s_{f,\pm},Q^2,s_{i,\pm}) = \Wc_{\df}(s_{f,\mp},Q^2,s_{i,\mp}) \, ,
\end{align}
which the double superscript indicates both variables are continued to their respective second sheets.
The on-shell representation Eq.~\eqref{eq:Wdf_def} ensures that the imaginary part of $\Wc_{\df}$ takes the form
\begin{align}
\label{eq:app.2Jto2.ImW}
    \mathrm{Im}\,\Wc_{\df}(s_f,Q^2,s_i) 
    & = 
    \Mc^{*}(s_f) \rho(s_f)  \Wc_{\df}(s_f,Q^2,s_i) 
    + 
     \Wc_{\df}^{*}(s_f,Q^2,s_i) \rho(s_i) \Mc(s_i) \nn \\
    & +
    \Mc^{*}(s_f) f(Q^2) \mathrm{Im} \,\Gc(s_f,Q^2,s_i) \, \Mc(s_i) \, ,
\end{align}
which shows there is an additional singular term arising from the triangle function.
This additional term implies that we cannot just continue both the external $\Mc$ functions in Eq.~\eqref{eq:Wdf_def}, but that we also need to continue $\Gc$.

We first write the imaginary part as the difference
\begin{align}
\label{eq:app.2Jto2.ImW_v2}
    2i\,\mathrm{Im} \, \Wc_{\df}(s_f,Q^2,s_i) & = \Wc_{\df}(s_{f,+},Q^2,s_{i,+}) - \Wc_{\df}^{*}(s_{f,+},Q^2,s_{i,+}) \, , \nn \\
    & = \Wc_{\df}(s_{f,+},Q^2,s_{i,+}) - \Wc_{\df}(s_{f,-},Q^2,s_{i,-}) \, ,
\end{align}
where in the second line we used the extension of the Schwartz reflection principle for multivariate functions, the edge-of-the-wedge theorem, to write the conjugated amplitude as a function of variables evaluated on the lower-half plane, i.e. $\Wc_{\df}^{*}(s_{f,+},Q^2,s_{i,+}) = \Wc_{\df}(s_{f,-},Q^2,s_{i,-})$.
Using the Schwartz reflection principle for the scattering amplitudes, Eqs.~\eqref{eq:app.2Jto2.ImW} and \eqref{eq:app.2Jto2.ImW_v2} give us the relation
\begin{align}
    \Wc_{\df}(s_{f,+},Q^2,s_{i,+}) - \Wc_{\df}(s_{f,-},Q^2,s_{i,-})
    & =
    2i\,\Mc(s_{f,-}) \rho(s_{f,+})  \Wc_{\df}(s_{f,+},Q^2,s_{i,+}) \nn \\
    & + 
     2i\,\Wc_{\df}(s_{f,-},Q^2,s_{i,-}) \rho(s_{i,+}) \Mc(s_{i,+}) \nn \\
    & +
    2i\,\Mc(s_{f,-}) f(Q^2) \mathrm{Im} \,\Gc(s_{f,+},Q^2,s_{i,+}) \, \Mc(s_{i,+}) \, .
\end{align}
We now impose the boundary conditions Eqs.~\eqref{eq:app.boundary} and \eqref{eq:app.2Jto2.WII}, again continuing to the upper-half planes, the $2+\Jc\to 2$ amplitude on the second sheets is given by
\begin{align}
    \Wc^{\mathrm{II},\mathrm{II}}_{\df}(s_f,Q^2,s_i) & =  \frac{1}{1+2i\,\Mc(s_{f}) \rho(s_{f})} \Wc_{\df}(s_{f},Q^2,s_{i}) \left[ \,1 -2i\, \rho(s_{i}) \Mc^{\mathrm{II}}(s_{i})\, \right] \nn \\
    & +   \Mc^{\mathrm{II}}(s_{f}) f(Q^2) \, \left[ \, \Gc(s_{f},Q^2,s_{i}) - 2i\, \mathrm{Im} \Gc(s_{f},Q^2,s_{i}) \, \right] \, \Mc^{\mathrm{II}}(s_{i}) \, ,
\end{align}
where we have extended the domain from near the real axis to the entire upper-half complex planes via Cauchy's theorem, as before for the $2\to 2$ amplitude.
We now use the on-shell form Eq.~\eqref{eq:Wdf_def}, as well as \eqref{eq:app.2to2.second_sheet} to construct an on-shell form
\begin{align}
    \Wc^{\mathrm{II},\mathrm{II}}_{\df}(s_f,Q^2,s_i) & = \Mc^{\mathrm{II}}(s_f)  \, \Big\{ \, \Ac_{22}(s_f,Q^2,s_i) + f(Q^2) \left[\, \Gc(s_f,Q^2,s_i) - 2i \, \mathrm{Im} \, \Gc(s_f,Q^2,s_i) \, \right] \, \Big\} \, \Mc^{\mathrm{II}}(s_i) \, ,
\end{align}
where the term in brackets is precisely the triangle function on the unphysical sheets as presented in Eq.~\eqref{eq:resonance.triangle_second_sheet}, using similar arguments as above.
As claimed in Sec.~\ref{sec:resonances}, this gives the analytic continuation of $\Wc_{\df}$ on to the second Riemann sheets in both variables, Eq.~\eqref{eq:resonance.2Jto2.WII}.
We comment that a similar procedure holds for arbitrary currents with the two hadrons in an arbitrary partial wave, noting that the Lorentz structure does not introduce any physical singularities in the $s_i/s_f$ planes.

\end{document}